\documentclass[preprint,12pt]{elsarticle}
\usepackage[utf8]{inputenc}
\usepackage{parskip}
\usepackage[left=2cm,right=2cm,top=2cm,bottom=2 cm]{geometry}
\usepackage{graphicx}
\usepackage{wrapfig}
\usepackage{amsmath}
\usepackage{caption}
\usepackage{comment}
\usepackage{ragged2e}
\usepackage{enumitem}
\usepackage[table]{xcolor}
\usepackage{multirow}
\usepackage{array}
\usepackage{subcaption}
\usepackage[numbers]{natbib}
\usepackage[colorlinks=true, linkcolor=blue, citecolor=blue]{hyperref}
\usepackage[noabbrev, nosort]{cleveref}
\usepackage{amsthm}

\theoremstyle{plain}  
\newtheorem{theorem}{Theorem}

\newtheorem{lemma}{Lemma}[section]

\newtheorem{corollary}{Corollary}

\theoremstyle{definition}  

\newcolumntype{s}{>{\columncolor{yellow!20}}c}
\renewcommand{\thefootnote}{\fnsymbol{footnote}}
\usepackage{amssymb}
\usepackage{mathrsfs}
\usepackage{bm}

\begin{document}

\begin{frontmatter}

\title{Elastic Landau states as martensite nuclei in austenite}

\author[inst1]{Mohd Tahseen}

\affiliation[inst1]{organization={Department of Aerospace Engineering},
            addressline={Indian Institute of Science},
            city={Bangalore},
            postcode={560012}, 
            country={India}}

\author[inst1]{Vivekanand Dabade}

\begin{abstract}

In this article, we investigate the geometric closure of austenite--martensite interfaces. Classical compatibility conditions permit stress-free interfaces between the two phases: the condition $\lambda_2 = 1$ admits planar interfaces, whereas the cofactor conditions additionally admit curved interfaces. However, neither, guarantees the geometric closure of these interfaces. We identify two-parameter families of martensitic deformations that overcome this limitation and enable the formation of closed, stress-free martensite nuclei embedded in an austenite matrix. A main result is the interpretation of the classical deformation modes of simple shear and pure shear as martensitic phase transformations. These correspond to cubic to monoclinic-II and cubic to orthorhombic transformations, respectively. The former naturally satisfies the \emph{extreme compatibility conditions}, which are stronger than the cofactor conditions, and gives rise to \emph{elastic Landau states}, a self-accommodating martensitic nuclei embedded in austenite. We show that these states can be assembled to form arbitrarily shaped, stress-free martensitic islands surrounded by austenite. Such states can arise even in conventional shape memory alloys that do not satisfy supercompatibility, provided hierarchical twinning produces an average deformation that is a simple shear. The theory provides a geometric mechanism for self-accommodation. Its predicted deformation gradients agree with those reconstructed from experimentally observed microstructures. This agreement suggests that these microstructures originate from simple shear.

\end{abstract}


\end{frontmatter}

\section{Introduction}
\label{sec:Intro}

Shape memory alloys constitute an important class of functional materials characterized by two distinctive properties: the shape memory effect and superelasticity. These unique properties have positioned shape memory alloys as promising materials for advanced biomedical technologies~\cite{MIT, ProstheticWrist,ultrasonic_ShapeRecovery}, adaptive robotic systems~\cite{origami_inspired, SMA_ContinuumRobot}, lightweight energy-efficient actuation~\cite{sma_actuators_review,WireFormSMA}, solid-state refrigeration~\cite{Li2016Elastocaloric,Song2026EnhancedElastocaloric}, and energy harvesting~\cite{vijay, kumar2019shape}. At the atomic scale, these macroscopic properties emerge from a reversible martensitic phase transformation. This transformation is a first-order, diffusionless structural transformation between a high-symmetry, high-temperature parent phase, known as austenite, to a low-symmetry, low-temperature product phase, known as martensite. Owing to symmetry breaking during the transformation, martensite can exist in multiple crystallographically equivalent variants. The spatial arrangement of these variants constitutes the martensitic microstructure, whose morphology plays a crucial role in determining the reversibility, hysteresis, and functional lifetime of the material. Consequently, understanding the mechanisms governing the nucleation and development of microstructures remains central to the design of high-performance shape memory alloys. 

A fundamental factor governing the reversibility of transformation is the geometric compatibility between the austenite and martensite phases. Since the transformation is diffusionless, the transformation strain must be accommodated through elastic deformation and the formation of coherent phase interfaces. Poor compatibility generates long-range elastic stresses and promotes plastic deformation, thereby increasing hysteresis and degrading the functional response under repeated cycling. In contrast, enhanced compatibility reduces the elastic energy associated with the transformation, promotes the reversible motion of phase boundaries, and improves the cyclic stability of the material. These considerations have led to extensive theoretical and experimental efforts aimed at understanding and engineering crystallographic compatibility in martensitic transformations~\cite{gu2018phaseengineering, James2019}.

A rigorous framework for analyzing such compatibility is provided by the geometrically nonlinear theory of martensite~\cite{Hane1998Symmetry, JAMES2000197, BALL200461}. Within this theory, coherent interfaces are characterized by rank-one connections between the deformation gradients of the adjoining phases. In general, the austenite–martensite interface is formed between austenite and a twinned martensitic laminate composed of two symmetry related variants. Accordingly, the average deformation gradient of the laminate must differ by a rank-one matrix from the identity deformation gradient of the austenite phase. In general, this compatibility is satisfied only for particular values of the laminate volume fraction. Two particularly important special cases have emerged from this framework. The first is the condition $\lambda_2=1$, under which a coherent interface between austenite and a single martensitic variant~\cite{Combinatorial_search}. The second is a set of three conditions called the cofactor conditions~\cite{JamesZhang2005}, which ensure that the laminate remains compatible with austenite for every volume fraction of its constituent variants. These compatibility conditions have been successfully exploited in the design of shape memory alloys exhibiting reduced hysteresis, enhanced reversibility, and superior functional fatigue resistance~\cite{Zarnetta2010,Song2013,chluba,Tong2019, zirconia_monoclinic}.

An important consequence of the cofactor conditions is that, for laminates of Type I and Type II twins, the elastic energy between austenite and a twinned martensitic laminate can be eliminated entirely~\cite{CHEN20132566}. This is achieved through the formation of triple junctions  (see Figure \ref{fig:Type_I})  for Type I twins and parallel domain walls (see Figure \ref{fig:Type_II}) for Type II twins. We collectively refer to these interface morphologies as \emph{triple clusters}. Triple clusters give rise to a continuous family of macroscopic deformation gradients, parameterized by the twin volume fraction, each compatible with austenite through stress-free interfaces. This flexibility allows the advancing phase interface to accommodate defects, precipitates, non-transforming inclusions, and other microstructural inhomogeneities while preserving compatibility.

In contrast to Type I and Type II twins, the cofactor conditions do not permit the formation of triple clusters for compound twins~\cite{CHEN20132566}. Consequently, compound twin laminates cannot eliminate the elastic energy at the austenite–martensite interface through the same mechanism, see Figure~\ref{fig:compound}. This limitation was recently overcome by Tahseen and Dabade~\cite{Tahseen2026ExtremeCompatibility}, who generalized the concept of triple clusters to compound twins by introducing the \emph{extreme compatibility conditions}, see Figure~\ref{fig:compound_triple}. Under these conditions, laminates of all twin types (Type I, Type II and Compound) admit triple clusters with austenite, providing the broadest known class of stress-free austenite–martensite interfaces.

Despite these advances, existing compatibility theories primarily address local interfaces between austenite and martensite. In contrast, the nucleation of martensite within austenite presents a substantially more restrictive geometric problem. A martensitic nucleus is bounded by a closed interface whose every segment must satisfy compatibility simultaneously. While a compatible habit plane eliminates elastic energy locally, it does not guarantee that compatible interfaces can be assembled into a closed boundary enclosing a martensitic region. When no stress-free nucleus exists, the elastic energy cannot be eliminated and enters the nucleation barrier. Knüpfer, Kohn and Otto quantified this barrier for the cubic-to-tetragonal transformation ~\cite{nucleation_barriers}. They proved that, in a geometrically linear setting, the minimal energy of a martensitic inclusion of volume $V$ scales as $V^{2/3}$ for small inclusions, where interfacial energy dominates, and as $V^{9/11}$ for large inclusions, where interfacial and elastic energy compete. In that case, no single variant is compatible with austenite, and self-accommodation requires all three variants in nearly equal volume fractions, so the elastic energy cannot vanish.

The question of where nucleation starts was addressed by Ball, Koumatos and Seiner~\cite{Ball2013Nucleation}. In CuAlNi, austenite nucleates at the corners of a mechanically stabilized single-variant martensite crystal even when the heat is applied elsewhere. Within nonlinear elasticity, for the lattice parameters and specimen geometries they consider, they showed that the martensite is a local minimizer with respect to localized variations in the interior and on faces and edges, but not at corners, where a localized microstructure can lower the energy. Together, these results indicate that, in the absence of an enclosed stress-free nucleus, nucleation in the bulk is penalized, while nucleation at corners, grain boundary junctions and defects is favored. The cofactor conditions change this picture. Under the cofactor conditions, Type-I twins permit the nucleation of austenite within martensite and vice versa, without incurring elastic energy (\cite{CHEN20132566}, Section 6). However, the nucleation sites and geometries differ: austenite nucleates along a line within a martensite band, whereas martensite nucleates at the boundary of an austenite crystal (Figures 6, 7 \cite{CHEN20132566}). Although these configurations eliminate the elastic penalty associated with the generic case \cite{nucleation_barriers}, they do not yield a martensitic region enclosed entirely by compatible interfaces within a single austenite crystal.

\begin{figure}[hbt!]
    \centering
    \includegraphics[scale=0.37]{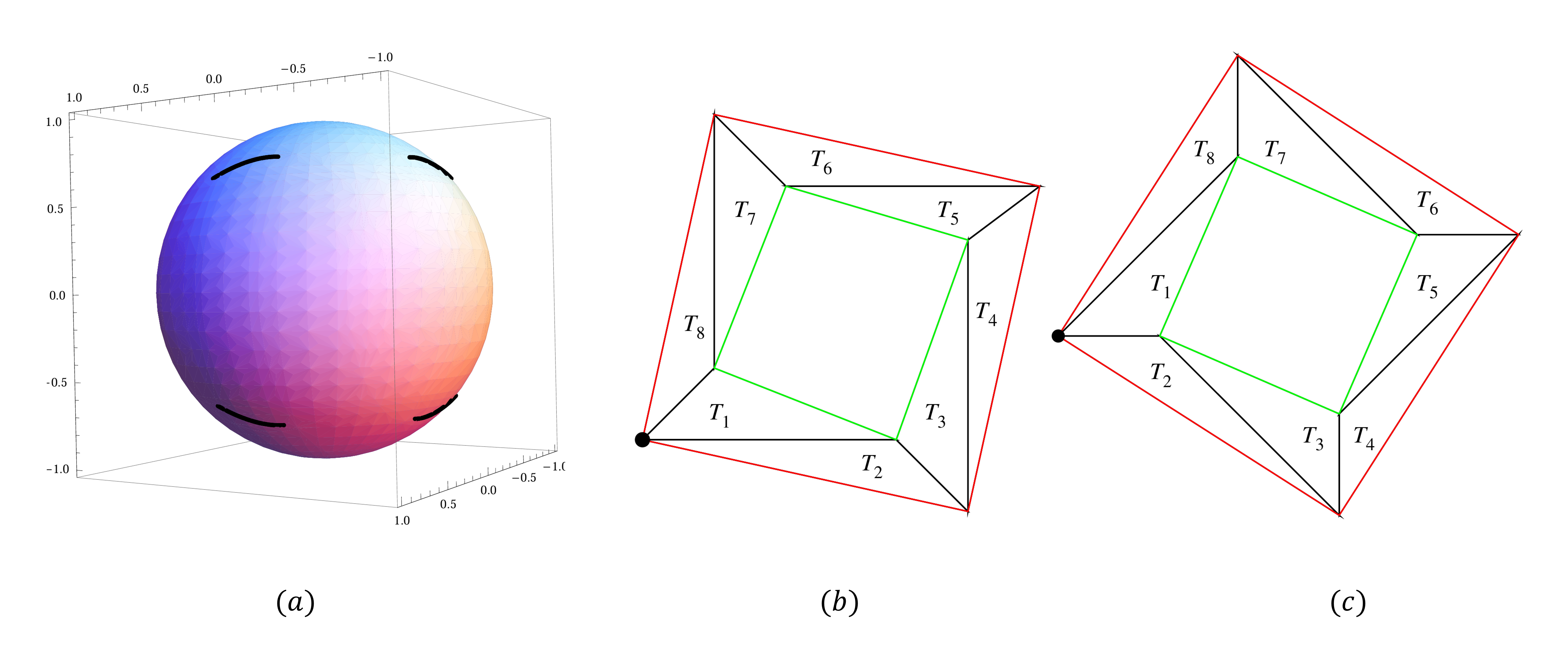}
    \caption{$(a)$ Admissible habit-plane normals obtained by varying the twin volume fraction in a crossing-twin microstructure, reproduced from \cite{non-classical}. The normals span disconnected arcs on the unit sphere, giving rise to smoothly curved, locally compatible austenite--martensite interfaces. The disconnected nature of the admissible habit-plane normals prevents the direct construction of a closed austenite--martensite interface. $(b)$ Undeformed configuration of the piecewise-affine stress-free martensitic inclusion proposed by Conti et al. \cite{Conti2017Piecewise} $(c)$ Corresponding deformed configuration. The martensitic phase occupies the annular region between the inner (green) and outer (red) squares, whereas the austenite phase occupies both the interior of the inner square and the exterior of the outer square.}
    \label{fig:intro}
\end{figure}

A further development in this context was the demonstration that compatible austenite--martensite interfaces can exhibit non-trivial geometry. Ball et al.~\cite{non-classical} showed that continuously varying the twin volume fraction within a crossing-twin microstructure produces a corresponding variation in the habit-plane orientation, resulting in smoothly curved yet locally compatible austenite--martensite interfaces. As illustrated in Figure~\ref{fig:intro}$(a)$, the admissible habit-plane normals span disconnected arcs on a unit sphere, corresponding to the range of interface orientations attainable through variations in the twin volume fraction. While this shows that compatibility is not restricted to planar phase boundaries, whether such interfaces can be assembled into a closed boundary surrounding a martensitic nucleus remains unresolved.

A different approach was later proposed by Conti et al.~\cite{Conti2017Piecewise}, who constructed a two-dimensional piecewise-affine stress-free martensitic inclusion using a rotated--squares construction . The undeformed and deformed configurations are illustrated in Figures~\ref{fig:intro}$(b)$ and \ref{fig:intro}$(c)$, respectively. In the deformed configuration, the martensitic phase occupies the annular region $T_1\cup T_2 \cup \ldots \cup T_8$ between the green inner square and the red outer square, while the regions enclosed by the green square and exterior to the red square remain in the austenite phase. 

While mathematically elegant, this construction possesses two features that limit its interpretation as a physically realistic nucleation mechanism. First, the martensitic variants are generated using the full orthogonal group rather than the proper orthogonal group prescribed by the principles of continuum mechanics~\cite{BALL200461}. As a result, the construction permits orientation-reversing transformations corresponding to interpenetration of matter. Second, the regions enclosed by the green square and exterior to the red outer square represent two austenitic regions that differ by a finite rotation. As the thickness of the annular martensitic region tends to zero, corresponding to the limiting case of a vanishing nucleus, these two rotated austenitic regions remain incompatible. Consequently, the construction does not describe the nucleation of martensite within a single, uniformly oriented austenite crystal.

The main aim of this work is to establish planar closure of austenite–martensite interfaces within the geometrically nonlinear theory of martensite. Specifically, we identify microstructures in which a stress-free martensitic nucleus is completely enclosed by compatible interfaces in the plane, so that nucleation incurs no elastic energy penalty. We begin by reinterpreting the classical simple shear and pure shear deformations of continuum mechanics as martensitic phase transformations. We show that simple shear corresponds to a cubic to monoclinic-II transformation, whereas pure shear corresponds to a cubic to orthorhombic transformation. Notably, the simple shear transformation satisfies the extreme compatibility conditions, allowing the construction of a stress-free martensitic nucleus consisting of four variants enclosed within the austenite matrix. The construction is invariant along the shear direction, so the nucleus is closed in the plane normal to the shear direction and emerges from a line, the common axis of the four variants. Owing to its close analogy with the magnetic Landau state in micromagnetics, which minimizes the magnetostatic energy through flux closure \cite{Dabade2019, HubertSchafer1998}, we refer to this microstructure as the \emph{elastic Landau state}, see Figure~\ref{fig:landau_schematic}. 

This construction naturally generalizes to a two-parameter family of martensitic transformations that admit stress-free martensitic nuclei with planar closure of the austenite-martensite interface. The corresponding microstructure, which we term the \emph{sheared Landau state}, consists of four martensitic variants arising from a cubic to triclinic transformation, see Figure~\ref{fig:sheared_landau_schematic}. Its morphology closely resembles the diamond and parallelogram microstructures observed experimentally~\cite{Saburi1979Crystallographic,Tan1990Observations,Heczko_diamond}. We give a complete crystallographic description of these nuclei by characterizing every interface and junction within the microstructure. Our analysis reveals that triple clusters play an essential role in achieving closure of the austenite–martensite interface in the constructions developed in this paper.

We further demonstrate that sheared Landau states serve as building blocks for constructing stress-free martensitic nuclei of arbitrary shape within austenite, see Figure~\ref{fig:vitali}. Finally, the deformation gradients predicted by our construction agree well with those reconstructed from the experimentally observed microstructures in~\cite{Li2019}, supporting the view that such self-accommodating microstructures can arise from a simple shear mechanism.

\emph{Notation}: $\mathbb{R}^3$ denotes the three-dimensional real vector space.
$\mathbb{R}^{3\times3}_{\mathrm{sym}+}$ denotes the set of all symmetric, positive-definite $3\times3$ matrices.
$\bm{SO(3)}$ is the special orthogonal group, i.e., the set of all $3\times3$ orthogonal matrices with unit determinant.
$\mathbb{S}^2$ denotes the unit sphere in $\mathbb{R}^3$. $\hat{\bm{e}}$ and its superscripts denote the symmetry axes relating pairs of compatible variants.
$\bm{\hat{u}_i}$ and $\bm{{\hat{e}}}_i$ denote the mutually orthonormal eigenvectors and the standard basis of $\mathbb{R}^3$, respectively.
The symbol $\operatorname{cof}\bm{U}$ denotes the cofactor matrix of $\bm{U}$.
Finally, $\bm{\mathscr{P}(a)}$ denotes the point group of the austenite phase, restricted to rotations.

\section{Cofactor conditions, Triple clusters and Extreme compatibility conditions}
\label{sec:tc_cc_ecc}
In this section, we review the role of triple clusters in enabling stress-free interfaces between austenite and martensite within the framework of the geometrically nonlinear theory of martensite. We employ the notation introduced by Tahseen and Dabade \cite{Tahseen2026ExtremeCompatibility}. Only the essential elements of the theory required for the present development are summarized here, a more comprehensive account is provided in \ref{app:glnm}.

Let $\bm{U,V} \in \mathbb{R}^{3\times3}_{sym+}$ be two martensitic variants. The general equation of compatibility between $\bm{U}$ and $\bm{V}$ is written as

\begin{equation}
    \label{twinning}
    \bm{QU}-\bm{V}=\bm{a}\otimes\bm{\hat{n}},
\end{equation}

where $\bm{a}\in\mathbb{R}^3\setminus\bm{\{0\}}$ is the twinning shear, $\bm{\hat{n}\in \mathbb{S}^2}$ gives the normal to the twinning plane and $\bm{Q} \in \bm{SO(3)}$. Since martensitic variants are related by crystallographic symmetry, there exists a $180^\circ$ rotation about an axis $\bm{\hat{e}}$ that maps one variant to the other.

\begin{equation}
\bm{V}=\bm{(-I+2\hat{e}\otimes\hat{e})\:U\:(-I+2\hat{e}\otimes\hat{e})}
    \label{symmetr_related_variants}.
\end{equation}

The twinning equation \eqref{twinning} possesses two distinct solutions, known as the Type-I and Type-II solutions. These are classified according to the orientation of the twinning elements relative to the symmetry axis $\hat{\bm e}$. Type-I twins are characterized by a twinning plane normal parallel to $\hat{\bm e}$, whereas Type-II twins are characterized by a shear direction parallel to $\hat{\bm e}$. In certain cases, each solution satisfies both symmetries simultaneously, giving rise to compound twins. In this, there exist two orthogonal axes $\hat{\bm e}^{\,I}$ and $\hat{\bm e}^{\,II}$ satisfying \eqref{symmetr_related_variants}. Explicit expressions for the twinning elements and the classification of solutions are well known and may be found in \cite{Tahseen2026ExtremeCompatibility,bhattacharya2003microstructure, CHEN20132566}. For completeness, the relevant expressions are summarized below.

\begin{equation}
\begin{aligned}
\text{Type I}\qquad
&\bm{\hat{n}}_{I}=\hat{\bm e},
\qquad
\bm{a}_{I}
=
2\left(
\frac{\bm{U}^{-1}\hat{\bm e}}
     {\left|\bm{U}^{-1}\hat{\bm e}\right|^2}
-\bm{U}\hat{\bm e}
\right),\\[2mm]
\text{Type II}\qquad
&\bm{\hat{n}}_{II}
=
\frac{2}{\rho}\left(
\hat{\bm e}
-\frac{\bm{U}^{2}\hat{\bm e}}
       {\left|\bm{U}\hat{\bm e}\right|^2}
\right),
\qquad
\bm{a}_{II}
=
\rho\:\bm{U}\hat{\bm e},
\end{aligned}
\label{typeI_typeII}
\end{equation}

\begin{equation}
\begin{aligned}
\text{Compound I}\qquad
\bm{\hat{n}}_C^{\,I}=\hat{\bm e}^{\,I},
\qquad
\bm{a}_C^{\,I}
&=
\xi\,\bm{U}\hat{\bm e}^{\,II},
\qquad
\xi
=
2\,
\frac{\hat{\bm e}^{\,II}\cdot \bm{U}^{-2}\hat{\bm e}^{\,I}}
     {\hat{\bm e}^{\,I}\cdot \bm{U}^{-2}\hat{\bm e}^{\,I}},
\\[2mm]
\text{Compound II}\qquad
\bm{\hat{n}}_C^{\,II}=\hat{\bm e}^{\,II},
\qquad
\bm{a}_C^{\,II}
&=
\eta\,\bm{U}\hat{\bm e}^{\,I},
\qquad
\eta
=
-2\,
\frac{\hat{\bm e}^{\,II}\cdot \bm{U}^{2}\hat{\bm e}^{\,I}}
     {\hat{\bm e}^{\,I}\cdot \bm{U}^{2}\hat{\bm e}^{\,I}},
\end{aligned}
\label{compound_twins}
\end{equation}

where $\rho$ in \eqref{typeI_typeII} is chosen so that $|\bm{\hat{n}}_{II}|$ = 1.

A classical austenite-martensite interface arises when a martensitic laminate, formed by mixing two martensitic variants with volume fraction $\mu \in [0,1]$, admits a rank-one connection with austenite. Within the crystallographic theory of martensite, this requirement is equivalent to the existence of a rotation $\bm{R}\in\bm{SO(3)}$, a vector $\bm{b}\in\mathbb{R}^3\setminus\bm{\{0\}}$ and a unit normal $\hat{\bm{m}}\in \mathbb{S}^2$ such that

\begin{equation}
    \label{crystallographic_theory_of_martensite}
    \bm{R}\bigl(\mu \bm{Q}\bm{U} + (1-\mu)\bm{V}\bigr) -\bm{I}
=
\bm{b}\otimes\hat{\bm{m}}.
\end{equation}

In general, \eqref{crystallographic_theory_of_martensite} can have upto eight distinct solutions for a given pair of variants $\bm{U}$ and $\bm{V}$ \cite{bhattacharya2003microstructure}. The cofactor conditions correspond to degeneracy of \eqref{crystallographic_theory_of_martensite}, under which the equation admits a solution for every $\mu \in [0,1]$. Equivalently, they constitute the necessary and sufficient conditions for the existence of a compatible austenite-martensite interface for every laminate volume fraction $\mu \in [0,1]$. These conditions are given by

\begin{equation}
\lambda_2 = 1,\; \text{where} \; \lambda_2 \; \text{is the middle eigenvalue of} \; \bm{U},
\tag{CC1}
\label{CC1}
\end{equation}

\begin{equation}
\bm{a}\cdot \bm{U}\,\mathrm{cof}(\bm{U}^2-\bm{I})\,\bm{\hat{n}}=0,
\tag{CC2}
\label{CC2}
\end{equation}

\begin{equation}
\operatorname{tr}(\bm{U}^2)-\det(\bm{U}^2)
-\frac{|\bm{a}|^2}{4}-2\ge 0.
\tag{CC3}
\label{CC3}
\end{equation}

Consequently, a compatible austenite-martensite interface exists for all laminate volume fractions, rather than a single volume fraction. The resulting increase in interfacial freedom is known to improve transformation reversibility, reduce thermal hysteresis, and has been successfully exploited in alloy design \cite{Song2013,chluba,Tong2019}.

\begin{figure}[!hbt]

    \centering
    \includegraphics[scale=0.27]{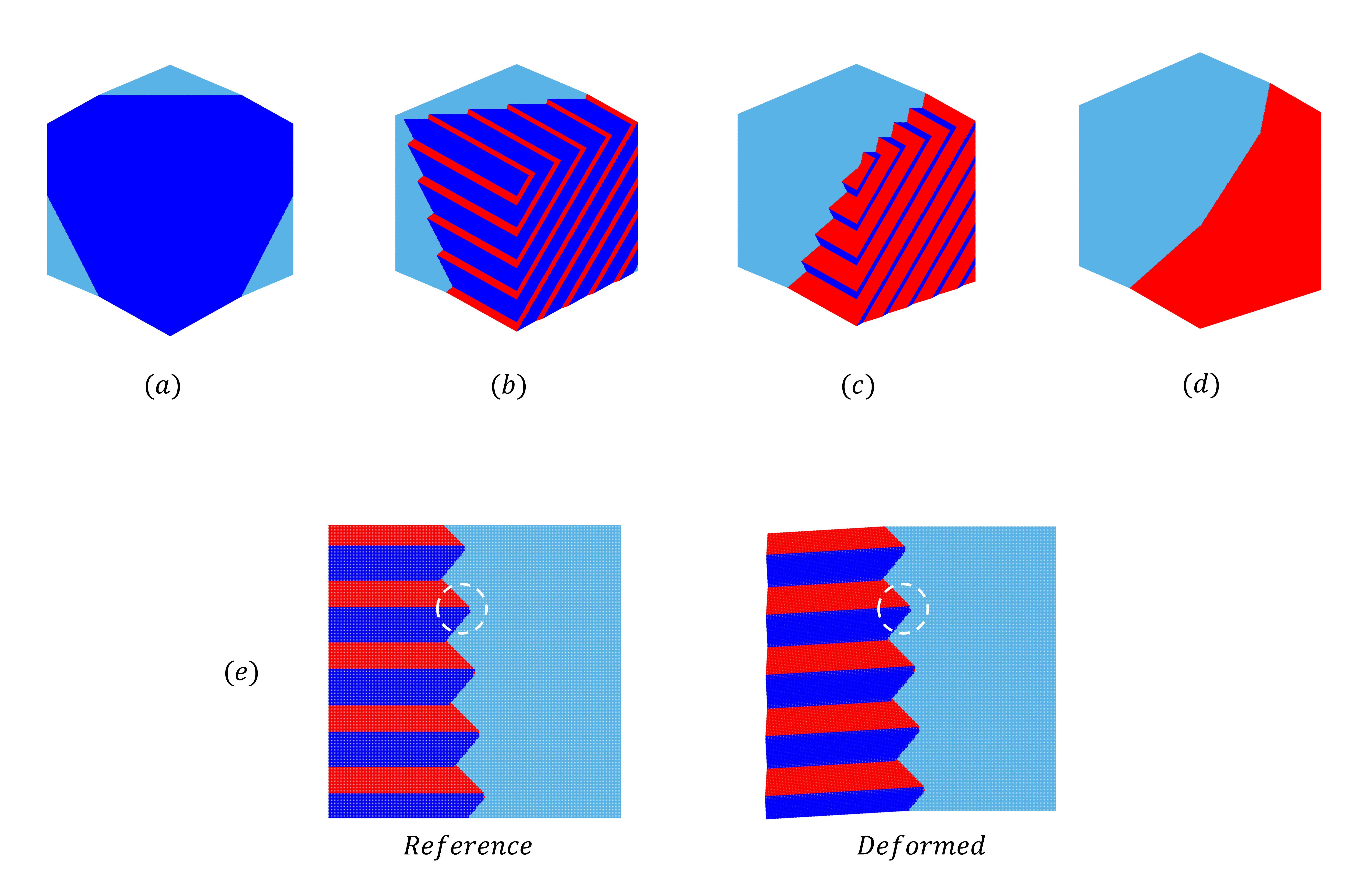}

\caption{Numerically generated zero-elastic-energy microstructures enabled by the cofactor conditions for Type-I twins. Red and blue denote the two martensitic variants forming a Type-I interface, while light blue denotes austenite. Panels $(a)-(d)$ show the microstructures corresponding to volume fractions $\mu=0,\;0.25,\;0.75$ and $1$, respectively. Panels $(e)$ shows cross-sectional views in the reference and deformed configurations. The dashed white circular region indicates a triple cluster of first kind formed by the two martensitic variants and austenite, with a volume fraction $\mu=0.5.$}
\label{fig:Type_I}
\end{figure}

\begin{figure}[!hbt]

    \centering
\includegraphics[scale=0.3]{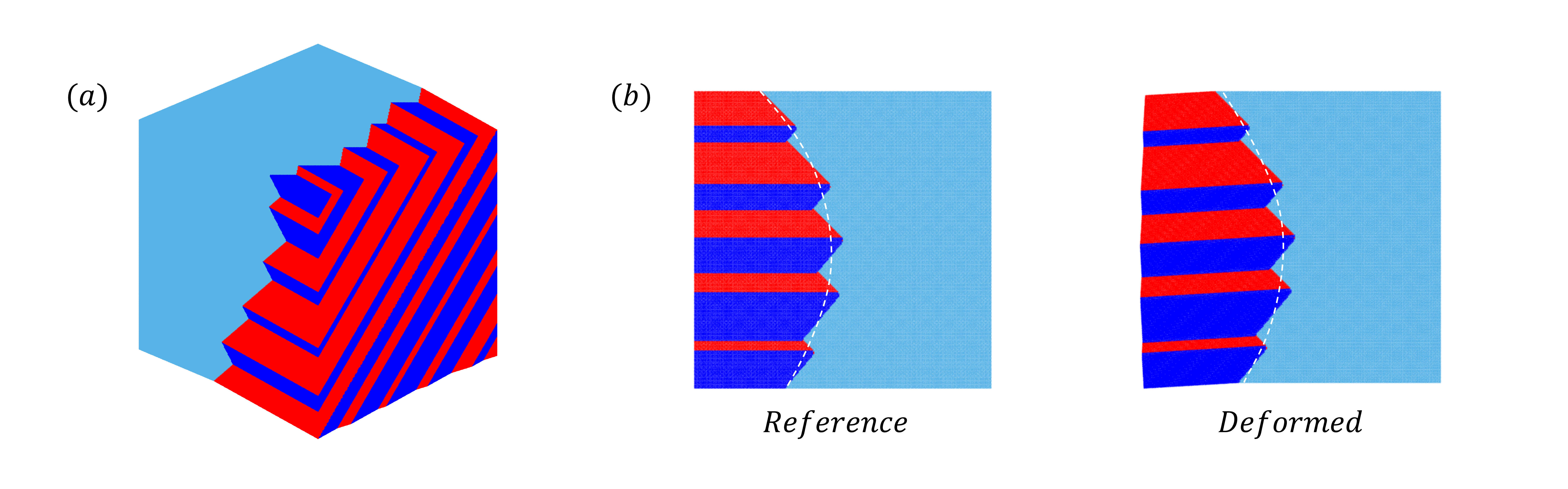}

\caption{A macroscopically curved austenite-martensite interface, generated by a spatially varying volume fraction of variants along the Type-I twin normal. Panels $(b)$ shows cross-sectional views in the reference and deformed configurations respectively. The dashed white line traces the macroscopically curved austenite–martensite interface.}
\label{fig:Type_I_variable}
\end{figure}

An important consequence of the cofactor conditions is the emergence of special microstructures involving austenite and two martensitic variants. Chen et al. \cite{CHEN20132566} showed that whenever the cofactor conditions are satisfied for Type-I and Type-II twins, the elastic energy at the interface can be completely eliminated. For Type-I twins, this is enabled by the meeting of two austenite--single-variant martensite interfaces with the twin interface, at a common three-fold junction, see Figures~\ref{fig:Type_I}$(e)$ and \ref{fig:Type_I}$(f)$. The compatibility relations that define this junction are

\begin{equation}
    \label{type_I_cluster}
    \begin{aligned}
        \bm{Q_u^IU}-\bm{I}&= \alpha \: \bm{a^I}\otimes\bm{\hat{n}^I_u},\\
        \bm{Q_v^IV}-\bm{I}&= \beta \: \bm{a^I}\otimes\bm{\hat{n}^I_v},\\
        \bm{Q_u^IU}-\bm{Q_v^IV}&= \gamma \:\bm{a^I}\otimes\bm{\hat{m}},
    \end{aligned}
\end{equation}

where $\bm{\hat{m}}$ is a unit normal along $(\alpha\bm{\hat{n}^{I}_u} - \beta\bm{\hat{n}_v^{I}})$ and $\alpha,\beta,\gamma$ are scalars. Microstructures of this kind are characterized by a common shear direction $\bm{a^I}$, with the interface normals lying in a common plane. We refer to such morphologies as \emph{triple clusters of first kind}. When assembled with a spatially varying volume fraction, these clusters generate macroscopically curved austenite–martensite interfaces, see Figure~\ref{fig:Type_I_variable}.

\begin{figure}[!hbt]

    \centering
    \includegraphics[scale=0.27]{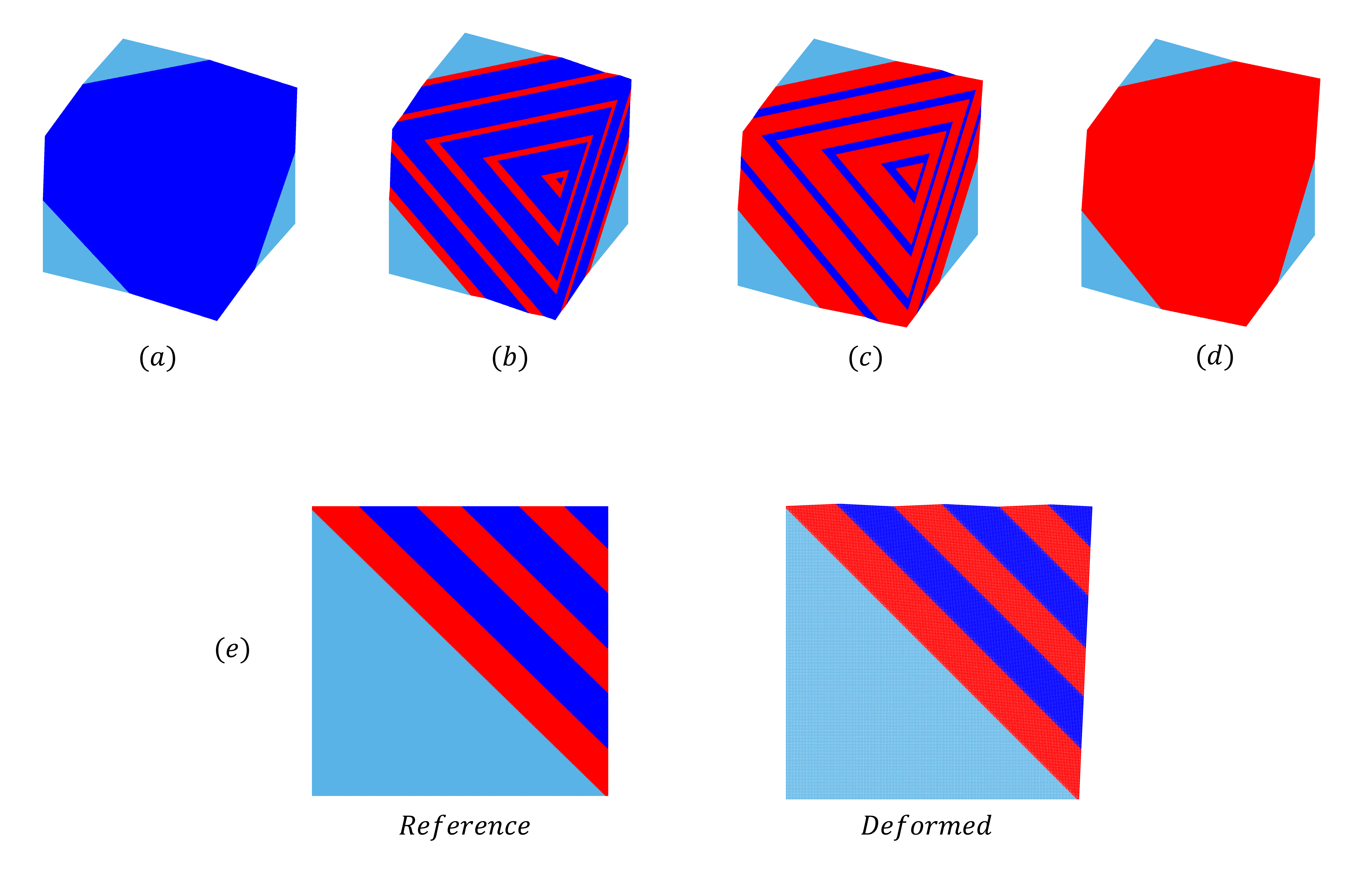}

\caption{Numerically generated zero-elastic-energy microstructures enabled by the cofactor conditions for Type-II twins. Red and blue denote the two martensitic variants forming a Type-II interface, while light blue denotes austenite. Panels $(a)-(d)$ show the microstructures corresponding to volume fractions $\mu=0,\;0.25,\;0.75$ and $1$, respectively. Panels $(e)$ shows cross-sectional views in the reference and deformed configurations. The layered configuration formed by the two martensitic variants and austenite constitutes a triple cluster of second kind.}
\label{fig:Type_II}
\end{figure}

\begin{figure}[!hbt]

    \centering
\includegraphics[scale=0.43]{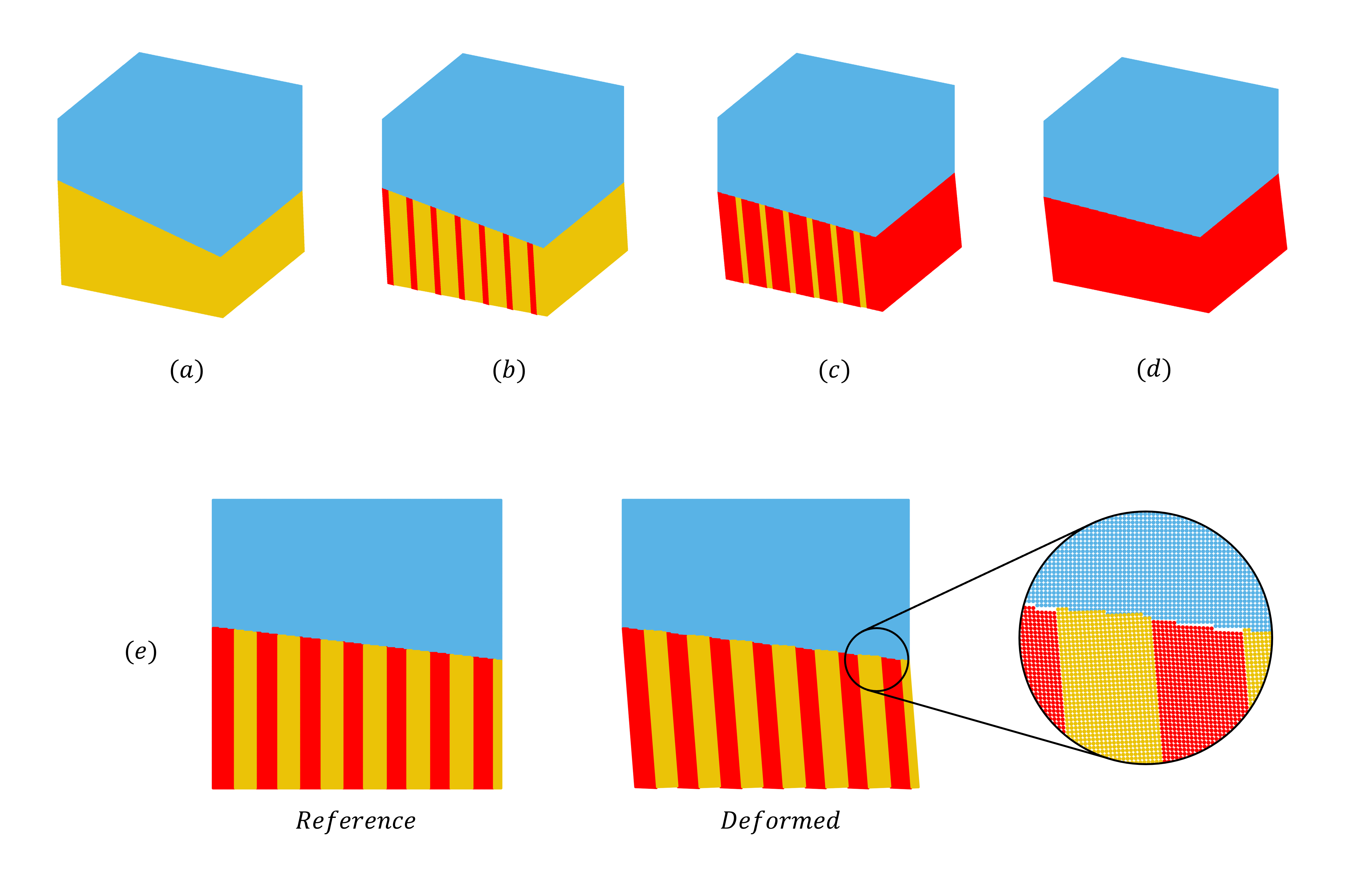}

\caption{Numerically generated microstructures enabled by the cofactor conditions for compound twins. Red and yellow denote the two martensitic variants, forming compound twins while light blue denotes austenite. Panels $(a)-(d)$ show the microstructures corresponding to volume fractions $\mu=0,\;0.25,\;0.75$ and $1$, respectively. Panels $(e)$ shows cross-sectional views in the reference and deformed configurations. The inset in $(e)$ provides a magnified view of the austenite–martensite interface. Unlike Type-I and Type-II twins satisfying the cofactor conditions, compound twins are unable to form compatible triple clusters with austenite. This incompatibility gives rise to elastic stresses near the austenite–martensite interface.}
\label{fig:compound}
\end{figure}

In contrast to Type-I twins, Type-II twins eliminate the elastic energy at the phase interface by forming austenite-–single-variant martensite interfaces that are parallel to the twin interface, resulting in a layered configuration, see Figures~\ref{fig:Type_II}$(e)$ and \ref{fig:Type_II}$(f)$. The compatibility relations defining this configuration are

\begin{equation}
    \label{type_II_cluster}
    \begin{aligned}
        \bm{Q_u^{II}U}-\bm{I}&=  \: \bm{a^{II}_u}\otimes\bm{\hat{n}^{II}},\\
        \bm{Q_v^{II}V}-\bm{I}&=  \: \bm{a_v^{II}}\otimes\bm{\hat{n}^{II}},\\
        \bm{Q_u^{II}U}-\bm{Q_v^{II}V}&= \:\bm{a^{II}}\otimes\bm{\hat{n}^{II}},
    \end{aligned}
\end{equation}

where $\bm{a^{II}}=(\bm{a_u^{II}} - \bm{a_v^{II}})$. Microstructures of this kind are characterized by a common interface normal $\bm{\hat{n}^{II}}$, with the shear vectors lying in a common plane. We refer to such morphologies as \emph{triple clusters of second kind}.

The notion of triple clusters was extended to compound twins by Tahseen and Dabade \cite{Tahseen2026ExtremeCompatibility}. This extension gives rise to the concept of \emph{extreme compatibility}, wherein all Type-I/Type-II and compound twin pairs can eliminate transition layers at austenite-martensite interfaces through the formation of triple clusters. We recall their main result below.

\begin{theorem}
\label{thm:EC}
Let $\bm{U,V}\in \mathbb{R}^{3\times 3}_{+\mathrm{sym}}$ be two variants of martensite such that $\bm{U}$ and $\bm{V}$ form compound twins with symmetry axes $\bm{\hat{e}^{I}}$ and $\bm{\hat{e}^{II}}$. Let $\lambda_1<1$, $\lambda_2=1$, $\lambda_3>1$ be the eigenvalues of $\bm{U}$ and $\bm{\hat u_1}$, $\bm{\hat u_2}$, $\bm{\hat u_3}$ be the corresponding eigenvectors such that $\bm{(\hat u_2,\hat e^{I},\hat e^{II})}$ forms an orthonormal basis. If there exists a triple cluster between $\bm{U}$, $\bm{V}$ and $\bm{I}$, then at least one of the following must hold:

\begin{align}
\lambda_1\sqrt{\lambda_3^2-1}\,
(\bm{\hat u}_3\!\cdot\!\bm{\hat e}^{I})
&=
-\lambda_3\sqrt{1-\lambda_1^2}\,
(\bm{\hat u}_1\!\cdot\!\bm{\hat e}^{I}),
\tag{EC1}
\label{EC1}\\
\lambda_1\sqrt{\lambda_3^2-1}\,
(\bm{\hat u}_1\!\cdot\!\bm{\hat e}^{I})
&=
-\lambda_3\sqrt{1-\lambda_1^2}\,
(\bm{\hat u}_3\!\cdot\!\bm{\hat e}^{I}),
\tag{EC2}
\label{EC2}\\
\sqrt{1-\lambda_1^2}\,
(\bm{\hat u}_3\!\cdot\!\bm{\hat e}^{I})
&=
- \sqrt{\lambda_3^2-1}\,
(\bm{\hat u}_1\!\cdot\!\bm{\hat e}^{I}),
\tag{EC3}
\label{EC3}\\
\sqrt{1-\lambda_1^2}\,
(\bm{\hat u}_1\!\cdot\!\bm{\hat e}^{I})
&=
-\sqrt{\lambda_3^2-1}\,
(\bm{\hat u}_3\!\cdot\!\bm{\hat e}^{I}).
\tag{EC4}
\label{EC4}
\end{align}
\end{theorem}

Several remarks are in order to better understand the theorem above. We present them below.

\begin{figure}[!hbt]

    \centering
    \includegraphics[scale=0.35]{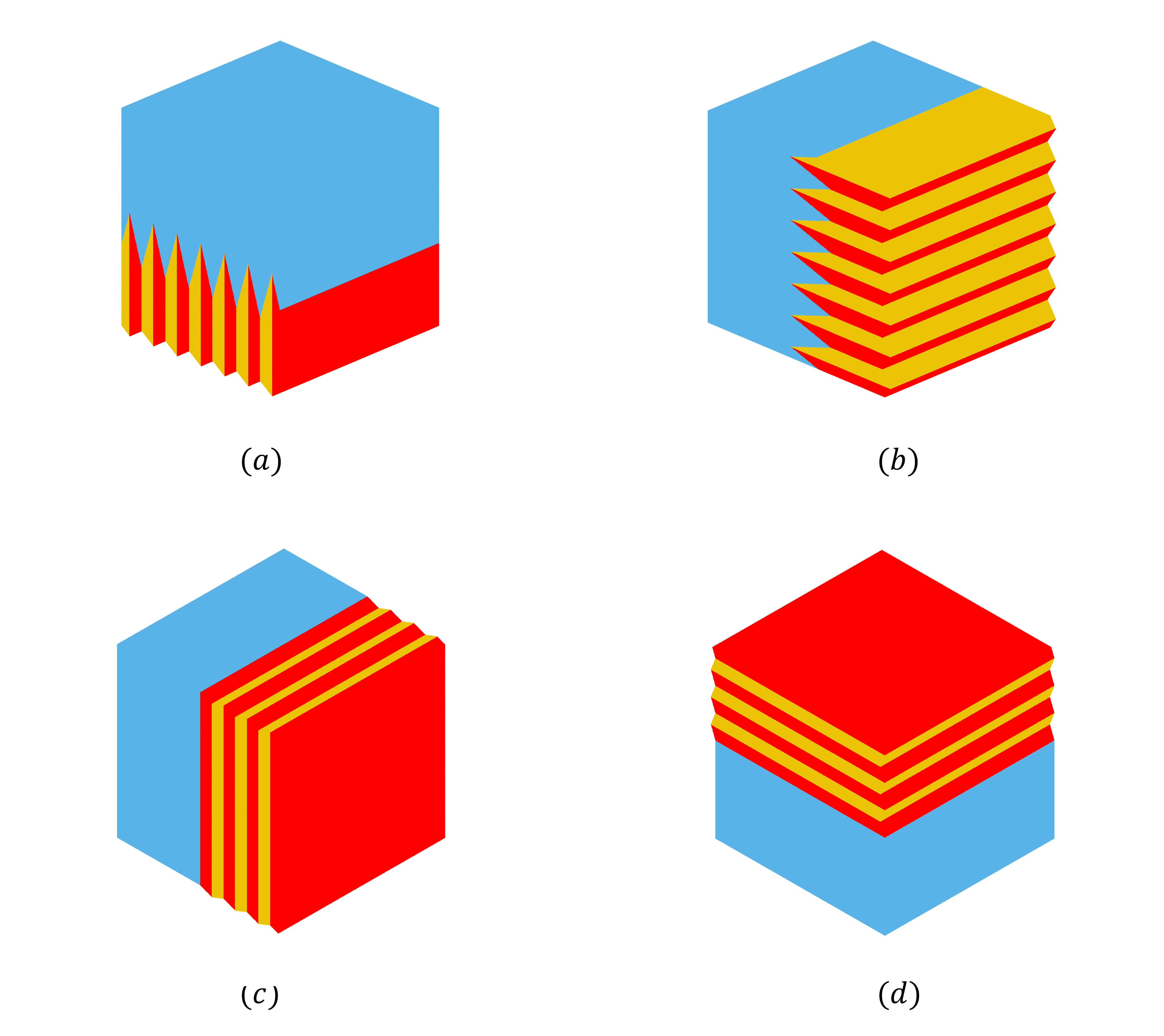}

\caption{Numerically generated microstructures enabled by the extreme compatibility conditions for compound twins. Red and yellow denote the two martensitic variants forming a compound twin, while light blue denotes austenite. Panels $(a)$ and $(b)$ show two distinct triple clusters of first kind satisfying \eqref{EC1} and \eqref{EC2} simultaneously. Panels $(c)$ and $(d)$ show two distinct triple clusters of second kind satisfying \eqref{EC3} and \eqref{EC4} simultaneously.}
\label{fig:compound_triple}
\end{figure}

\begin{enumerate}[label=\arabic*)]
\item Conditions \eqref{EC1} and \eqref{EC2} are associated with triple clusters of the first kind, while conditions \eqref{EC3} and \eqref{EC4} are associated with triple clusters of the second kind between $\bm{U,V}$ and $\bm{I}$.
\item If \eqref{EC1} and \eqref{EC2} are simultaneously satisfied, two distinct triple clusters of first kind exist between $\bm{U,V}$ and $\bm{I}$, corresponding to two distinct sets of compatibility equations of the form \eqref{type_I_cluster}, see Figures~\ref{fig:compound_triple}$(a)$ and \ref{fig:compound_triple}$(b)$. Similarly, if \eqref{EC3} and \eqref{EC4} are simultaneously satisfied, two distinct triple clusters of second kind exist between $\bm{U,V}$ and $\bm{I}$, corresponding to two distinct sets of compatibility equations of the form \eqref{type_II_cluster}, see Figures~\ref{fig:compound_triple}$(c)$ and \ref{fig:compound_triple}$(d)$. Furthermore, if either of the two cases above holds, the matrices $\bm{U}$ and $\bm{V}$ must commute.
\item Simultaneous satisfaction of \eqref{EC1} and \eqref{EC4} is not possible, as it leads to the degeneracy of transformation. Similarly, simultaneous satisfaction of \eqref{EC2} and \eqref{EC3} is not possible. Therefore, whenever two distinct triple clusters form between $\bm{U,V}$ and $\bm{I}$, then they must necessarily be of the same kind.
\item It is possible for a triple cluster to be simultaneously of the first and second kind. It happens if either \eqref{EC1} and \eqref{EC3} or \eqref{EC2} and \eqref{EC4} hold simultaneously. The resulting cluster satisfies the defining properties of both kinds of clusters, i.e. the shear vectors are parallel, and the interface normals coincide, see the pairs $\bm{(F_3,F_4)}$ and $\bm{(F_1,F_2)}$ in Figures~\ref{fig:landau_schematic}(b) and Figures~\ref{fig:landau_schematic}(c) respectively. A necessary condition for the existence of such a triple cluster is that $\det(\bm{U})=1$. In this case, there exists a unique triple cluster between $\bm{U}$, $\bm{V}$, and $\bm{I}$.
\end{enumerate}

The last case is particularly significant for the present work, as the condition $\det(\bm{U})=1$ is also a necessary condition for self-accommodation \cite{bhattacharya2003microstructure,Bhattacharya1992}, the concept which lies at the heart of this paper. We explore this connection in the next section.

Before concluding this section, we recall a useful observation due to Della Porta~(\emph{Proposition 3.1, \cite{DELLAPORTA201927}}). The compatibility relations defining triple clusters of the first and second kind, given by \eqref{type_I_cluster} and \eqref{type_II_cluster} respectively, are invariant under rigid rotations. Specifically, if $(\bm U,\bm V)$ forms a triple cluster with austenite, then for any $\bm {R\in SO(3)}$, the transformed variants,
$\bm R\bm U\bm R^T$ and $\bm R\bm V\bm R^T$, also form a triple cluster of the same kind with austenite. The corresponding compatibility relations are obtained by transforming every rotation matrix as 
\[\bm Q\mapsto \bm R\bm Q\bm R^T,\]
every shear vector as 
\[\bm a\mapsto \bm R\bm a,\]
and every interface normal as 
\[\hat{\bm n}\mapsto \bm R\hat{\bm n}.
\]
Consequently, the transformed cluster satisfies compatibility equations of the same form as the original cluster. In particular, by choosing $\bm R\in \bm{\mathscr{P}{(a)}}$, all symmetry-related variant pairs that admit triple clusters of the same kind can be identified starting from a given pair. This procedure allows us to systematically generate all triple clusters associated with a given transformation.

\section{Martensitic transformations as shear deformations}
\label{sec:shear}
In this section, we establish a connection between shear transformations and the crystallography of martensitic phase transformations. In particular, we show that a simple shear transformation can be interpreted as a cubic to monoclinic-II transformation. Likewise, a pure shear transformation can be interpreted as a cubic to orthorhombic transformation. In the former case, the extreme compatibility conditions are automatically satisfied, giving rise to self-accommodated microstructures, which we refer to as \emph{elastic Landau states}. These elastic Landau states enable the complete closure of the austenite–martensite interface.

\subsection{Simple shear as martensitic transformation}
\label{sec:shear_simple}

Consider a simple shear transformation given by 
\begin{equation*}
    \bm{F}=\bm{I}+ t\;\hat{\bm{e}}_1 \otimes \hat{\bm{e}}_2=\begin{bmatrix}
        1 & t & 0\\
        0 & 1 & 0\\
        0 & 0 & 1\\
    \end{bmatrix},
\end{equation*}

where $t\neq0$ is the magnitude of shear. Then, the transformation can be rewritten as
\begin{equation}
\label{simple_shear}
    \bm{F}-\bm{I} \:=\: t\;\hat{\bm{e}}_1 \otimes\hat{\bm{e}}_2.
\end{equation}

Equation \eqref{simple_shear} implies that there is a rank-1 connection between $\bm{F}$ and $\bm{I}$. We interpret $\bm{F}$ as the deformation gradient of a martensitic variant and determine the corresponding stretch tensor $\bm{U}$ through the polar decomposition of $\bm{F}$.

\begin{equation}
\bm{U}
=
\sqrt{\bm{F}^{T}\bm{F}}
=
\begin{bmatrix}
\alpha(t) & \beta(t) & 0 \\
\beta(t) & \delta(t) & 0 \\
0 & 0 & 1
\end{bmatrix},
\label{polar_t}
\end{equation}

where,

\begin{equation*}
\alpha(t) = \frac{2}{\sqrt{4+t^2}}, \quad
\beta(t) = \frac{t}{\sqrt{4+t^2}}, \quad
\delta(t) = \frac{2+t^2}{\sqrt{4+t^2}}.
\end{equation*}

The matrix in equation \eqref{polar_t} is of the same form as the stretch tensors of variants for cubic to monoclinic-II transformations, see \eqref{app:cubicmonoII}. In particular, $\bm{U}$ in \eqref{polar_t}, corresponds to $\bm{U_8}$ in \eqref{eq:22} with $d=1$. The eigenvalues of $\bm{U}$ are

\begin{equation*}
    1, \frac{1}{2} \left(\sqrt{4+t^2}\pm t\right).
\end{equation*}

Without loss of generality, we assume $t>0$, which fixes the choice of sign and yields

\begin{equation*}
\lambda_1=\frac{\sqrt{4+t^2}-t}{2} < \lambda_2=1<
\lambda_3=\frac{\sqrt{4+t^2}+t}{2}.
\end{equation*}

The transformation is volume preserving, since $\det(\bm U)$ = $\lambda_1\lambda_2\lambda_3=1$. The corresponding normalized eigenvectors are

\begin{align*}
\notag
    \bm{\hat{u}_1} &= \frac{1}{\sqrt{\left(\sqrt{4+t^2}+t\right)^2+4}}\left\{-\sqrt{4+t^2}-t,\;2,\;0\right\},\quad
    \bm{\hat{u}_2} = \{0,0,1\},\\
     \bm{\hat{u}_3} &= \frac{1}
     {\sqrt{\left(t-\sqrt{4+t^2}\right)^2+4}}
     \left\{\sqrt{4+t^2}-t,\;2,\;0 \right\}.
\end{align*}

Consider a symmetry related variant of $\bm U$,
\begin{equation*}
    \begin{bmatrix}
\alpha(t) & -\beta(t) & 0 \\
-\beta(t) & \delta(t) & 0 \\
0 & 0 & 1
\end{bmatrix},
\end{equation*}

corresponding to $\bm{U_{11}}$ of \eqref{eq:22}. The variants $\bm {U_8}$ and $\bm {U_{11}}$ are related by $180^\circ$ rotations about either of the axes $\bm{\hat{e}_1}$ and $\bm{\hat{e}_2}$. $\bm {U_8}$ and $\bm {U_{11}}$ therefore form a compound twin pair%
\begingroup
\renewcommand{\thefootnote}{\fnsymbol{footnote}}
\footnote{The compound twins in this case are termed \emph{conventional compound twins}, since both rotations relating the corresponding variants belong to the point group of the austenite.}
\endgroup .

Choosing $\bm U=\bm {U_8}$, $\bm V=\bm {U_{11}}$, $\bm{\hat e^{I}}=\bm{\hat{e}_1}$, and $\bm{\hat e^{II}}=\bm{\hat{e}_2}$, together with $\bm{\hat u_1}$ and $\bm{\hat u_3}$ as defined above, we find that the hypotheses of \autoref{thm:EC} are satisfied. Furthermore, conditions \eqref{EC2} and \eqref{EC4} are satisfied simultaneously. Therefore, $\bm{U_{8}}$ and $\bm{U_{11}}$ form a triple cluster with austenite that is simultaneously of the first and second kind (\emph{see remark 4 following \autoref{thm:EC}}).

Consider another variant,
\begin{equation*}
    \begin{bmatrix}
\alpha(t) & 0 & \beta(t)\\
0 & 1 & 0\\
\beta(t) & 0 & \delta(t)\\
\end{bmatrix},
\end{equation*}

corresponding to $\bm{U_{4}}$ of \eqref{eq:22}. The variants $\bm {U_8}$ and $\bm {U_{4}}$ are related by a $180^\circ$ rotation about the axis $\frac{1}{\sqrt{2}}(\bm{\hat{e}_2}-\bm{\hat{e}_3})$. $\bm {U_8}$ and $\bm {U_{4}}$ therefore form a Type-I/II twin pair.

Choosing $\bm U=\bm U_8$, $\bm{\hat{e}}=\frac{1}{\sqrt{2}}(\bm{\hat{e}_2}-\bm{\hat{e}_3})$, and Type-I solution for $\bm{a}$ and $\bm{\hat{n}}$ from \eqref{typeI_typeII}, we find that the cofactor conditions are satisfied. This enables $\bm U_8$ and $\bm U_4$ to form a triple cluster of first kind with austenite.

Now consider the variant,
\begin{equation*}
    \begin{bmatrix}
    1 & 0 & 0\\
0 & \delta(t) & \beta(t)\\
0 & \beta(t) & \alpha(t)\\
\end{bmatrix},
\end{equation*}

corresponding to $\bm{U_{5}}$ of \eqref{eq:22}. The variants $\bm {U_8}$ and $\bm {U_{5}}$ are related by a $180^\circ$ rotation about the axis $\frac{1}{\sqrt{2}}(\bm{\hat{e}_1}-\bm{\hat{e}_3})$ and therefore form a Type-I/II twin pair.

Choosing $\bm U=\bm U_8$, $\bm{\hat{e}}=\frac{1}{\sqrt{2}}(\bm{\hat{e}_1}-\bm{\hat{e}_3})$ and Type-II solution for $\bm{a}$ and $\bm{\hat{n}}$ from \eqref{typeI_typeII}, we verify that the cofactor conditions are satisfied for this choice. Therefore, $\bm U_8$ and $\bm U_5$ form a triple cluster of second kind with austenite.

We conclude that, in a simple shear transformation, triple clusters are enabled for Type-I, Type-II, and compound twins simultaneously. This is precisely the feature that characterizes extreme compatibility. Hence, a simple shear transformation satisfies the extreme compatibility conditions.

The variants considered above were specifically chosen to establish the extreme compatibility conditions for the simple shear transformation. An immediate consequence of this enhanced compatibility is that it permits the formation of geometrically closed austenite–martensite interfaces. Such a construction can be achieved using a different set of variants belonging to the same transformation, as demonstrated by the following simple geometric construction.

Let $\bm{F}=\bm{I}+ t\;\hat{\bm{e}}_1 \otimes \hat{\bm{e}}_2$ as in \eqref{simple_shear}. Define rotations, 
\begin{equation*}
    \bm{R_1}= \bm {I},
    \quad
    \bm{R_2} = R\;[\pi,\bm{\hat{{e}}_1}],
    \quad
    \bm{R_3} = R\;[\pi,\frac{1}{\sqrt{2}}(\bm{\hat{e}_2+\hat{e}_3})],
    \quad
    \bm{R_4} = R\;[\pi,\frac{1}{\sqrt{2}}(\bm{\hat{e}_2-\hat{e}_3})],
\end{equation*}

where $R\;[\theta,\bm{\hat{n}}]$ denotes the rotation matrix corresponding to a rotation of angle $\theta$ about the unit vector $\bm{\hat{n}}$.

Defining $\bm{F_i}=\bm{R_i^TFR_i}$ and applying the transformations $\bm {R_i^T(.)R_i}$, $i=1,2,3,4,$ to equation \eqref{simple_shear}, we get

\begin{equation}\label{eq:landau}
    \begin{aligned}
        \bm{F_1-I} &= t\;\bm{\hat{e}_1 \otimes\hat{e}_2} \\
        \bm{F_2-I} &= t\;\bm{\hat{e}_1 \otimes-\hat{e}_2} \\
        \bm{F_3-I} &= t\;\bm{\hat{e}_1 \otimes\hat{e}_3} \\
        \bm{F_4-I} &= t\;\bm{\hat{e}_1 \otimes-\hat{e}_3}.
    \end{aligned}
\end{equation}

The deformation gradients $\bm {F_i}$ in \eqref{eq:landau} are pairwise rank one connected and are each rank one connected to $\bm I$. Applying the polar decomposition to the deformation gradients, $\bm {F_i}$, we obtain the associated transformation stretch tensors. Specifically, $\bm {F_1},\bm {F_{2}},\bm {F_3}$ and $\bm {F_4}$ correspond to the martensitic variants $\bm {U_8},\bm {U_{11}},\bm {U_{3}}$ and $\bm {U_{4}}$ of \eqref{eq:22} respectively.

\begin{figure}[hbt!]
    \centering
    \includegraphics[scale=0.40]{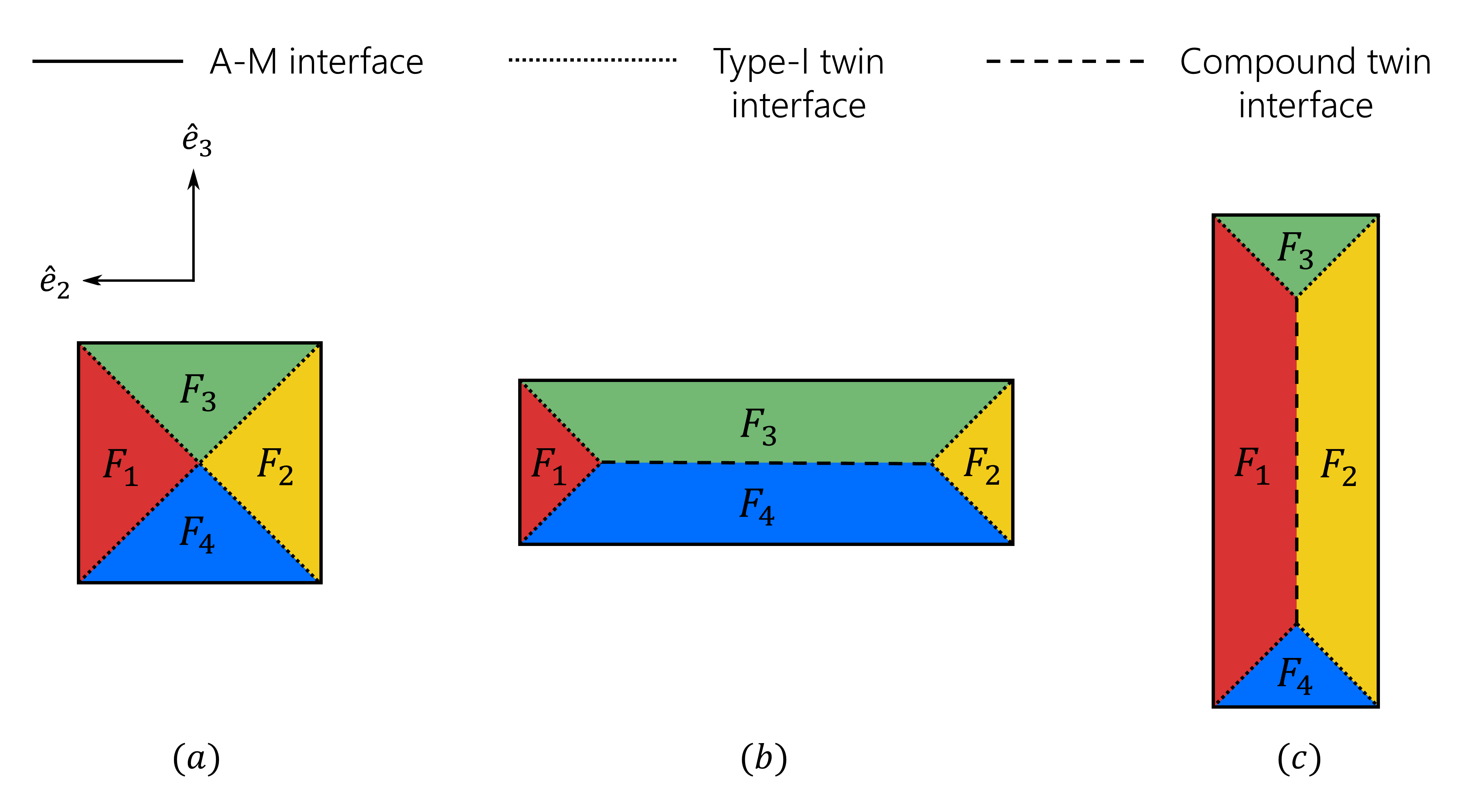}
    \caption{$(a)$ An elastic Landau state comprised of four variants, forming a self-accommodating nucleus of martensite. The variant pairs (red, yellow) and (blue, green) form compound twins, all other variant pairs form Type-I twins $(b),(c)$ growth of martensitic nucleus into the austenite matrix by forming a compound twin interface between opposite pair of variants. Each corner constitutes a triple cluster of the first kind, while the layered configuration of compound twins with austenite in $(b)$ and $(c)$ constitutes triple clusters that are simultaneously of the first and second kinds.}
    \label{fig:landau_schematic}
\end{figure}

The gradients $\bm {F_1},\bm {F_{2}},\bm {F_3}$ and $\bm {F_4}$ give rise to a stress-free self-accommodated nucleus of martensite embedded in austenite, as shown in Figure~\ref{fig:landau_schematic}. In the absence of a more precise term, we refer to this nucleus as the \emph{elastic Landau state}, by analogy with the Landau state in micromagnetics, where flux-closure domain structures emerge to minimize the demagnetization energy. Likewise, the present microstructure adopts a self-accommodated arrangement of martensitic variants that eliminates the elastic strain energy. 

The pairwise rank-one connections between the deformation gradients constituting the \emph{elastic Landau state} can be written as

\begin{equation}
\begin{aligned}
\bm{F_1-F_2} &= 2t\;\bm{\hat{e}_1 \otimes \hat{e}_2},\\
\bm{F_1-F_3} &= \sqrt{2}t\;\bm{\hat{e}_1 \otimes} (\frac{\bm{\hat{e}_2 - \hat{e}_3}}{\sqrt{2}}),\\
\bm{F_1-F_4} &= \sqrt{2}t\;\bm{\hat{e}_1 \otimes} (\frac{\bm{\hat{e}_2 + \hat{e}_3}}{\sqrt{2}}),\\
\bm{F_2-F_3} &= \sqrt{2}t\;\bm{\hat{e}_1 \otimes} -(\frac{\bm{\hat{e}_2 + \hat{e}_3}}{\sqrt{2}}),\\
\bm{F_2-F_4} &= \sqrt{2}t\;\bm{\hat{e}_1 \otimes} -(\frac{\bm{\hat{e}_2 - \hat{e}_3}}{\sqrt{2}}),\\
\bm{F_3-F_4} &= 2t\;\bm{\hat{e}_1 \otimes \hat{e}_3}.
\end{aligned}
\end{equation}

For the variant pair $\bm{(F_1,F_2)}$, the underlying stretch-tensor pair is $\bm{(U_8,U_{11})}$. As shown earlier, these two stretch tensors are related by two distinct $180^\circ$ rotations about the axes $\bm{\hat e_1}$ and $\bm{\hat e_2}$. Hence, the corresponding interface is a compound twin.

Now consider the variant pair $\bm{(F_1,F_4)}$. The underlying stretch-tensor pair is $\bm{(U_8,U_4)}$, which are related by a single $180^\circ$ rotation in $\bm{\mathscr{P}(a)}$ about the axis along $\bm{(\hat e_2+\hat e_3)}$.
In this case, the interface normal coincides with the axis of the $180^\circ$ rotation, and hence $\bm{(F_1,F_4)}$ form a Type-I twin. By the same argument, the pairs $\bm{(F_1,F_3)}$, $\bm{(F_2,F_3)}$, and $\bm{(F_2,F_4)}$  form Type-I twins, while $\bm{(F_3,F_4)}$ forms a compound twin.

This construction highlights the central role of triple clusters in enabling the closure of the austenite--martensite interface. The \emph{elastic Landau state} shown in Figure~\ref{fig:landau_schematic}(a) comprises four triple clusters of the first kind, located at the four corners of the nucleus. Although the nucleus in Figure~\ref{fig:landau_schematic}(a) employs only Type-I triple clusters, its formation relies on the extreme compatibility conditions satisfied by the transformation. The cofactor conditions alone are insufficient to achieve this stress-free construction. The enhanced compatibility, however, permits the nucleus to elongate further in a stress-free manner by introducing an additional compound interface between either pair of opposite variants, as illustrated in Figures~\ref{fig:landau_schematic}(b) and \ref{fig:landau_schematic}(c). The resulting compound interface gives rise to a triple cluster involving the corresponding pair of compound twins and the austenite. These share the same shear direction and have parallel interface normals, thereby constituting a triple cluster that is simultaneously of the first and second kind. It is worth noting that, despite the transformation satisfying the cofactor conditions for Type-II twin solution in \eqref{typeI_typeII}, triple clusters of the second kind are not directly employed in the construction of the nucleus shown in Figure~\ref{fig:landau_schematic}.

Furthermore, within the nucleus, the four pairwise rank-one connected deformation gradients $\bm{F_1},\bm{F_2},\bm{F_3}$, and $\bm{F_4}$ meet along a common junction line directed along $\bm{\hat e_1}$ in Figure~\ref{fig:landau_schematic}(a), satisfying the hypotheses of a \emph{crossing twin}, \ref{app:crossing_twins}~\cite{Bhattacharya1997}. A distinctive feature of this crossing-twin junction is that all four interfaces meeting at the junction are Type-I interfaces. Such junctions were investigated in \cite{Tahseen2026ExtremeCompatibility}. As the nucleus grows from Figure~\ref{fig:landau_schematic}(a) to Figures~\ref{fig:landau_schematic}(b) and \ref{fig:landau_schematic}(c), this fourfold junction evolves into two threefold junctions of martensitic variants. Each threefold junction comprises two Type-I interfaces and one compound interface, with the compound interface common to both junctions, and satisfies the \emph{triplet condition}  (TC Ih, \ref{lem:triplet})~\cite{DellaPorta2022Triplet}. Thus, the simple-shear transformation not only allows greater geometric freedom in the formation of austenite--martensite interfaces through triple clusters, but also enhances compatibility among the martensitic variants.

Finally, applying the transformation $\bm{R_i^T (.)R_i}$ to \eqref{simple_shear} for all $\bm{R_i\in \mathscr{P}(a)}$ generates twelve deformation gradients, which can be classified into three distinct \emph{elastic Landau states}, each comprising four martensitic variants. The resulting deformation gradients are of the form,

\begin{equation*}
    \bm{F_i} = \bm I +t\;\bm{\hat{e}_j\otimes \pm\;\hat{e}_k},\quad i\in \{1,2, \cdots, 12\}, \quad j, k \in \{1,2,3\}, \quad j \neq k.
\end{equation*}

The interface normals within each state $j$ are coplanar, while the corresponding planes $(\perp \bm{\hat{e}}_j)$ for the three states are mutually orthogonal, thereby completing the classification of such stress-free nuclei for this transformation.

\subsection{Pure shear as martensitic transformation}
\label{sec:pure_shear}

Consider a pure shear transformation given by 
\begin{equation}\label{eq:pure_shear}
    \bm{G}=\bm{I}+ t\;(\hat{\bm{e}_1} \otimes \hat{\bm{e}_2} + \hat{\bm{e}_2} \otimes \hat{\bm{e}_1})=\begin{bmatrix}
        1 & t & 0\\
        t & 1 & 0\\
        0 & 0 & 1\\
    \end{bmatrix},
\end{equation}

where $0<t<1$.

Unlike $\bm F$ in a simple shear transformation \eqref{simple_shear}, the deformation gradient $\bm G$ in a pure shear transformation is a positive definite symmetric matrix with eigenvalues,

\begin{equation*}
    \lambda_1=1-t < \lambda_2=1<\lambda_3=1+t.
\end{equation*}

The corresponding  eigenvectors are

\begin{equation*}
    \bm{\hat{v}_1} = \frac{1}{\sqrt{2}} (\bm{\hat{e}_2 - \hat{e}_1}),
    \quad \bm{\hat{v}_2} = \{0,0,1\},
    \quad \bm{\hat{v}_3} = \frac{1}{\sqrt{2}} (\bm{\hat{e}_1 + \hat{e}_2}).
\end{equation*}

The matrix $\bm G$ is of the same form as the stretch tensors of variants for cubic to orthorhombic transformation, see (\ref{app:cubicortho}). In particular $\bm G$ corresponds to $\bm {U_6}$ in \eqref{eq:19} with $a=d=1$.

Let $\bm {G_1} = \bm{G}$ and 

\begin{equation}
    \bm{G_2} = \bm{I}- t\;(\hat{\bm{e}}_1 \otimes \hat{\bm{e}}_2 + \hat{\bm{e}}_2 \otimes \hat{\bm{e}}_1)=\begin{bmatrix}
        1 & -t & 0\\
        -t & 1 & 0\\
        0 & 0 & 1\\
    \end{bmatrix},
\end{equation}

which corresponds to $\bm{U_5}$ in \eqref{eq:19}. The stretch tensors $\bm{G_1}$ and $\bm{G_2}$ are related by $180^\circ$ rotations about the $\bm{\hat{e}_1}$ and $\bm{\hat{e}_2}$ axes. Therefore, $\bm{G_1}$ and $\bm{G_2}$ form a compound twin pair.

Taking $\bm U=\bm {G_1}$, $\bm{\hat e^{I}}=\bm{\hat{e}_1}$, and $\bm{\hat e^{II}}=\bm{\hat{e}_2}$, the cofactor conditions are satisfied by both the compound twin solutions in \eqref{compound_twins}. In fact, for this transformation, the cofactor conditions are satisfied by all compound twin pairs, corresponding to variant pairs $(\bm{U_1,U_2})$, $(\bm{U_3,U_4})$, $(\bm{U_5,U_6})$ in \eqref{eq:19}.

However, none of these pairs satisfies any of the conditions \eqref{EC1}$-$\eqref{EC4}. Thus, despite satisfying the cofactor conditions, the compound twins cannot form triple clusters in this transformation.

Moreover, the Type-I and Type-II twin solutions also fail to satisfy the cofactor conditions. Consequently, no triple clusters can arise in this transformation. Nevertheless, the condition $\lambda_2=1$ still permits stress-free interfaces between the pure martensitic variants and the austenite.

We conclude our discussion of pure shear transformation here and revisit the simple shear transformation in the next section, where the ideas developed in this section are placed in a more general setting.

\section{Generalization of the elastic Landau state}
\label{sec:new_generalized_shear}

In \autoref{sec:shear_simple}, we established an analogy between simple shear deformations and martensitic phase transformations. The key observation was that certain rotations in the cubic point group leave the shear direction invariant (up to sign), while simultaneously generating multiple interface normals lying in a common plane. Together with the volume preserving condition $\det(\bm{F})=1$, this geometric feature formed the basis for constructing the elastic Landau state.

The aim of this section is to generalize the preceding construction. The simple shear considered in \autoref{sec:shear_simple} constituted a one-parameter family of deformations, characterized solely by the shear magnitude, $t$. The construction presented in this section introduces a two-parameter family of simple shear deformations that preserves the geometric mechanism underlying the stress-free closure of austenite--martensite interface. In particular, this construction yields a family of eight compatible deformation gradients sharing a common shear direction. The corresponding transformation stretch tensors belong to a cubic to triclinic transformation, representing the maximal symmetry reduction among martensitic phase transformations. A notable feature of this family is that both Type-I twins and compound twins%
\begingroup
\renewcommand{\thefootnote}{\fnsymbol{footnote}}
\footnote{The compound twins in this case are \emph{non-conventional compound twins}, belonging to the more general class of compound domains, discussed in more detail later in the text.}
\endgroup 
form stress-free triple clusters with austenite. Furthermore, the eight-gradient family contains two distinguished subsets of four deformation gradients each that form \emph{sheared elastic Landau states}, closely related to the diamond microstructures reported in the martensite literature. These states can be assembled to form arbitrarily shaped martensitic islands embedded in austenite.

We begin by identifying the vectors in $\bm {\mathbb{S}^2}$ with the largest stabilizer subgroup under the action of the
cubic point group, $\bm {\mathscr{P}(a)}$. These are precisely the coordinate directions $\bm{\pm \hat{e}_i}$,
$i=\{1,2,3\}$, each of which is stabilized by four elements in $\bm {\mathscr{P}(a)}$. Consequently, the set of cubic
rotations satisfying $\bm {R\hat{e}_i}=\bm {\pm \hat{e}_i}$ contains eight elements. Choosing the
shear direction to be $\bm {\hat{e}_1}$, we consider deformation gradient of the form

\begin{equation}
\label{eq:general_shear}
\bm{F}=\bm{I+} t\; \bm {{\hat{e}_1}\otimes\hat{n}},
\end{equation}

where $t \neq 0$ is the shear magnitude.

To ensure that the transformation is volume preserving, we set
\begin{equation*}
    \det(\bm F)=1 +t\;\bm{\hat{e}_1 \cdot\hat{n}}=1,
\end{equation*}
which requires $\bm{\hat{e}_1 \cdot\hat{n}}=0$. We therefore parameterize the normal as
\begin{equation*}
\bm{\hat{n}}
=
\left(0,p,\sqrt{1-p^2}\right),
\qquad
p\in(0,1)\setminus\left\{{1}/{\sqrt{2}}\right\},
\end{equation*}
for some parameter $p$. The limiting cases $p=\{0,1\}$ correspond to the special case discussed in \autoref{sec:shear_simple}. The value $p=1/\sqrt{2}$ is excluded; the reason for its exclusion is discussed later in \autoref{sec:crystallography_sheared}.

Consider the rotations
\begin{equation*}
\begin{aligned}
\bm{R_1} &= \bm I, &
\bm{R_2} &= R[\pi,\bm{\hat e_1}], &
\bm{R_3} &= R\!\left[\frac{\pi}{2},\bm{\hat e_1}\right], &
\bm{R_4} &= R\!\left[\frac{3\pi}{2},\bm{\hat e_1}\right], \\
\bm{R_5} &= R[\pi,\bm{\hat e_2}], &
\bm{R_6} &= R[\pi,\bm{\hat e_3}], &
\bm{R_7} &= R[\pi,\frac{1}{\sqrt{2}}(\bm{\hat e_2}+\bm{\hat e_3})], &
\bm{R_8} &= R[\pi,\frac{1}{\sqrt{2}}(\bm{\hat e_2}-\bm{\hat e_3})].
\end{aligned}
\end{equation*}

where $R\;[\theta,\bm{\hat{n}}]$ denotes the rotation matrix corresponding to a rotation of angle $\theta$ about the unit vector $\bm{\hat{n}}$. Each of the rotations, $ \{\bm{{R_i}}\}_{i=1}^{8}$ satisfies 
\begin{equation*}
    \bm {R_i\hat{e}_1}=\bm {\pm \hat{e}_1}.
\end{equation*} 

For $i=\{1,\ldots,8\},$ define

\begin{equation*}
    \bm{F_i}=\bm{R_i^T FR_i}.
\end{equation*}

The corresponding deformation gradients are

\begin{equation}
\label{eq:eight_gradients}
\bm{F_i} = \bm I+t\,\bm{\hat e_1 \otimes} \bm{\hat n_i},
\end{equation}

where

\begin{equation*}
\begin{aligned}
\bm{\hat n_1} &= \left(0,p,\sqrt{1-p^2}\right), &
\bm{\hat n_2} &= \left(0,-p,-\sqrt{1-p^2}\right),\\
\bm{\hat n_3} &= \left(0,-\sqrt{1-p^2},p\right), &
\bm{\hat n_4} &= \left(0,\sqrt{1-p^2},-p\right),\\
\bm{\hat n_5} &= \left(0,-p,\sqrt{1-p^2}\right), &
\bm{\hat n_6} &= \left(0,p,-\sqrt{1-p^2}\right),\\
\bm{\hat n_7} &= \left(0,-\sqrt{1-p^2},-p\right), &
\bm{\hat n_8} &= \left(0,\sqrt{1-p^2},p\right).
\end{aligned}
\end{equation*}

The deformation gradients $\bm{F_i}$ are pairwise rank-one connected, and each is rank-one connected to $\bm I$. They share a common shear vector, $\bm{\hat{e}_1},$ while the corresponding normals, $\bm{\hat{n}_i}$ lie on a unit circle in the $\bm{\hat{e}_2-\hat{e}_3}$ plane, forming four diametrically opposite pairs. 

\begin{figure}[hbt!]
    \centering
    \includegraphics[scale=0.40]{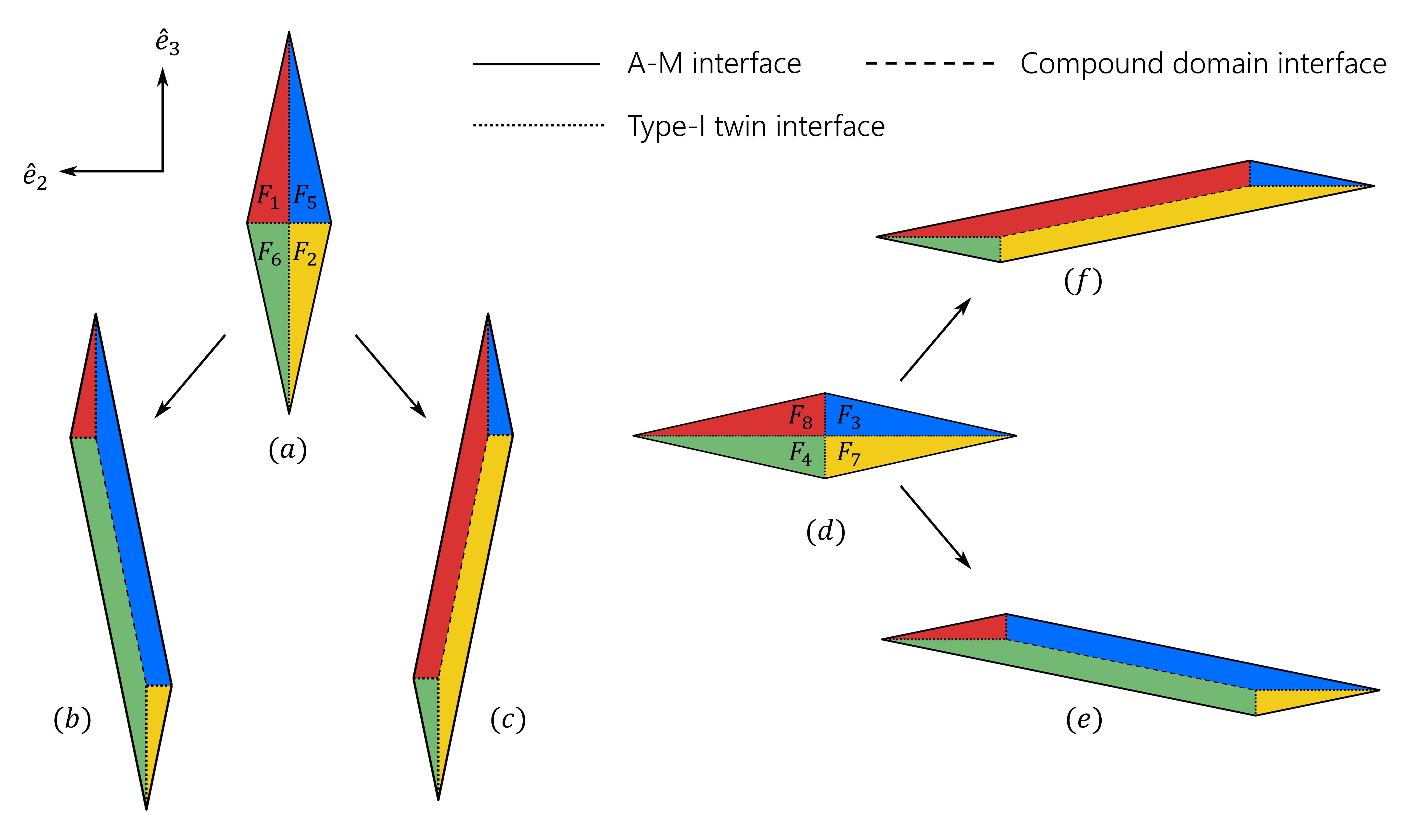}
    \caption{$(a),(d)$ Sheared Landau states comprising four variants, forming self-accommodating martensitic nuclei. The variant pairs (red, yellow) and (blue, green) form compound domains, all other variant pairs form Type-I twins $(b),(c),(e),(f)$ growth of martensitic nucleus into the austenite matrix by forming a compound twin interface between opposite pair of variants. Each corner constitutes a triple cluster of the first kind, while the layered configuration of compound domains with austenite in $(b),(c),(e)$ and $(f)$ constitutes triple clusters that are simultaneously of the first and second kinds. Here $p=0.2$ and $t=0.4$.}
    \label{fig:sheared_landau_schematic}
\end{figure}

The eight deformation gradients naturally partition into two four-element groups according to whether their interface normals are of the form $(0,\pm p,\pm \sqrt{1-p^2})$ or $(0,\pm \sqrt{1-p^2},\pm p)$. Specifically, the first group consists of $\bm{(F_1,F_2,F_5,F_6)}$, while the second consists of $\bm{(F_3,F_4,F_7,F_8)}$. Each of these groups assembles into a \emph{sheared Landau state} (see Figure~\ref{fig:sheared_landau_schematic}), closely resembling the diamond microstructure reported in the experimental literature (see Figures \ref{fig:Li_et_al} and \ref{fig:CuAlNi})~\cite{Tan1990Observations,Li2019,Schwabe2021}. 

Like the classical elastic Landau state introduced in \autoref{sec:shear_simple}, each diamond can extend into a parallelogram by introducing an additional rank-one interface between the opposite pair of deformation gradients. A characterization of all the rank-one interfaces arising in these sheared Landau states is presented in \autoref{sec:crystallography_sheared}. The sheared Landau states give rise to a two-parameter family of deformation gradients, parametrized by the geometric parameter $p$, and the shear magnitude $t$, that realize the geometric closure of austenite--martensite interfaces in a stress-free manner. 

Finally, we note that the above construction was carried out with the common shear direction as $\bm{\hat e_1}$. In general, the common shear direction may be chosen along any of the three cubic axes $\bm{\hat e_1},\bm{\hat e_2}$ or $\bm{\hat e_3}$. Each choice gives rise to two four-element groups, resulting in a total of six such groups for this transformation.

\subsection{Crystallography of the sheared Landau state}
\label{sec:crystallography_sheared}

Consider the family of simple shear transformations
\begin{equation}
\label{eq:generalized_shear}
    \bm {F_1} = \bm I + t\:\bm{\hat{e}_1 \otimes} \left( 0,p,\sqrt{1-p^2}\right)= \begin{bmatrix}
        1 & tp & t\sqrt{1-p^2}\\
        0 & 1 & 0\\
        0 & 0 & 1\\
    \end{bmatrix},
\end{equation}

where $t\in \mathbb{R}\setminus \{\bm0\}$ and $p\in(0,1)\setminus\left\{{1}/{\sqrt{2}}\right\}$.

We interpret $\bm{F_1}$ as the deformation gradient of a martensitic variant and determine the corresponding stretch tensor $\bm{U_1}$ through the polar decomposition of $\bm{F_1}$.

\begin{equation}
\bm{U_1}
=
\sqrt{\bm{F_1}^{T}\bm{F_1}}
=
\begin{bmatrix}
\alpha(t) & \beta(t,p) & \gamma(t,p) \\
\beta(t,p) & \delta(t,p) & \varepsilon(t,p) \\
\gamma(t,p) & \varepsilon(t,p) & \omega(t,p)
\end{bmatrix},
\label{polar_pt}
\end{equation}

where,

\begin{equation*}
\begin{alignedat}{2}
\alpha(t) &=
\frac{2}{\sqrt{t^2+4}}
&\qquad
\beta(t,p) &=
\frac{pt}{\sqrt{t^2+4}},\\[1ex]
\gamma(t,p) &=
\frac{\sqrt{1-p^2}\,t}{\sqrt{t^2+4}}
&\qquad
\delta(t,p) &=
1+\frac{p^2\left(t^2-\sqrt{t^2+4}+2\right)}{\sqrt{t^2+4}},\\[1ex]
\varepsilon(t,p) &=
\frac{p\sqrt{1-p^2}\left(t^2-\sqrt{t^2+4}+2\right)}{\sqrt{t^2+4}}
&\qquad
\omega(t,p) &=
1+\frac{(p^2-1)\left(-t^2+\sqrt{t^2+4}-2\right)}{\sqrt{t^2+4}}.
\end{alignedat}
\end{equation*}

The matrix in equation \eqref{polar_pt} generates 24 distinct positive-definite stretch tensors under conjugation by the rotations of the cubic point group, $\bm{\mathscr{P}(a)}$. Therefore, when interpreted as a martensitic transformation, $\bm{U_1}$ corresponds to a cubic-to-triclinic transformation. The eigenvalues of $\bm{U_1}$ are

\begin{equation*}
    1,\frac{t^2\pm t\sqrt{t^2+4} +4}{2 \sqrt{t^2+4}}.
\end{equation*}

Without loss of generality, we assume $t>0$, which fixes the choice of sign and yields

\begin{equation*}
\lambda_1=\frac{t^2-t\sqrt{t^2+4} +4}{2 \sqrt{t^2+4}} < \lambda_2=1<
\frac{t^2+t\sqrt{t^2+4} +4}{2 \sqrt{t^2+4}}.
\end{equation*}

The transformation is volume preserving by construction, so $\det(\bm {U_1})$ = $\lambda_1\lambda_2\lambda_3=1$. The corresponding normalized eigenvectors are

\begin{align*}
\notag
    \bm{\hat{u}_1} &= \left\{\frac{-\sqrt{t^2+4}-t}{\sqrt{2 t \left(\sqrt{t^2+4}+t\right)+8}},\frac{\sqrt{2} p}{\sqrt{t \left(\sqrt{t^2+4}+t\right)+4}},\sqrt{\frac{2-2 p^2}{t \left(\sqrt{t^2+4}+t\right)+4}}\right\},\\
    \bm{\hat{u}_2} &= \left\{0,-\sqrt{1-p^2},p\right\},\\
     \bm{\hat{u}_3} &= \left\{\frac{1}{\sqrt{\frac{1}{2} t \left(\sqrt{t^2+4}+t\right)+2}},p \sqrt{\frac{t}{2\sqrt{t^2+4}}+\frac{1}{2}},\frac{1}{2} \sqrt{\left(2-2 p^2\right) \left(\frac{t}{\sqrt{t^2+4}}+1\right)}\right\}.
\end{align*}

Consider the symmetry related variant $\bm {U_5}$, corresponding to stretch tensor of $\bm{F_5}$ in \eqref{eq:eight_gradients}.

\begin{equation*}
\bm{U_5}=
    \begin{bmatrix}
\alpha(t) & -\beta(t,p) & \gamma(t,p) \\
-\beta(t,p) & \delta(t,p) & -\varepsilon(t,p) \\
\gamma(t,p) & -\varepsilon(t,p) & \omega(t,p)
\end{bmatrix}.
\end{equation*}

The variants $\bm {U_1}$ and $\bm {U_{5}}$ are related by a $180^\circ$ rotation about $\bm{\hat{e}_2}$. $\bm {U_1}$ and $\bm {U_{5}}$ therefore form a Type-I/II twin pair.

Choosing $\bm U=\bm U_1$, $\bm V=\bm U_{5}$, and $\bm{\hat e}=\bm{\hat{e}_2}$, together with $\bm{\hat u_1}$ and $\bm{\hat u_3}$ as defined above, we find that the cofactor conditions are satisfied by the Type-I solution  in \eqref{typeI_typeII}. Therefore, the variant pair $\bm{U_{1}}$ and $\bm{U_{5}}$ forms a triple cluster of first kind with austenite, with the associated deformation gradients $\bm{F_1}$ and $\bm{F_5}$ respectively (see Figure~\ref{fig:sheared_landau_schematic}). In fact, each corner of the sheared Landau states shown in Figure~\ref{fig:sheared_landau_schematic} corresponds to a triple cluster of the first kind involving Type-I twins and austenite.

Now, consider the symmetry related variant $\bm {U_2}$, corresponding to stretch tensor of $\bm{F_2}$ in \eqref{eq:eight_gradients}.

\begin{equation*}
\bm{U_2}=
    \begin{bmatrix}
\alpha(t) & -\beta(t,p) & -\gamma(t,p) \\
-\beta(t,p) & \delta(t,p) & \varepsilon(t,p) \\
-\gamma(t,p) & \varepsilon(t,p) & \omega(t,p)
\end{bmatrix}.
\end{equation*}

The variants $\bm {U_1}$ and $\bm {U_{2}}$ are related by a $180^\circ$ rotation about $\bm{\hat{e}_1}$. Since a cubic to triclinic transformation possesses 24 distinct transformation stretch tensors, each related to every other by a unique cubic symmetry operation, one might at first conclude that the variants $\bm {U_1}$ and $\bm {U_{2}}$ form a Type-I/II pair, since compound twins are characterized by the existence of two distinct $180^\circ$ rotations relating the corresponding variants, see \autoref{thm:compound_domains}. However,
%
%
%
there is a second $180^\circ$ rotation relating these pairs, whose axis is not a symmetry axis of the parent cubic lattice. This second rotation is about the axis
\begin{equation*}
    \bm{\hat e_c}=\{0,p,\sqrt{1-p^2}\},
\end{equation*}
which is precisely the normal to the interface between the deformation gradients $\bm{F_1}$ and $\bm{F_2}$. Variant pairs for which the $180^\circ$ rotation axes relating them do not necessarily belong to the cubic point group form the broader class of domains (\autoref{thm:Type_I_II_domains_check}, \autoref{thm:compound_domains_check})~\cite{CHEN20132566,Tahseen2026ExtremeCompatibility}. Accordingly, the pair $\bm {(U_1},\bm {U_{2})}$ constitutes a compound domain, or equivalently, a non-conventional compound twin~\cite{PITTERI1998225}.

Choosing $\bm U=\bm{U_1}$, $\bm V=\bm {U_{2}}$, $\bm{\hat e^{I}}=\bm{\hat{e}_1}$, and $\bm{\hat e^{II}}=\bm{\hat{e}_c}$, together with $\bm{\hat u_1}$ and $\bm{\hat u_3}$ as defined above, we find that the hypotheses of \autoref{thm:EC} are satisfied. Furthermore, conditions \eqref{EC2} and \eqref{EC4} are satisfied simultaneously, establishing that the triple cluster formed  by the pair $\bm{(F_1,F_2)}$ with austenite is simultaneously of the first and second kinds. Similarly the variant pairs corresponding to deformation gradient pairs $\bm{(F_3,F_4), (F_5,F_6)}$ and $\bm{(F_7,F_8)}$ in Figure~\ref{fig:sheared_landau_schematic} form compound domains and constitute triple clusters of the dual nature with austenite. All the pairwise rank-one connections between the deformation gradients constituting the sheared Landau states in Figure~\ref{fig:sheared_landau_schematic} can be explicitly written as 

\begin{minipage}{0.45\textwidth}
\begin{equation*}
\begin{aligned}
\bm{F_1}-\bm{F_2} &= 2t\;\bm{\hat e_1} \bm{\otimes} (0,p,\sqrt{1-p^2}),\\
\bm{F_1}-\bm{F_5} &= 2tp\;\bm{\hat e_1} \bm{\otimes} \bm{\hat e_2},\\
\bm{F_1}-\bm{F_6} &= 2t\sqrt{1-p^2}\;\bm{\hat e_1} \bm{\otimes} \bm{\hat{e}_3},\\
\bm{F_2}-\bm{F_5} &= -2t\sqrt{1-p^2}\;\bm{\hat e_1} \bm{\otimes} \bm{\hat{e}_3},\\
\bm{F_2}-\bm{F_6} &= -2tp\;\bm{\hat e_1} \bm{\otimes} \bm{\hat e_2},\\
\bm{F_5}-\bm{F_6} &= 2t\;\bm{\hat e_1} \bm{\otimes} (0,-p,\sqrt{1-p^2}),
\end{aligned}
\end{equation*}
\end{minipage}
\hfill
\begin{minipage}{0.45\textwidth}
\begin{equation}
\begin{aligned}
\bm{F_3}-\bm{F_4} &= 2t\;\bm{\hat e_1} \bm{\otimes} (0,-\sqrt{1-p^2},p),\\
\bm{F_3}-\bm{F_7} &= 2tp\;\bm{\hat e_1} \bm{\otimes} \bm{\hat e_3},\\
\bm{F_3}-\bm{F_8} &= -2t\sqrt{1-p^2}\;\bm{\hat e_1} \bm{\otimes} \bm{\hat e_2},\\
\bm{F_4}-\bm{F_7} &= 2t\sqrt{1-p^2}\;\bm{\hat e_1} \bm{\otimes} \bm{\hat e_2},\\
\bm{F_4}-\bm{F_8} &= -2tp\;\bm{\hat e_1} \bm{\otimes} \bm{\hat e_3},\\
\bm{F_7}-\bm{F_8} &= -2t\;\bm{\hat e_1} \bm{\otimes} (0,\sqrt{1-p^2},p).
\end{aligned}
\end{equation}
\end{minipage}

The pairs $\bm{(F_1,F_2)}$, $\bm{(F_5,F_6)}$, $\bm{(F_3,F_4)}$ and $\bm{(F_7,F_8)}$ form compound domains, with the underlying variants related by two distinct $180^\circ$ rotations: one about the cubic symmetry axis $\bm{\hat e_1}$ and the other about the respective interface normal. Together with austenite, these pairs form triple clusters that are simultaneously of the first and second kinds. All the remaining pairwise connections correspond to Type-I twin pairs, for which the underlying variants are related by a single $180^\circ$ rotation about the corresponding interface normal. Together with austenite, these pairs form triple clusters of the first kind.

However, unlike the simple shear transformation considered in \autoref{sec:shear_simple}, this transformation does not satisfy the cofactor conditions for the Type-II solution in \eqref{typeI_typeII}. Consequently, triple clusters of the second kind involving Type I/II pairs and austenite are not possible in this transformation.

The martensitic compatibility relations, however, retain the structure of those in the classical Landau state discussed in \autoref{sec:shear_simple}. In particular, within each sheared Landau state, a fourfold crossing-twin junction formed by Type-I interfaces, shown in Figures~\ref{fig:sheared_landau_schematic}(a) and \ref{fig:sheared_landau_schematic}(d), evolves into two threefold junctions, each comprising two Type-I interfaces and one compound interface, as shown in Figures~\ref{fig:sheared_landau_schematic}(b),(c) and \ref{fig:sheared_landau_schematic}(e),(f), respectively. Each of these threefold junctions satisfies the \emph{triplet condition}~(TC 1h)~\cite{DellaPorta2022Triplet}, thereby completing the crystallographic characterization of the interfaces and junctions within the sheared elastic Landau state.

We now return to the excluded value $p=1/\sqrt{2}$. At this value, the two distinct sheared Landau states $\bm{(F_1, F_2, F_5,F_6)}$ and $\bm{(F_3, F_4, F_7,F_8)}$ degenerate into a single state. In particular, the eight distinct normals $\bm{\hat{n}_1},...,\bm{\hat{n}_8}$ defined in \eqref{eq:eight_gradients} reduce to four with 
\[
\bm{\hat n_1}=\bm{\hat n_8}, \qquad \bm{\hat n_2} = \bm{\hat n_7}, \qquad \bm{\hat n_3}=\bm{\hat n_5}, \qquad\bm{\hat n_4}=\bm{\hat n_6}.
\]

The deformation gradients $\bm{(F_1, F_2, F_3,F_4)}$ then generate a microstructure that is a $90^\circ$ rigid rotation  about the $\bm{\hat{e}_1}$ axis of the microstructure shown in Figure~\ref{fig:landau_schematic}(a). In this case, the transformation reduces from cubic-to-triclinic to cubic-to-monoclinic-II. Consequently, this nucleus inherits all the crystallographic properties of the microstructure in Figure~\ref{fig:landau_schematic}(a), and doesn't give rise to a qualitatively new microstructure.

\subsection{Self-accommodation in the sheared Landau state}

We now investigate the self-accommodation properties of the sheared Landau state. A self-accommodating microstructure is an arrangement of martensitic variants whose average deformation is macroscopically compatible with the parent austenite, so that the boundary of the transformed region undergoes no net displacement. Bhattacharya~\cite{Bhattacharya1992} showed that the conditions for the existence of a self-accomodating microstructure depend only on the symmetry of the parent phase. In particular, for transformations with a cubic austenite, the condition $\det(\bm U)=1$ is necessary and sufficient for self-accommodation. Since the deformation gradients comprising the sheared Landau state satisfy $\det(\bm U)=1$ by construction, it follows that the sheared Landau state is self-accommodating. Consequently, the corresponding deformation gradients are kinematically admissible and therefore satisfy the minors relations of Morrey~\cite{Morrey1966}, see~\ref{app:minors}.

We now examine self-accommodation in the context of the diamond to parallelogram transition introduced earlier in this section. Consider the deformation gradients $\bm{F_1,F_2,F_5}$ and $\bm{F_6}$ comprising the sheared Landau state in \eqref{eq:eight_gradients}. Their matrix representations are given by

\begin{equation*}
\begin{aligned}
\bm {F_1}=
\begin{pmatrix}
1 & pt & t\sqrt{1-p^2}\\
0 & 1 & 0\\
0 & 0 & 1
\end{pmatrix},
\quad
\bm {F_2} =
\begin{pmatrix}
1 & -pt & -t\sqrt{1-p^2}\\
0 & 1 & 0\\
0 & 0 & 1
\end{pmatrix},\\[2ex]
\bm {F_5}=
\begin{pmatrix}
1 & -pt & t\sqrt{1-p^2}\\
0 & 1 & 0\\
0 & 0 & 1
\end{pmatrix},
\quad
\bm {F_6} =
\begin{pmatrix}
1 & pt & -t\sqrt{1-p^2}\\
0 & 1 & 0\\
0 & 0 & 1
\end{pmatrix}.
\end{aligned}
\end{equation*}

Initially, the diamond microstructure consists of the four deformation gradients $\bm{F_1}, \bm{F_2}, \bm{F_5}$, and $\bm{F_6}$, each occupying an equal volume fraction,
\begin{equation*}
\mu_{\bm{1}}=\mu_{\bm{2}}=\mu_{\bm{5}}=\mu_{\bm{6}}=\frac{1}{4}.
\end{equation*}

Consequently,
\begin{equation*}
\frac{1}{4}\left(\bm{F_1}+\bm{F_2}+\bm{F_5}+\bm{F_6}\right)=\bm{I}.    
\end{equation*}

Thus, the initial diamond microstructure is self-accommodating. As the microstructure evolves into a parallelogram, the additional interface between $\bm{F_1}$ and $\bm{F_2}$ propagates, increasing the total volume fraction of this pair at the expense of the pair $(\bm{F_5},\bm{F_6})$. Since this growth occurs symmetrically about the additional interface, the opposite variants remain in equal proportion throughout the evolution. Consequently, the volume fractions satisfy
\begin{equation*}
\mu_{\bm{1}}=\mu_{\bm{2}},\qquad
\mu_{\bm{5}}=\mu_{\bm{6}}.
\end{equation*}

Defining
\begin{equation*}
    \mu=\mu_{\bm{1}}+\mu_{\bm{2}},
\end{equation*}

it follows that
\begin{equation*}
    \mu_{\bm{1}}=\mu_{\bm{2}}=\frac{\mu}{2},
\qquad
\mu_{\bm{5}}=\mu_{\bm{6}}=\frac{1-\mu}{2}.
\end{equation*}

Consequently, the average deformation gradient is
\begin{equation*}
    \bm{F}_{\mathrm{avg}}
=\frac{\mu}{2}\left(\bm{F_1}+\bm{F_2}\right)
+\frac{1-\mu}{2}\left(\bm{F_5}+\bm{F_6}\right).
\end{equation*}

\begin{equation*}
  =\mu \left(\frac{1}{2}\bm{F_1}+\frac{1}{2}\bm{F_2}\right)
+(1-\mu)
\left(\frac{1}{2}\bm{F_5}+\frac{1}{2}\bm{F_6}\right)
=
\mu\bm{I} + (1-\mu)\bm{I} = \bm I.
\end{equation*}

Therefore, the sheared Landau state remains self-accommodating throughout the entire diamond to parallelogram transformation. While four variants are required to nucleate the initial diamond microstructure, subsequent self-accommodated growth is achieved solely through the symmetric propagation of one pair of  opposite variants.

\subsection{Generalization to other shear directions}
\label{sec:further generalization}
The construction of the sheared Landau states developed earlier in this section assumes the shear direction to be along the cubic edge direction $\bm{\hat e_1}$, corresponding to the $\langle100\rangle$ family. In general, one may choose the shear direction along any non-trivial symmetry direction of the cubic crystal, including the face-diagonal $\langle110\rangle$ and the body-diagonal $\langle111\rangle$ families. For each choice of shear direction, the corresponding deformation gradients are generated by the action of the projective stabilizer of the chosen direction within the cubic point group $\bm{\mathscr{P}(a)}$.

As a representative example of $\langle110\rangle$ family, consider the shear direction
\[
\bm {\hat{a}_f} = \frac{1}{\sqrt{2}}\left(\bm{\hat e_1 + \hat{e}_2}\right).
\]
The projective stabilizer of this direction consists of four rotations $\bm{R_i^f},~i=\{1,2,3,4\}$ in $\bm{\mathscr P(a)}$ satisfying
\[
\bm{R_i^f \hat{a}_f}=\bm{\pm\hat{a}_f}.
\]
These are
\[
\bm{R_1^f} = \bm I, \qquad 
\bm{R_2^f} = R[\pi,\frac{1}{\sqrt{2}}\bm{(\hat e_1 + \hat{e}_2)}], \qquad
\bm{R_3^f} = R\!\left[\pi, \frac{1}{\sqrt{2}}\bm{(\hat e_1 - \hat{e}_2)}\right], \qquad
\bm{R_4^f} = R\!\left[\pi,\bm{\hat e_3}\right].
\]

Choosing the unit habit-plane normal as
\[
\bm{\hat n_f}(p)
=
\left(
p,
-p,
 \sqrt{1-2p^2}
\right),
\qquad
0 < p <\frac1{\sqrt2},
\]
ensures that
\[
\bm{\hat{a}_f}\bm{\cdot\hat n_f}=0.
\]

Consider the deformation gradients generated by the action of stabilizing rotations,

\begin{equation}
\label{eq:exp_shear}
\bm{F_i^f}=\bm{I}+t\,\bm{R_i^f}\bm{\hat{a}_f}\otimes \bm{R_i^f\hat{n}_f},
\qquad
i=1,\ldots,4,
\end{equation}

where each deformation gradient satisfies
\[
\det\bm{F_i^f}=1.
\]
Expanding these expressions, we have

\begin{equation}
\begin{aligned}
\bm{F_1^f}
&=\bm{I}+ t\:\bm{\hat a_f}
\otimes
\left(
p,-p, \sqrt{1-2p^2}
\right),\\
\bm{F_2^f}
&=\bm{I}+ t\:\bm{\hat a_f}
\otimes
\left(
-p,p, -\sqrt{1-2p^2}
\right),\\
\bm{F_3^f}
&=\bm{I}+ t\:\bm{\hat a_f}
\otimes
\left(
-p,p, \sqrt{1-2p^2}
\right),\\
\bm{F_4^f}
&=\bm{I}+ t\:\bm{\hat a_f}
\otimes
\left(
p,-p, -\sqrt{1-2p^2}
\right).
\end{aligned}    
\end{equation}

The four deformation gradients $\bm{F_i^f}$ again form two pairs of antiparallel habit plane normals and constitute a sheared Landau state, which  inherits the same geometric and crystallographic characteristics as those described in \autoref{sec:crystallography_sheared}. Sheared Landau states associated with the $\langle110\rangle$ shear directions are consistent with the diamond-shaped martensitic nuclei observed experimentally in NiMnGa system, see~\autoref{sec:exp}. For $p\in\{0,1/\sqrt2\}$, the four deformation gradients degenerate into two and closure of the austenite–martensite interface fails, leaving behind an unbounded lamellar twin microstructure. Furthermore, the transformation symmetry degenerates from cubic-to-triclinic to cubic-to-monoclinic-I, when $p=0$, and to cubic-to-monoclinic-II, when $p=1/\sqrt{2}$.

Likewise, the shear direction may be chosen along any axis in the $\langle111\rangle$ family. As a representative example, consider the shear direction
\[
\bm {\hat{a}_b} = \frac{1}{\sqrt{3}}\left(\bm{\hat e_1 + \hat{e}_2} + \bm{\hat e_3} \right).
\]

Unlike the $\langle110\rangle$ directions, whose projective stabilizers contain four elements, the projective stabilizer of a $\langle111\rangle$ direction contains six rotations satisfying
\[
\bm{R_i^b \hat{a}}_b=\bm{\pm\hat{a}}_b.
\]
These are

\begin{equation*}
\begin{aligned}
\bm{R_1^b} &= \bm I, &
\bm{R_2^b} &=
R\!\left[\frac{2\pi}{3},
\frac{1}{\sqrt3}(\bm{\hat e_1+\hat e_2+\hat e_3})\right],\\
\bm{R_3^b} &=
R\!\left[\frac{4\pi}{3},
\frac{1}{\sqrt3}(\bm{\hat e_1+\hat e_2+\hat e_3})\right],&
\bm{R_4^b} &=
R\!\left[\pi,
\frac{1}{\sqrt2}(\bm{\hat e_1-\hat e_2})\right],\\
\bm{R_5^b} &=
R\!\left[\pi,
\frac{1}{\sqrt2}(\bm{\hat e_1-\hat e_3})\right],&
\bm{R_6^b} &=
R\!\left[\pi,
\frac{1}{\sqrt2}(\bm{\hat e_2-\hat e_3})\right].
\end{aligned}
\end{equation*}

A convenient parameterization of the unit habit-plane normal is
\[
\hat{\bm n_b}(p)
=
\frac{1}{\sqrt{2(1+p+p^2)}}(1,p,-1-p), \qquad p \notin\{0,-1\}.
\]
The six stabilizing rotations generate six planar habit-plane normals, giving rise to six deformation gradients of the form
\begin{equation}
\label{eq:hex_shear}
\bm{F_i^b}=\bm{I}+t\,\bm{R_i^b}\bm{\hat{a}_b}\otimes \bm{R_i^b\hat{n}_b},
\qquad
i=1,\ldots,6,
\end{equation}

where each deformation gradient satisfies
\[
\det\bm{F_i^b}=1.
\]

Expanding these expressions, we have

\begin{equation}
\label{eq:kite}
\begin{aligned}
\bm{F_1^b}
&=
\bm I + t\:\bm{\hat a_b} \otimes
\frac{1}{\sqrt{2(1+p+p^2)}}
\left(1,p,-1-p\right),
\\
\bm{F_2^b}
&=
\bm I + t\:\bm{\hat a_b} \otimes
\frac{1}{\sqrt{2(1+p+p^2)}}
\left(-1-p,1,p\right),
\\
\bm{F_3^b}
&=
\bm I + t\:\bm{\hat a_b} \otimes
\frac{1}{\sqrt{2(1+p+p^2)}}
\left(p,-1-p,1\right),
\\
\bm{F_4^b}
&=
\bm I + t\:\bm{\hat a_b} \otimes
\frac{1}{\sqrt{2(1+p+p^2)}}
\left(p,1,-1-p\right),
\\
\bm{F_5^b}
&=
\bm I + t\:\bm{\hat a_b} \otimes
\frac{1}{\sqrt{2(1+p+p^2)}}
\left(-1-p,p,1\right),
\\
\bm{F_6^b}
&=
\bm I + t\:\bm{\hat a_b} \otimes
\frac{1}{\sqrt{2(1+p+p^2)}}
\left(1,-1-p,p\right).
\end{aligned}
\end{equation}

\begin{figure}[hbt!]
    \centering
    \includegraphics[scale=0.33]{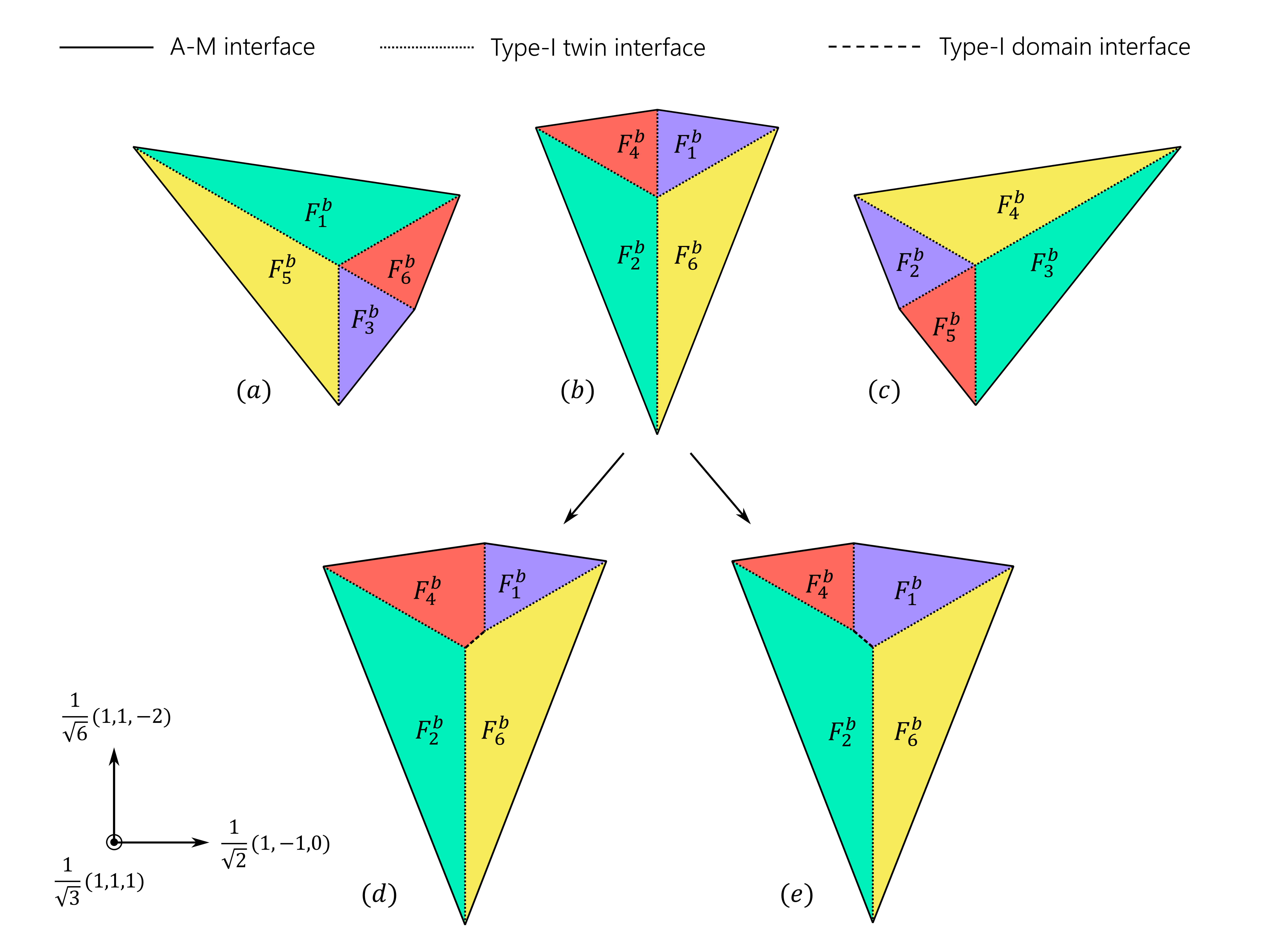}
    \caption{$(a),(b),(c)$ Kite-shaped martensitic nuclei embedded in the austenite matrix, arising from the $\langle111\rangle$ family of shear directions, defined by \eqref{eq:kite}. All interfaces between neighboring variants are Type-I interfaces with interface normals belonging to the $\langle110\rangle_A$ family. $(d),(e)$ Growth of a kite microstructure by introducing an additional interface between the opposite pair of variants. This interface is non-crystallographic and the opposite variant pair forms a Type-I domain. Here, $p =0.6$.}
    \label{fig:kite}
\end{figure}

Unlike the $\langle110\rangle$ family, the $\langle111\rangle$ family of shear directions does not naturally produce pairs of antiparallel habit plane normals, in general. Consequently, this family does not give rise to a diamond-shaped microstructure. Instead, the six compatible deformation gradients can be arranged into three distinct kite-shaped microstructures, each comprising four compatible deformation gradients, see Figure~\ref{fig:kite}. 

It can be verified using \autoref{thm:Type_I_II_domains_check},~\autoref{thm:compound_domains_check} that none of the variant pairs underlying the six deformation gradient pairs forms a compound domain. As a result, triple clusters of dual nature, which simultaneously satisfy the geometric conditions for both triple clusters of the first and second kinds, are absent in the $\langle111\rangle$ family. Each corner of a kite microstructure instead forms a triple cluster of the first kind. Similar to the sheared Landau state, the kite microstructure can evolve by introducing an additional interface between the two opposite variants, see Figures~\ref{fig:kite}$(b),(d),(e)$. The two opposite variants form Type-I domains, connected by a non-crystallographic interface. Together with the austenite, these domains also constitute a triple cluster of the first kind. Furthermore, conjugation of any of the six deformation gradients by the cubic point group $\bm{\mathscr{P}(a)}$, generates 24 distinct variants, establishing that the transformation is cubic to triclinic.

The compatibility between the martensitic variants in a kite microstructure differs from that of the diamond nuclei discussed earlier. The four-fold crossing twin formed by Type-I interfaces within the kite nucleus, see Figure~\ref{fig:kite}$(b)$, gives way to two three-fold junctions, see Figures~\ref{fig:kite}$(d),(e)$. Each of these three-fold junctions comprises two Type-I twin interfaces and a Type-I domain interface, which is common to both the junctions. Consequently, each of the three-fold junctions satisfies the triplet condition (TC I, \ref{lem:triplet}).

\begin{figure}[hbt!]
    \centering
    \includegraphics[scale=0.35]{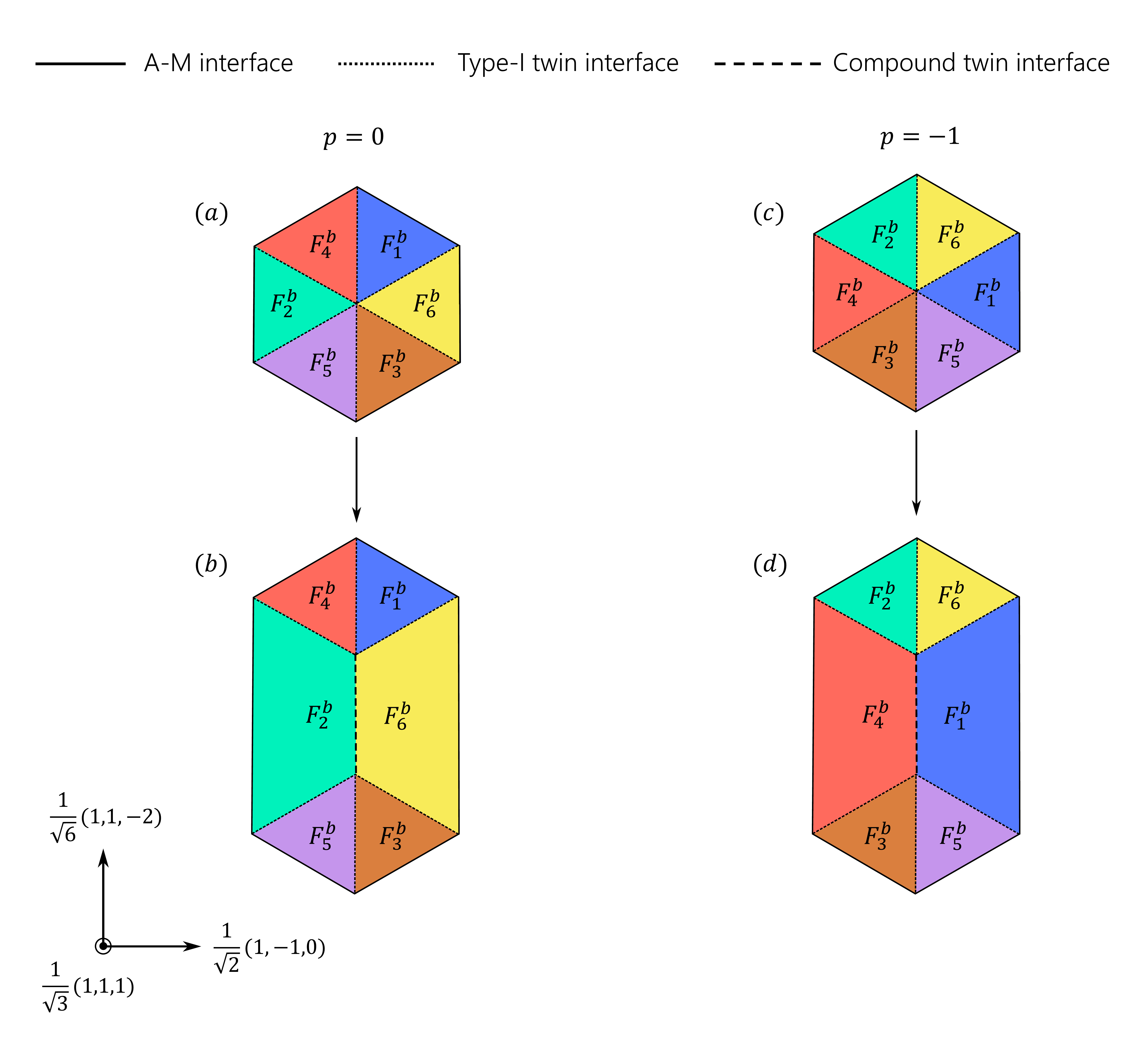}
    \caption{$(a),(c)$ Hexagon-shaped martensitic nuclei embedded in the austenite matrix, arising from the $\langle111\rangle$ family of shear directions, when $(a)\;p=0$ and $(c)\;p=-1$. All interfaces between the neighboring variants are Type-I interfaces with interface normals belonging to the $\langle110\rangle_A$ family. $(b),(d)$ Growth of a kite microstructure by introducing an additional (conventional) compound twin interface between the opposite pair of variants.}
    \label{fig:kite_special}
\end{figure}

At the two special values, $p=0$ and $p=-1$, the closure mechanism of the kite microstructure changes. Unlike the generic case, where closure is achieved using just four of the six deformation gradients, closure at these special values requires all six deformation gradients. Moreover, the six habit-plane normals occur in three antiparallel pairs, resulting in a closed hexagonal martensitic nucleus, as shown in Figures~\ref{fig:kite_special}~$(a),(c)$. The hexagonal nucleus can subsequently extend by introducing an interface between any pair of opposite variants, see Figures~\ref{fig:kite_special}~$(b),(d)$.

This additional interface is a conventional compound interface. Together with the austenite, the corresponding pair of variants forms a triple cluster of dual nature, satisfying simultaneously the geometric conditions for triple clusters of the first and second kinds. At the corners of the hexagonal nucleus, the triple clusters with austenite remain of the first kind. Importantly, despite this change in microstructural geometry, the underlying transformation remains cubic-to-triclinic without any degeneracy of the deformation gradients.

The martensitic compatibility also changes at these special values. In particular, the six-fold junction of Type-I twin interfaces in the hexagonal nucleus in Figures~\ref{fig:kite_special}~$(a),(c)$ gives way to two four-fold junctions in Figures~\ref{fig:kite_special}~$(b),(d)$. Each four-fold junction consists of three Type-I twin interfaces and one compound interface and satisfies the hypotheses of a crossing twin~(\ref{app:crossing_twins}).

Collectively, the constructions presented in this section establish a unified framework for generating stress--free austenite--martensite interface closures for shear directions belonging to the $\langle100\rangle$, $\langle110\rangle$, and $\langle111\rangle$ families. These are precisely the symmetry directions of the cubic crystal whose projective stabilizers have order at least three. In each case, the construction generates a two-parameter family of simple-shear deformation gradients, parameterized by the shear magnitude $t$ and the habit-plane orientation $p$, and the resulting transformation is cubic-to-triclinic (expect for some special values of $p$). While the $\langle100\rangle$ and $\langle110\rangle$ families give rise to sheared Landau states with diamond-shaped microstructures, the $\langle111\rangle$ family instead produces kite-shaped microstructures and hexagonal nuclei, owing to the different structure of its projective stabilizer.

\section{Martensitic islands of arbitrary shape}
\label{sec:vitali}
In \autoref{sec:new_generalized_shear}, we investigated the closure of austenite–martensite interfaces and constructed martensitic nuclei with different geometries corresponding to different shear directions, each with zero elastic energy. The goal of this section is to show that the existence of such a closure nucleus allows us to construct martensitic islands of arbitrary shape embedded in an austenite matrix, while maintaining zero elastic energy. This result is motivated by the work of James and Kinderlehrer~\cite{James1990} in micromagnetics, where the existence of a single prototype flux-closure microstructure (Landau state) was shown to imply the existence of zero-magnetostatic-energy magnetization patterns in arbitrary bounded domains.

For the construction presented here, we use the sheared Landau state discussed in \autoref{sec:crystallography_sheared} as the prototype nucleus, although any of the nuclei constructed in \autoref{sec:new_generalized_shear} could serve the same purpose.

\section*{Setting}

Let the free energy density be normalized so that
$\bm \Phi(\cdot,\theta_c)\ge 0$, with equality exactly when the
deformation gradient lies on the austenite well $\bm{SO(3)}$ or on one of
the martensite wells $\bm{SO(3) U_k}$, see \ref{app:glnm}. The
sheared Landau state is built from the four gradients 

\[
\mathcal{K} = \{\bm{F_1,F_2,F_5,F_6}\},
\qquad
\bm {F_j}=\bm I + t\,\bm{\hat e_1}\otimes\bm{\hat n_j},
\qquad t>0,\ 0<p<1,
\]

with habit-plane normals

\[
\bm{\hat n_1}=(0,\,p,\,\sqrt{1-p^2}),\qquad
\bm{\hat n_2}=-\bm{\hat n_1},\qquad
\bm{\hat n_5}=(0,\,-p,\,\sqrt{1-p^2}),\qquad
\bm{\hat n_6}=-\bm{\hat n_5}.
\]

Each $\bm{F_j}$ is a simple shear of magnitude $t$ in the
direction $\bm{\hat e_1}$, and every habit normal is perpendicular to
$\bm{\hat e_1}$. Since all four shears act along the same direction $\bm{\hat e_1}$,
any deformation whose gradient takes values in $\mathcal{K}$ has the form

\begin{equation}
\bm y(\bm x)=\bm x + u(x_2,x_3)\,\bm{\hat e_1}.
\label{eq:def_y}
\end{equation}

\begin{figure}[hbt!]
    \centering
    \includegraphics[scale=0.30]{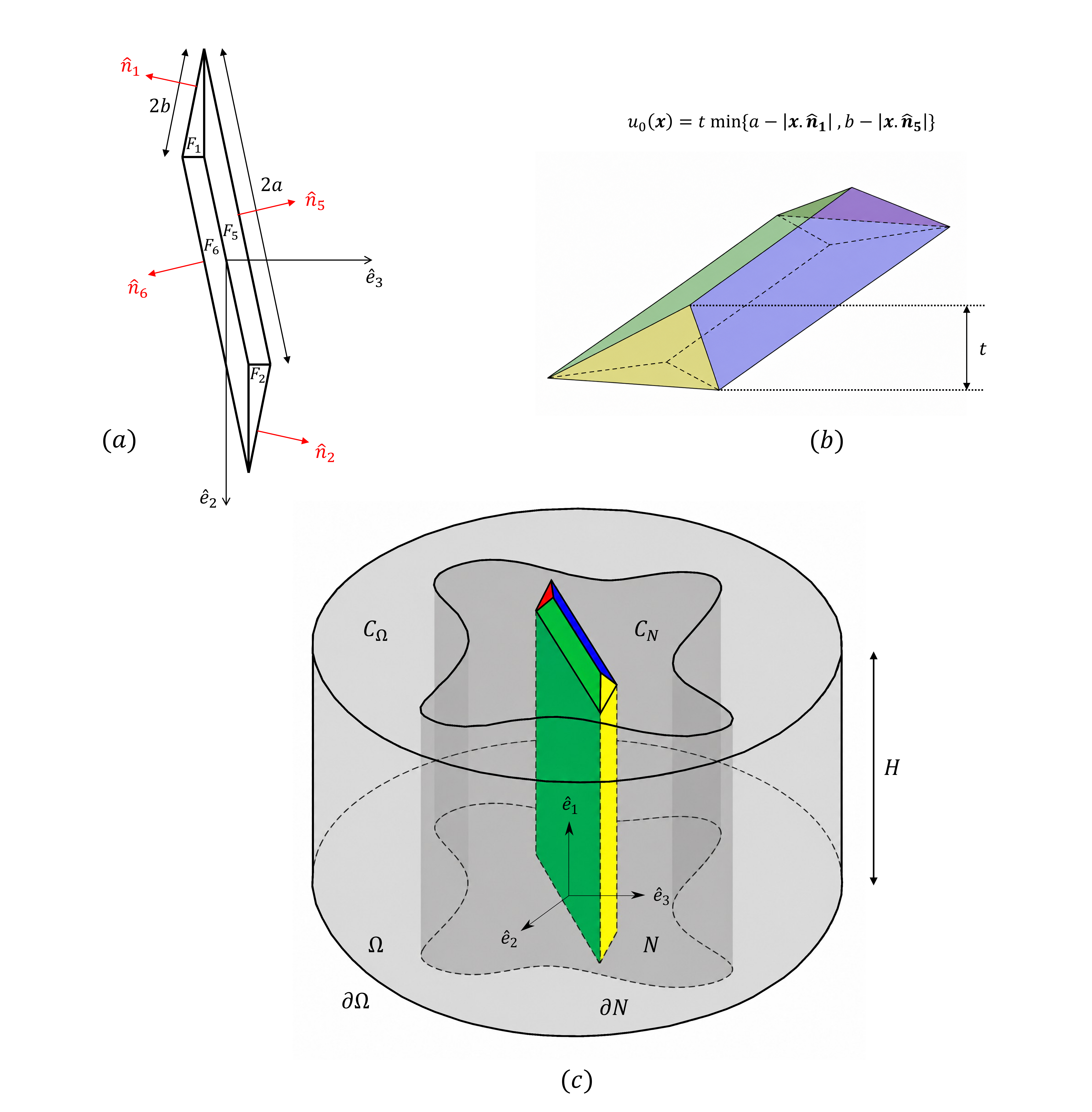}
    \caption{Geometry of the sheared Landau nucleus and its embedding in the body.
$(a)$~Cross-section $L$ in the $\bm{\hat{e}_2}$--$\bm{\hat{e}_3}$ plane: a parallelogram
bounded by the habit planes $\bm x\cdot\bm{\hat{n}_1}=\pm a$ and $\bm x\cdot\bm{\hat{n}_5}=\pm b$, with $a>b$. The region carries the four variants $\bm{ F_1, F_2, F_5, F_6}$, and the habit-plane normals $\bm{\hat{n}_1},\bm{\hat{n}_2},\bm{\hat{n}_5},\bm{\hat{n}_6}$ are shown in red.
$(b)$~The tent-shaped displacement
$u_0(\bm x)=t\min\{a-|\bm x\cdot\bm{\hat{n}_1}|,\,b-|\bm x\cdot\bm{\hat{n}_5}|\}$ over $L$. It is continuous and piecewise linear, and
vanishes on $\partial L$.
$(c)$~The body $\mathcal C_\Omega$ (yellow), of height $H$ and
cross-section $\Omega$ with boundary $\partial\Omega$, contains a nucleus $\mathcal C_N$ (cyan) whose cross-section $N$ has arbitrary shape and boundary $\partial N$. The nucleus is filled with scaled copies of the
Landau prism $\mathcal C_L$; one copy is shown, with its four
variant regions colored. The surrounding austenite is undeformed.}
    \label{fig:vitali}
\end{figure}

Since the deformation is uniform along $\bm{\hat e_1}$, we shall assume that the \emph{body} is the cylinder

\begin{equation}
\mathcal C_\Omega=\{\bm x \bm{\in\mathbb{R}^3}:\ (x_2,x_3)\in\Omega,\ 0<x_1<H\},
\label{eq:body}
\end{equation}

over a bounded open region $\Omega$ on the $\bm{\hat e_2}$--$\bm{\hat e_3}$ plane. Let the height of the cylinder be $H$, with lateral surface $\partial\Omega\times(0,H)$ and end faces $\Omega\times\{0\}$ and
$\Omega\times\{H\}$, see Figure~\ref{fig:vitali}$(c)$.

We will now proceed to construct a martensite nucleus

\begin{equation}
\mathcal C_N=\{\bm x \in \bm{\mathbb{R}^3}:\ (x_2,x_3)\in N,\ 0<x_1<H\}
\label{eq:nucleus}
\end{equation}

over an open region $N\subset\Omega$ of arbitrary shape, the
cross-section of the nucleus. The austenite occupies
$\mathcal C_\Omega\setminus\mathcal C_N$. The boundary of the nucleus
consists of its \emph{lateral interface}

\begin{equation}
\Sigma=\partial N\times(0,H),
\label{eq:Sigma}
\end{equation}

a closed cylindrical surface across which the nucleus meets the
austenite, and its \emph{end faces}

\begin{equation}
E=N\times\{0\}\ \cup\ N\times\{H\},
\label{eq:E}
\end{equation}

where the nucleus reaches the surface of the body, see Figure~\ref{fig:vitali}$(c)$.

The deformation
gradient of \eqref{eq:def_y} is given as

\begin{equation}
\bm{\nabla y}=\bm I+\bm{\hat e_1}\otimes\nabla u ,
\label{eq:gradform}
\end{equation}

and $\bm{\nabla y}$ belongs to $\mathcal{K}$  when

\begin{equation}
\nabla u\in\{t\,\bm{\hat n_1}, \, t\, \bm{\hat n_2}, \, t\, \bm{\hat n_3}, \, t\, \bm{\hat n_4} \}.
\label{eq:slopes}
\end{equation}

We work with the displacement $u(x_2,x_3)$ rather than the deformation gradient because the gradient cannot detect rigid shifts of individual patches. An assembly that is compatible can be made discontinuous by such shifts without any change in its gradients or rank-one conditions. With finitely many patches the shifts can be removed one interface at a time. With the infinitely many patches of the covering this may not be possible. Solving for a continuous displacement $u$ removes all shifts at once.

The problem thus reduces to finding a scalar displacement $u(x_2,x_3)$ on the cross-sectional plane. We require $u(x_2,x_3)=0$ outside $N$ and $\nabla u$ to take only the values in $\{t\,\bm{\hat n_1}, \, t\, \bm{\hat n_2}, \, t\, \bm{\hat n_3}, \, t\, \bm{\hat n_4} \}$ inside $N$. Once such a $u(x_2,x_3)$ is found on the cross-sectional plane, it extends to the whole of the body $\mathcal C_\Omega$ through \eqref{eq:def_y}.  Every point of $\mathcal C_N$ is displaced along $\bm{\hat e_1}$ by the value of $u(x_2,x_3)$, while $u$ vanishes outside $\mathcal C_N$, so the
points surrounding the martensite nucleus are not displaced at all. Together, these define the deformation $\bm y(\bm x)$ of the whole body
$\mathcal C_\Omega$.

\begin{theorem}[Interface closure for nuclei of arbitrary
cross-section]
\label{thm:closure}
Let $\mathcal C_\Omega$ be the body and let $N\subset\Omega$ be any open
region of arbitrary shape, the cross-section of the nucleus
$\mathcal C_N$. Then there exists a continuous deformation
$\bm y(\bm x):\mathcal C_\Omega \to \mathbb{\bm R}^3$ of the body, with bounded strains,
such that:
\begin{enumerate}

\item[(i)] \emph{(surrounding austenite)} $\bm y(\bm x)=\bm x$ at every point of $\mathcal C_\Omega\setminus\mathcal C_N$. This includes every point of the lateral interface $\Sigma$. The austenite surrounding the nucleus is a single, undeformed, uniformly oriented crystal. The nucleus is kinematically compatible with it along the whole of $\Sigma$,

\item[(ii)] \emph{(martensitic nucleus)} inside $\mathcal C_N$ the deformation gradient equals one of the four variants $\bm {F_1,F_2,F_3,F_4}$ at every point outside a set of zero volume. Opposite variants occupy equal volume fractions of the nucleus, $\mu_1=\mu_2$ and $\mu_5=\mu_6$. These fractions are set by the aspect ratio of the Landau state used in the construction. The nucleus is therefore self accommodating. The rhombus gives equal fractions $1/4$.

\item[(iii)] \emph{(zero energy)} $\bm{\Phi}(\bm{\nabla y},\theta_c)=0$ except on a set of zero volume. The total free energy of the body therefore vanishes, and $\bm y (\bm x)$ minimizes it among all deformations of $\mathcal C_\Omega$. The nucleus is stress free and fully closed by austenite along its lateral interface.
\end{enumerate}
\end{theorem}

\section*{Proof}

The proof has four steps. The first three take place entirely in the
cross-sectional plane $\Omega$ and construct the scalar displacement
$u$. In the first, we construct the elastic Landau state $L$ and show
that the elastic Landau nucleus $\mathcal C_L$ closes perfectly. In the
second, we fill the arbitrary cross-section $N$ with non-overlapping
scaled copies of $L$. In the third, we place a correspondingly scaled
displacement on each copy and verify that the assembly is a
continuous and well behaved function. In the fourth and final step, we extend $u$ to
the whole of the body $\mathcal C_\Omega$ through \eqref{eq:def_y}
and read off properties $(i)$ to $(iii)$.

\medskip
\emph{Step 1: The elastic Landau state $L$ and the elastic Landau
nucleus $\mathcal C_L$.}
Fix two half-widths $a\ge b>0$ (the case $a\le b$ is symmetric) and
let $L$ be the parallelogram bounded by the four habit planes
$\bm x\cdot \bm{\hat n_1} =\pm a$ and $\bm x \cdot \bm{\hat n_5}=\pm b$, see Figure~\ref{fig:vitali}$(a)$.

\begin{equation}
L=\bigl\{(x_2,x_3):\ |\bm x \cdot \bm{\hat n_1}|<a \ \text{and}
\
|\bm x\cdot\bm{\hat n_5}|<b\bigr\}.
\label{eq:rhombus}
\end{equation}

The conditions $|\bm x\cdot \bm{\hat n_1}|<a$ and $|\bm x\cdot\bm{\hat n_5}|<b$ each define an infinite strip. The strips are not parallel, since $0<p<1$. The region $L$ is their intersection. For $a=b$, $L$ is the rhombus of Figure~\ref{fig:sheared_landau_schematic}$(a)$ and for $a\ne b$, $L$ is an elongated parallelogram of Figures~\ref{fig:sheared_landau_schematic}$(b),(c)$, the shape through which the nucleus grows into the austenite. On $L$ define the tent-shaped displacement

\begin{equation}
u_0(\bm x)=t\,\min\bigl\{a-|\bm x\cdot\bm{\hat n_1}|,\ b-|\bm x\cdot\bm{\hat n_5}|\bigr\} = t\times\bigl(\text{distance from }\bm x\text{ to }\partial L\bigr).
\label{eq:tent}
\end{equation}

The function $u_0$ is shown in Figure~\ref{fig:vitali}$(b)$. It is continuous and piecewise linear, positive inside $L$ and zero on $\partial L$. It has constant gradient $t\,\bm{\hat n_j}$ on each of four regions, as required by \eqref{eq:slopes}. Recall that if $a\ge b$, the regions at
the sides $\bm x\cdot\bm{\hat n_1}=\pm a$ are triangles carrying the compound pair $\bm {F_1,F_2}$. The regions at the sides $\bm x\cdot\bm{\hat n_5}=\pm b$ are trapeziums carrying the compound pair $\bm{F_5,F_6}$. The two trapeziums meet along the ridge $\bm x\cdot\bm{\hat n_5}=0$, $|\bm x\cdot\bm{\hat n_1}|\le a-b$. This ridge is the
compound twin interface, shown as a dashed line in Figure~\ref{fig:sheared_landau_schematic}$(b)$. In
the symmetric case $b>a$, the roles of the two pairs are exchanged. The trapeziums then carry $\bm{F_1,F_2}$ and the ridge is $\bm x\cdot\bm{\hat n_1}=0$, $|\bm x\cdot\bm{\hat n_5}|\le b-a$, as in Figure~\ref{fig:sheared_landau_schematic}$(c)$. When
$a=b$ the ridge shrinks to the center point and the interfaces are the diagonals of the rhombus Figure~\ref{fig:sheared_landau_schematic}$(a)$. Because $u_0$ is a single continuous function, every internal interface is 
kinematically compatible. The variants $\bm {F_1}$ and
$\bm{F_2}$ each occupy the fraction $b/(4a)$ of $L$, and
$\bm {F_5}$ and $\bm{F_6}$ each the fraction $(2a-b)/(4a)$. All
four fractions equal $1/4$ when $a=b$.

Note that the slope of $u_0$ never exceeds $t$:
\begin{equation}
|u_0(\bm x)-u_0(\bm x')|\le t\,|\bm x-\bm x'|
\qquad \forall \quad \bm x,\bm x' \in L.
\label{eq:lip0}
\end{equation}

We now define the \emph{elastic Landau nucleus} as the cylinder over
the elastic Landau state $L$:

\begin{equation}
\mathcal C_L=\{\bm x\in\mathbb{R}^3:\ (x_2,x_3)\in L,\ 0<x_1<H\}.
\label{eq:prototype}
\end{equation}

Substituting $u_0$ into \eqref{eq:def_y} defines a deformation
$\bm y_0$ of $\mathcal C_L$. This is precisely the sheared Landau state of Figures~\ref{fig:sheared_landau_schematic}$(a)-(c)$. The four prisms of $\mathcal C_L$ over the four regions of $L$ carry the four variants. The planes over the internal interfaces carry the twin interfaces, of Type-I between adjacent variants and, for $a\ne b$, compound between the opposite pair along the ridge. The four lateral faces of $\mathcal C_L$ over the sides of $L$ carry the four habit planes. Since $u_0$ vanishes on the boundary of $L$, the deformation matches the undeformed austenite exactly along the entire lateral surface $\partial L\times(0,H)$ of $\mathcal C_L$. This is interface closure for the elastic Landau nucleus.

\medskip
\emph{Step 2: Filling the cross-section $N$ with scaled
copies of $L$.}
The shape of $L$ is now held fixed. That is, the aspect ratio $b/a$ is fixed, and only the size and position of the copies vary. Consider the family of all copies $c+\varepsilon\overline L$ of $L$ that fit
inside $N$. Here, the center is shifted to any point $c$ of the plane and the size is scaled by any factor $\varepsilon>0$. There is no rotation and no distortion. Copies of arbitrarily small size fit
around every point of $N$. A classical covering theorem of Vitali (\ref{app:vitali}), stated here in the form it takes for scaled copies of a fixed shape, guarantees that one can
select from this family a countable collection of copies
\[
L_i=c_i+\varepsilon_i\overline L\subset N,\qquad i=1,2,3,\ldots,
\]
that do not overlap one another and that together fill $N$
completely except for a leftover set of zero area:

\begin{equation}
\Bigl|\,N\setminus\bigcup\nolimits_i L_i\,\Bigr|=0 .
\label{eq:vitali}
\end{equation}

Intuitively, one places a large copy of $L$ in the interior of $N$. Progressively smaller copies then go into the remaining gaps. The covering theorem confirms that nothing of positive area is ever left over. The covering of the cross-section gives a covering of the nucleus. The cylinders over the copies $L_i$ are scaled copies of the elastic
Landau nucleus $\mathcal C_L$ of \eqref{eq:prototype}. They do not overlap, and they fill $\mathcal C_N$ except for a set of zero volume. This step is the direct counterpart of the construction of James and Kinderlehrer~\cite{James1990} for magnetic domains.

\medskip
\noindent\emph{Step 3: Extension of the displacement field to the entire nucleus.}

Define 
\begin{equation}
u(\bm x)=\varepsilon_i\,
u_0\!\Bigl(\frac{\bm x-c_i}{\varepsilon_i}\Bigr)
\ \ \text{on each }L_i,
\qquad
u(\bm x)=0\ \ \text{at every other point of the plane} .
\label{eq:assembly}
\end{equation}

This is well defined because the copies do not overlap and each
scaled tent vanishes on the boundary of its own copy. Scaling a tent changes its size but not its slopes. By the chain rule, the gradient of $\varepsilon\,u_0((\bm x-c)/\varepsilon)$ at a point equals the gradient of $u_0$ at the corresponding point of $L$. Hence, every copy carries the same four gradient values $t\,\bm{\hat n_j}$ as $L$ itself. It also carries the same twin plane and habit planes. It closes in the same way on its own boundary. Equivalently, the cylinder over every $L_i$ carries a scaled copy of the elastic Landau state.

It remains to check that gluing infinitely many tents of different sizes to one another and to the flat exterior produces a continuous well-behaved function, with no jumps and no unbounded slopes developing in the assembly. We show that the assembled function obeys the same bound \eqref{eq:lip0} as a single tent:

\begin{equation}
|u(\bm x)-u(\bm x')|\le t\,|\bm x-\bm x'|
\qquad \forall \quad\bm x,\bm x' .
\label{eq:lipglobal}
\end{equation}

If $\bm x$ and $\bm x'$ lie in the same copy, or if both lie outside all copies (where $u=0$), the bound is immediate. Suppose $\bm x$ lies in the copy $L_i$ and $\bm x'$ outside it. The segment from $\bm x$ to $\bm x'$ crosses the boundary of $L_i$ at a point $\bm w$. Also, since $\bm x'$ lies in a second copy $L_j$, the segment then crosses the boundary of $L_j$ at a point $\bm w'$. If $\bm x'$ lies in no
copy, take $\bm w'=\bm x'$. The copies are convex and do not overlap. Hence, the points $\bm x,\bm w,\bm w',\bm x'$ lie on the segment in this order. Both $\bm w$ and $\bm w'$ are points where $u$ vanishes. Applying \eqref{eq:lip0} inside each copy now gives

\[
|u(\bm x)-u(\bm x')|
=|u(\bm x)-u(\bm w)+u(\bm w')-u(\bm x')|
\le t\,|\bm x-\bm w|+t\,|\bm w'-\bm x'|
\le t\,|\bm x-\bm x'|.
\]

The last step is true because the two stretches ($\bm x-\bm w, \,\bm w'-\bm x' $) are disjoint parts of the same segment. Hence, \eqref{eq:lipglobal} holds for all $\bm x,\bm x'$ in
the plane $\Omega$. Since $u$ is Lipschitz continuous, it is differentiable at almost every point (Rademacher's theorem). Inside the copies its gradient takes the four values $t\,\bm{\hat n_j}$. Elsewhere $u$ vanishes identically. This includes the leftover set in \eqref{eq:vitali}, the boundaries of the copies, and the whole of the exterior of $N$. Therefore,

\begin{equation}
\begin{aligned}
\nabla u&\in \{t\,\bm{\hat n_1}, \, t\, \bm{\hat n_2}, \, t\, \bm{\hat n_3}, \, t\, \bm{\hat n_4} \}
&&\text{at almost every point of }N,\\
\nabla u&=0
&&\text{at almost every point outside }N .
\end{aligned}
\label{eq:inclusion}
\end{equation}

\medskip
\noindent\emph{Step 4: The deformation of the body and verification
of $(i)$--$(iii)$.}
Extend the assembled displacement $u$ to the whole of the body
through \eqref{eq:def_y}:
\[
\bm y(\bm x)=\bm x+u(x_2,x_3)\,\bm{\hat e_1},\qquad \bm x\in\mathcal C_\Omega .
\]
Since $u$ is continuous with bounded slope, $\bm y(\bm x)$ is a continuous deformation of the body with bounded strains. Property $(i)$ holds because $u=0$ outside $N$ and on $\partial N$. 
Property $(ii)$ follows from \eqref{eq:gradform} and
\eqref{eq:inclusion}. Every copy $L_i$ inherits the fractions of Step~1. By \eqref{eq:vitali}, these add up over the copies to the same fractions of $\mathcal C_N$. At almost every point of $\mathcal C_\Omega$ the gradient is either $\bm I$, which lies on the austenite well $\bm{SO(3)}$, or one of the
$\bm {F_j}$, which lie on the martensite wells $\bm{SO(3)U_k}$
by the polar decompositions of \autoref{sec:crystallography_sheared}. Hence,
$\bm{\Phi}(\bm{\nabla y},\theta_c)=0$ at almost every point of the body, and the total free energy $\int_{\mathcal C_\Omega}\bm {\Phi}(\bm{\nabla y},\theta_c)\,d\bm x$
vanishes. The nucleus is stress-free and fully closed by austenite along its lateral interface.
$\blacksquare$

Two comments on the construction. First, the cylinder $\mathcal C_\Omega$ is a matter of convenience. For a bounded body of arbitrary shape, the same displacement $u$ defines a deformation by restriction of \eqref{eq:def_y}. All conclusions continue to hold.
Each variant occupies the fractions $\mu_j$ of every full cross
section of the nucleus. The end faces of the nucleus are displaced along $\bm{\hat e_1}$. This is the surface relief familiar where martensite plates meet a free surface. Second, the copies used in the covering need not all have the same aspect ratio. A single nucleus may contain copies of several shapes, for instance the growth sequence of Figures~\ref{fig:sheared_landau_schematic}$(a)-(c)$. The ratios must stay within fixed bounds, so that the family remains admissible for the Vitali theorem. Opposite variants still occupy equal fractions, so the nucleus remains self-accommodating. Only the split between the two compound pairs depends on the mixture.

\section{Relevance to experiments}
\label{sec:exp}
The sheared Landau states developed in this work are constructed from a two-parameter family of deformation gradients, imposing stringent constraints on the lattice parameters for their existence. Consequently, they are unlikely to arise directly from pure martensitic variants in most known materials. However, real materials possess an additional degree of freedom through hierarchical twinning, whereby finely laminated microstructures generate average deformation gradients that satisfy compatibility conditions unavailable to the constituent variants. With each successive level of lamination, the constraints on lattice parameters required to satisfy a given compatibility condition become progressively weaker. 

A well-known example is provided by wedge-like microstructures in cubic to tetragonal transformations, where each side of the wedge consists of a laminate whose average deformation gradient is compatible both across the midrib as well as with the surrounding austenite, even though such a configuration cannot be realized using pure variants alone in this transformation~\cite{Bhattacharya1991}. We propose that the same compatibility mechanism underlies the formation of experimentally observed diamond microstructures. Within this framework, the experimentally observed diamond microstructures may be interpreted as laminate realizations of the sheared Landau states developed here.

\begin{figure}[hbt!]
    \centering
    \includegraphics[scale=0.40]{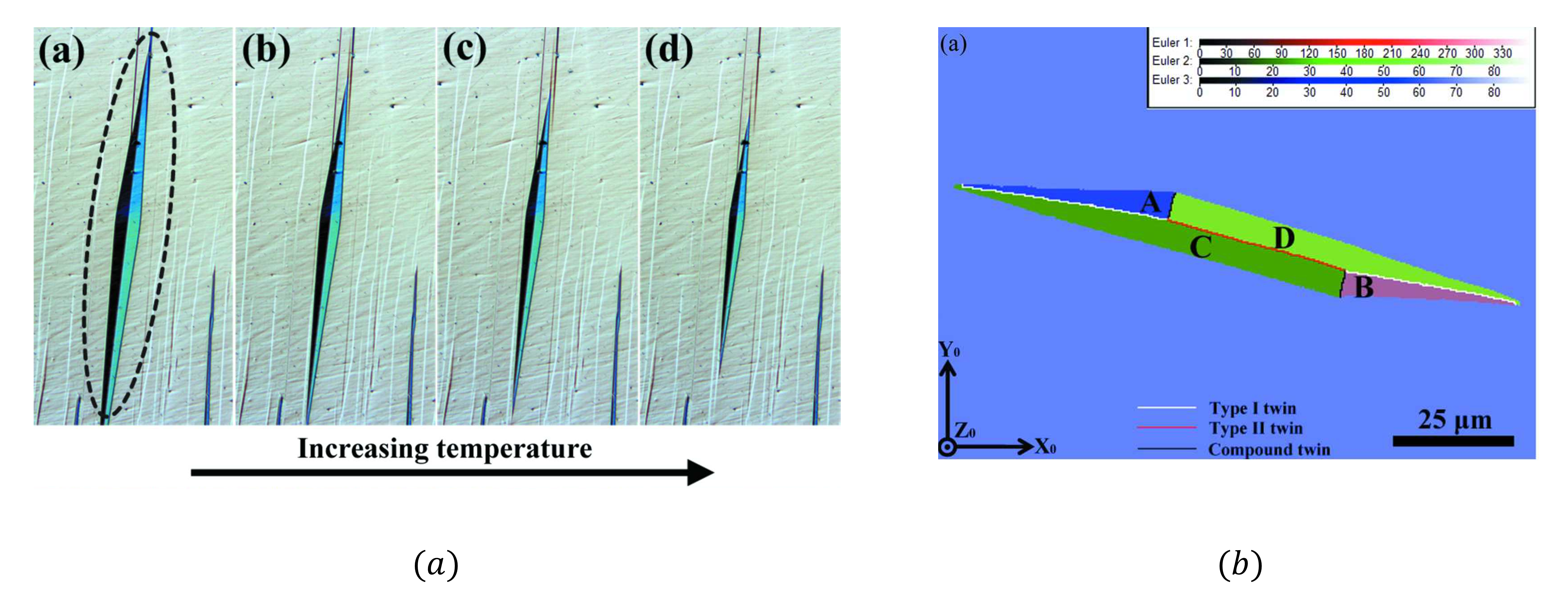}
    \caption{$(a)$ Optical micrographs acquired under polarized light showing the microstructural evolution of a diamond-shaped martensite nucleus embedded in the austenite matrix with increasing temperature. $(b)$ EBSD all-Euler orientation map of a diamond-shaped martensite nucleus consisting of four distinct martensitic variants (A--D) embedded in the austenite matrix. Reproduced from \cite{Li2019}.}
    \label{fig:Li_et_al}
\end{figure}

Of particular relevance to the present work, Li et al.~\cite{Li2019} investigated the crystallography of diamond-shaped martensitic nucleus in polycrystalline Ni$_{53}$Mn$_{22}$Ga$_{25}$ alloy using spatially resolved EBSD, determining the deformation gradients of the four constituent variants and identifying the transformation shear to be along the $[10\bar{1}]_A$ direction, see Figure~\ref{fig:Li_et_al}. Expressed in the crystallographic basis $\left(
[1\,0\,\bar{1}]_A,\;
[1\,0\,1]_A,\;
[0\,\bar{1}\,0]_A \right)$ employed by Li et al., the deformation gradient of the variant C is 

\[
\begin{pmatrix}
1.000883 & 0.093765 & 0.004193 \\
0        & 0.996474 & 0        \\
0        & 0.003796 & 1.002006
\end{pmatrix}.
\]

We compare this with the deformation gradient in \eqref{eq:generalized_shear}, corresponding to shear along $\bm{\hat{e}_1}$ direction and expressed in the standard Cartesian basis, since both deformation gradients are represented in bases aligned with their respective shear directions. The comparison shows that the experimentally determined deformation gradient exhibits the same characteristic simple-shear structure, with diagonal entries close to unity and a dominant shear contribution in the $(1,2)$ component, suggesting that the experimentally observed deformation may be interpreted as a small perturbation of the ideal simple-shear mechanism considered in \eqref{eq:generalized_shear}.

\begin{figure}[hbt!]
    \centering
    \includegraphics[scale=0.40]{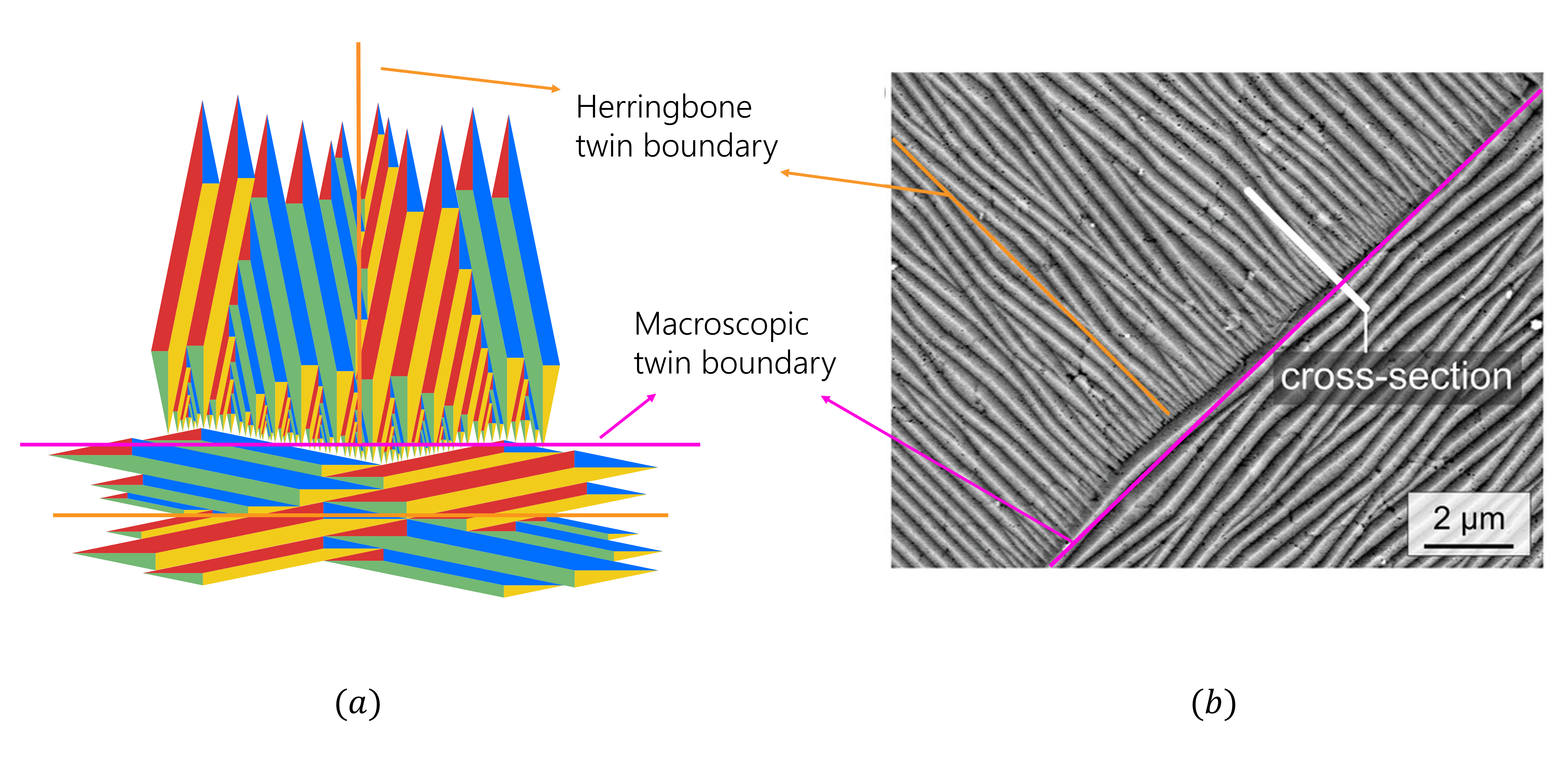}
    \caption{$(a)$ Schematic illustrating the hierarchical arrangement of self-accommodated herringbone laminates separated by a macroscopic twin boundary. The herringbone twin boundaries (orange) within each laminate and the macroscopic twin boundary (pink), formed between Landau states corresponding to different shear directions in the ⟨110⟩ family, are indicated. (b) SEM micrograph showing two herringbone laminates separated by a macroscopic twin boundary. Reproduced from \cite{Schwabe2021}.}
    \label{fig:schwabe_et_al}
\end{figure}

A second experimental observation relevant to the present work is provided by Schwabe et al~\cite{Schwabe2021}, who proposed a hierarchical building-block description of martensitic microstructures spanning from nanoscale to macroscale in Ni\textsubscript{48}Mn\textsubscript{33}Ga\textsubscript{19} thin films. In particular, they identified diamond-shaped nuclei as fundamental building blocks that coalesce into parallelogram nuclei and subsequently into herringbone laminates and macrotwins, thereby generating progressively more complex hierarchical morphologies, see Figure~\ref{fig:schwabe_et_al}$(b)$.

\begin{figure}[hbt!]
    \centering
    \includegraphics[scale=0.40]{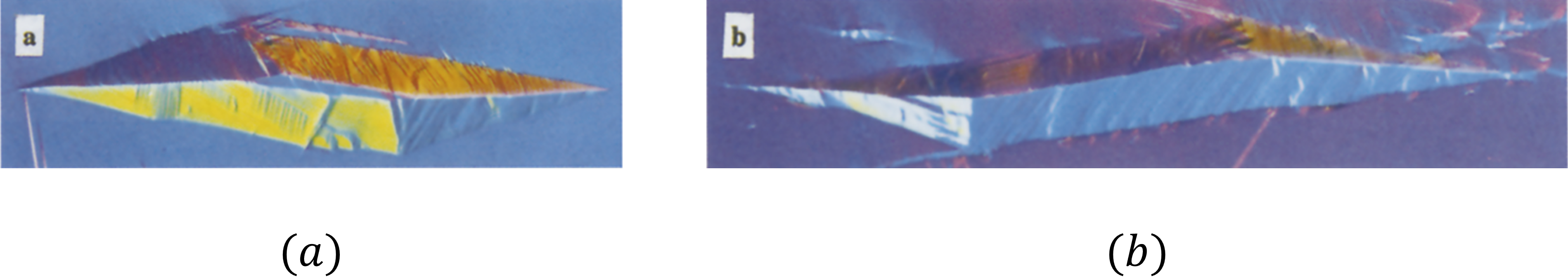}
    \caption{Optical micrographs showing parallelogram shaped martensitic nuclei embedded within the austenite matrix in a CuAlNi alloy. Reproduced from \cite{Tan1990Observations}.}
    \label{fig:CuAlNi}
\end{figure}

Within the present framework, such hierarchical morphologies may be interpreted as arising from the interaction of multiple sheared Landau states, with individual diamond nuclei evolving into parallelogram nuclei through either of the two admissible extension modes before combining to form larger hierarchical microstructures that undergo progressive refinement towards the macroscopic twin boundaries, see Figure~\ref{fig:schwabe_et_al}$(a)$.

Overall, agreement between the observed morphologies and the experimentally determined average deformation gradients suggests that the proposed compatibility mechanism provides a plausible explanation for the formation of diamond microstructures. Furthermore, the repeated observation of similar diamond morphologies across several material systems (see Figure~\ref{fig:CuAlNi}) supports the view that the proposed simple-shear mechanism underlies the formation of these microstructures. Collectively, these observations suggest that the sheared Landau states developed in this work provide a unified kinematic framework for understanding both the nucleation of diamond microstructures and their subsequent evolution into more complex hierarchical martensitic morphologies.

\section{Conclusion and Future work}
In this work, we established the planar  closure of the austenite--martensite interface through the construction of stress-free martensitic nuclei. To this end, we developed a crystallographic analogy between classical simple and pure shear deformations and martensitic phase transformations, identifying simple shear with the cubic to monoclinic-II transformation and pure shear with the cubic to orthorhombic transformation. The former satisfies the extreme compatibility conditions, allowing both Type-I/II and compound twins to form stress-free triple clusters with austenite.Our analysis indicates that triple clusters, in the form guaranteed by the extreme compatibility conditions, provide the mechanism by which the austenite--martensite interface is closed in the plane normal to the shear direction. Triple clusters enable the construction of a four-variant elastic Landau state, analogous to the flux-closure Landau state in micromagnetics, representing a stress-free martensitic nucleus enclosed by austenite. Whereas the cofactor conditions permit zero elastic energy nucleation of austenite within martensite along a lower-dimensional set, such as a line, the present construction establishes the analogous possibility for martensite nucleation within austenite.  A further generalization of this construction leads to three distinct two-parameter families of deformation gradients, corresponding to the shear directions $\langle100\rangle$, $\langle110\rangle$ and $\langle111\rangle$. These families give rise to elastic Landau states, sheared Landau states and kite microstructures, providing a geometric mechanism for stress-free martensitic nucleation.

The proposed nucleation mechanism also provides a natural crystallographic interpretation of experimentally observed diamond microstructures. We argued that while the idealized sheared Landau states are constructed from pure martensitic variants, the same compatibility mechanism may operate in real materials through compatible laminate averages generated by hierarchical twinning. The close agreement between experimentally determined average deformation gradients and the simple-shear family developed here, together with the repeated observation of diamond morphologies across several material systems, supports the hypothesis that simple shear is the fundamental geometric mechanism underlying the formation of diamond-shaped nuclei.

The present work also uncovers, to the best of our knowledge, the first example of non-conventional twins in a cubic to triclinic transformation. Cubic to triclinic transformations remain largely unexplored, partly due to the cumbersome algebra associated with the large number of variants. In the present work, the existence of these non-conventional compound twins is revealed through the construction of compound triple clusters that are simultaneously of the first and second kinds, a feature that cannot be realized by Type-I/II twins. A more detailed study of the crystallography of cubic to triclinic transformations and their admissible twinning systems warrants further investigation.

Finally, the analysis presented here is based entirely on crystallographic compatibility and does not account for energetic effects. An open question in the framework of nonlinear elasticity concerns the experimentally observed transition from diamond-shaped nuclei to parallelogram-shaped martensitic plates during growth. While the present work explains the crystallographic compatibility of both morphologies and establishes why four deformation gradients are required for stress-free nucleation whereas only two are sufficient for subsequent growth, it does not explain the energetic mechanism responsible for this morphological transition. Incorporating elastic and interfacial energy into the present framework may provide a criterion for the stability of the competing morphologies and determine the critical nucleus size at which the diamond-to-parallelogram transition becomes energetically favorable.

 A further open question is whether fully enclosed three-dimensional martensitic nuclei with stress-free austenite–martensite interfaces can exist. Such nuclei would require closure not only in the plane normal to the shear direction but also along the shear direction, including compatible end caps. The existence of such nuclei and the crystallographic compatibility conditions required for their construction remain problems for future investigation.

\section*{Declaration of competing interest}
The authors declare that they have no known competing financial interests or personal relationships that could have appeared to influence the work reported in this paper.

\section*{Acknowledgments}
The authors gratefully acknowledge the financial support provided by the Prime Minister’s Research Fellowship (PMRF ID: 0203695).

\newpage
\bibliography{references}

\newpage

\appendix

\section{Geometrically Non-Linear Theory of Martensite}
\label{app:glnm}

Let $\bm{\Omega \in \mathbb{R}^3} $ be a region occupied by a crystal in a reference configuration, when the crystal structure is austenite. As the crystal deforms upon phase transformation, a point $\bm{x \in \Omega}$ is displaced to $\bm{y(x) \in \mathbb{R}^3}$. The entire deformation of the crystal is given by a mapping $\bm{y: \Omega\rightarrow\mathbb{R}^3}$.\\ We assume that the helmholtz free energy density of the material ${\Phi}$ depends on the gradient of the deformation, $\bm{\nabla y(x) \in \mathbb{R}^{3\times3}_{+}}$ as well as the absolute temperature $\theta$. The total thermoelastic energy of the crystal is given by

\begin{equation}
\label{eq:1}
    \bm{\int_{\Omega}} \Phi(\bm{\nabla y(x)}, \theta) \,\bm{dx}.
\end{equation}

The free energy density $\Phi$ has two important invariance characteristics. First is that of the principle of frame indifference, i.e. for any rigid rotation $\bm{ Q \in SO(3)}$, we have 

\begin{equation}
\label{eq:2}
\Phi(\bm{Q\nabla y(x)}, \theta) = \Phi(\bm{\nabla y(x)}, \theta).
\end{equation}

The second invariance comes from the symmetry of the crystalline material. We restrict ourselves to rotational symmetries only i.e. the symmetries which belong to the Ericksen-Pitteri neighborhood \cite{pitteri_zanzotto}. If $ \bm{R \in SO(3)}$ is a symmetry operation for crystal, then

\begin{equation}
\label{eq:3}
\Phi(\bm{\nabla y(x) R}, \theta) = \Phi(\bm{\nabla y(x)}, \theta).
\end{equation}

We can combine \eqref{eq:2} and \eqref{eq:3} to write

\begin{equation}
\label{eq:4}
\Phi(\bm{R^T\nabla y(x) R}, \theta) = \Phi(\bm{\nabla y(x)}, \theta).
\end{equation}

The free energy density $\Phi$ defined on $\bm{\Omega}$ can be related to the lattice level deformations using the Cauchy-Born rule \cite{ericksen1984cauchy}. Therefore, we can understand $\bm{\nabla y(x)}$ as the gradient of the deformation which maps the lattice of austenite to that of the martensite. The symmetry of austenite and martensite lattices then requires that there is a set of transformation stretch matrices $\bm{U}_1, \bm{U}_2,\cdots, \bm{U}_N$ $\bm{\in \mathbb{R}^{3\times3}_{+sym}}$ that are related by symmetry $\bm{U}_i = \bm{R}_i\bm{^TU_1R}_i,\;i=1,2,....N,$ where $\bm{R}_i\bm{ \in \mathscr{P}(a)}$, the point group of austenite. $\bm{U}_1, \bm{U}_2,\cdots, \bm{U}_N$ are called the variants of martensite and $N$ is given by the ratio of orders of the point groups of austenite and martensite $\bm{\mathscr{P}(a)/\mathscr{P}(m)}$. We assume that the dependence of $\bm{U}_i$ on temperature is small enough to neglected. $\bm{U}_i$ are then the energy wells of the martensite i.e. if $\theta_c$ is the transformation temperature, we have

\begin{equation}
\label{eq:5}
\Phi(\bm{U}_1, \theta) = \cdots = \Phi(\bm{U}_N, \theta) \;\leq\; \Phi(\bm{F}, \theta),\quad \theta \leq \theta_c\:,
\end{equation}

$\bm{F}$ is any element of $\bm{\mathbb{R}^{3\times3}}$. At the transformation temperature, $\theta = \theta_c$, austenite and martensite equally minimize the energy i.e. $\bm{I}$ is also a minimizer of free energy. For temperatures above the transformation temperature, only austenite minimizes the energy.

\begin{equation}
\label{eq:6}
\Phi(\bm{U}_1, \theta) = \cdots = \Phi(\bm{U}_N, \theta)=\Phi(\bm{I},\theta) \;\leq\; \Phi(\bm{F}, \theta),\quad \theta = \theta_c\:,
\end{equation}

\begin{equation}
\label{eq:7}
\Phi(\bm{I}, \theta)\;\leq\; \Phi(\bm{F}, \theta),\quad \theta \geq \theta_c.
\end{equation}

Using the principle of frame indifference, we can construct the full set of minimizers of free energy $\Phi$ at the transformation temperature $\theta_c$ as

\begin{equation}
\label{eq:8}
\bm{SO(3)U}_1\bm{\cup SO(3)U}_2 \cdots \bm{\cup SO(3)U}_N \bm{\cup SO(3)I}.
\end{equation}
\bigskip

\subsection{Kinematic Compatibility}

\bigskip
\label{app:kin_comp}

The general equation of compatibility between two distinct variants of martensite $\bm{U}$ and $\bm{V}$ is written as 
\begin{equation}
\label{eq:9}
   \bm{ QV - U = a \otimes \hat{n}}.
\end{equation}

In \eqref{eq:9}, $\bm{U,V \in \mathbb{R}^{3\times3}_{+\scriptstyle{sym}}} $ are known and the equation needs to be solved for  $\bm{Q \in SO(3)}$, $\bm{a \in \mathbb{R}^3}$ and $\bm{\hat{n} \in S^2 }$ that satisfy \eqref{eq:9}. The following classic result by Ball and James (1987) provides the solution this equation.

\begin{theorem}
    \label{thm: main_solution}
    (Ball and James \cite{Ball1987}, Bhattacharya \cite{bhattacharya2003microstructure}) Given matrices $\bm{F}$, $\bm{G}$ $\bm{\in \mathbb{R}^{3\times3}_+}$, such  that $\bm{C=G^{-T}F^{T}FG^{-1} \neq I},$ and the eigenvalues of $\bm{C},\:\lambda_i$ satisfy ${\lambda_1 \leq1, \;\lambda_2=1,\;\lambda_3\geq1}$, then for $\bm{Q\in SO(3)}$, there are exactly two solutions to the equation $\bm{QF-G = a \otimes \hat{n}}$, given by

\begin{subequations}\label{eq:BJSolutions}
\begin{equation}
    \bm{a} = \rho\left(\sqrt{\frac{\lambda_3(1-\lambda_1)}{\lambda_3 - \lambda_1}} \bm{\hat{u}}_1\; +\; \kappa \sqrt{\frac{\lambda_1(\lambda_3 -1)}{\lambda_3 - \lambda_1}}\bm{\hat{u}}_3 \right),
\end{equation}

\begin{equation}
\label{eq:10}
    \bm{\hat{n}} = \frac{\sqrt{\lambda_3} - \sqrt{\lambda_1}}{\rho\sqrt{\lambda_3 - \lambda_1}}\left(-\sqrt{1-\lambda_1}\:\bm{G^T\hat{u}}_1\; +\; \kappa \sqrt{\lambda_3 -1}\:\bm{G^T\hat{u}}_3 \right).
\end{equation}
\end{subequations}

where ${\kappa = \pm1}$, $\rho\neq0$ is chosen to make $\bm{\left|{\hat{n}}\right|} =1$  and $\bm{\hat{u}}_i$ are the normalized eigenvectors of $\bm{C}$ corresponding to the eigenvalues $\lambda_i$. 
\end{theorem} 

For each choice of $\kappa\in\{1,-1\}$, $\bm{Q}$ can be determined since $\bm{F}$ and $\bm{G}$ are known.

\begin{corollary}
    
\label{corollary:identity_solutions}
Let $\bm{F = F^T = U}$ be a variant of martensite and $\bm{G=I}$ in the theorem above. If the eigenvalues of $\bm{U}$ satisfy $0<\lambda_1 \leq1, \;\lambda_2=1,\;\lambda_3\geq1$, then the solutions to $\bm{QU-I = a\otimes \hat{n}}$ are given by 

\begin{subequations}\label{eq:AMinterface}

\begin{equation}
    \bm{a} = \frac{\rho}{\sqrt{\lambda_3^2 - \lambda_1^2}}\left(\lambda_3\sqrt{1-\lambda_1^2}\;\bm{\hat{u}}_1\; +\; \kappa\; \lambda_1\sqrt{\lambda_3^2 -1}\;\bm{\hat{u}}_3 \right),
\end{equation}

\begin{equation}
\label{eq:11}
    \bm{\hat{n}} = \frac{\lambda_3 - \lambda_1}{\rho\sqrt{\lambda_3^2 - \lambda_1^2}}\left(-\sqrt{1-\lambda_1^2}\;\bm{\hat{u}}_1\; +\; \kappa \sqrt{\lambda_3^2 -1}\;\bm{\hat{u}}_3 \right).
\end{equation}

\end{subequations}

where $\bm{\hat{u}}_i$ are the normalized eigenvectors of $\bm{U}$ corresponding to the eigenvalues $\lambda_i$.\eqref{eq:AMinterface} gives the solutions of two possible interfaces between a single variant of martensite and austenite.
\end{corollary} 

\begin{theorem}
\label{thm:Mallards}
    (Gurtin \cite{Gurtin1983}, Bhattacharya \cite{bhattacharya2003microstructure}) Let $\bm{R}$ be a $180^{\circ}$ rotation about some axis $\bm{\hat{e}}$ and the matrices $\bm{F}$ and $\bm{G}$ as in \autoref{thm: main_solution} satisfy
\begin{enumerate}
    \item $\bm{F = QGR}$ for some rotation $\bm{Q \in SO(3)},$
    \item $\bm{F^T F \neq G^TG.}$
\end{enumerate}
Then there are two solutions to the equation $\bm{QF-G=a\otimes \hat{n}}$, given by

\begin{subequations}
\begin{equation}
    \bm{a_I} = 2\left(\frac{\bm{G^{-T}\hat{e}}}{\bm{\left|G^{-T}\hat{e}\right|^2}} - \bm{G\hat{e}}\right), \quad\quad \bm{\hat{n}_I = \hat{e}}.
\end{equation}

\begin{equation}
\label{eq:12}
    \bm{a_{II}} = \rho \bm{G\hat{e}}, \quad\quad  \bm{\hat{n}_{II}}=\frac{2}{\rho}\left(\bm{\hat{e} - \frac{G^TG\hat{e}}{\left|G\hat{e}\right|^2}}\right).
\end{equation}
\end{subequations}

\end{theorem} 

\begin{corollary}
\label{corollary:type12_solutions}
If $\bm{U}$ and $\bm{V}$ in \eqref{eq:9} satisfy $\bm{V = (-I + 2\;\hat{e}\otimes \hat{e})U(- I + 2\;\hat{e}\otimes \hat{e})}$, then by \autoref{thm:Mallards}, we have

\begin{subequations}\label{eq:TypeI/II_Solns}

\begin{equation}
     \bm{a_I} = 2\left(\bm{\frac{U^{-1}\hat{e}}{\left|U^{-1}\hat{e}\right|^2} - U\hat{e}}\right), \quad\quad \bm{\hat{n}_I = \hat{e}},
\end{equation}

\begin{equation}
\label{eq:13}
    \bm{a_{II}} = \rho \bm{U\hat{e}}, \quad\quad  \bm{\hat{n}_{II}}=\frac{2}{\rho}\bm{\left(\hat{e} - \frac{U^2\hat{e}}{\left|U\hat{e}\right|^2}\right)}.
\end{equation}
\end{subequations}

where $\rho\neq0$ is chosen to make $\bm{\left|{\hat{n}_{II}}\right| =}1$. In this case $(\bm{a_I,\hat{n}_1})$ generate a Type I solution and $(\bm{a_{II},\hat{n}_{II}})$ generate a Type II solution.
\end{corollary}

\begin{theorem}
\label{thm:compound_domains}
    (Chen et al \cite{CHEN20132566}, Proposition 1)  Let $\bm{U \in \mathbb{R}^{3\times3}_{\scriptstyle{+sym}}}$ be a variant of martensite and define $\bm{V = (-I + 2\;\hat{e}_1\otimes \hat{e}_1)U(- I + 2\;\hat{e}_1\otimes \hat{e}_1)}$ such that $\bm{V \neq U}$. There is a second unit vector $\bm{\hat{e}_2 \nparallel \hat{e}_1}$, satisfying $\bm{V = (-I + 2\;\hat{e}_2\otimes \hat{e}_2)U(- I + 2\;\hat{e}_2\otimes \hat{e}_2)}$ if and only if $\bm{\hat{e}_1}$ is perpendicular to an eigenvector of $\bm{U}$. Furthermore, $\bm{\hat{e}_1,\hat{e}_2}$ and that eigenvector(normalized) form an orthonormal basis of $\bm{\mathbb{R}^3}$. In this case $\bm{U}$ and $\bm{V}$ are said to from compound \textit{domains} and the solutions to \eqref{eq:9} are given by

    \begin{subequations}
        
\begin{equation}
    \bm{a^c_I} = \zeta\: \bm{U\hat{e}}_2,    \quad\quad  \bm{\hat{n}^c_1 = \hat{e}_1}, \quad\quad \text{where}\;\; \zeta = 2\bm{\frac{\hat{e}_2.{U_1}^{-2}\hat{e}_1}{\hat{e}_1.{U_1}^{-2}\hat{e}_1}} \:,
\end{equation}

\begin{equation}
\label{eq:14}
    \bm{a^c_{II}} = \eta\: \bm{U\hat{e}}_1,    \quad\quad  \bm{\hat{n}^c_1 = \hat{e}_2}, \quad\quad \text{where}\;\; \eta= -2\bm{\frac{\hat{e}_2.{U_1}^{2}\hat{e}_1}{\hat{e}_1.{U_1}^{2}\hat{e}_1}}.
\end{equation}

\end{subequations}

\end{theorem}

Chen et al.~\cite{CHEN20132566} showed that the general solutions to the compatibility equation between martensitic variants \eqref{eq:9} are classified as Type I/II domains or compound domains. This classification extends the notion of crystallographic twins to the broader class of domains by allowing the $180^\circ$ rotations relating compatible variants to be non-crystallographic.

\begin{theorem}
\label{thm:Type_I_II_domains_check}
(Chen et al \cite{CHEN20132566}, Proposition 12)
Let $\bm{A}=\bm{A^T}$ and $\bm{B}=\bm{B^T}$ be $3\times3$ positive-definite matrices satisfying
\[
\bm{B}=\bm{R}\bm{A}\bm{R^T}
\]
for some $\bm{R\in O(3)}$. Suppose $\bm{A}$ and $\bm{B}$ are compatible in the sense that there is a matrix $\bm{Q\in SO(3)}$ such that
\begin{equation}
\bm{Q}\bm{B}-\bm{A}=\bm{a}\otimes\bm{n},
\label{A.1}
\end{equation}
where $\bm{a},\bm{n}\in\mathbb{R}^3$. Then there is a unit vector $\hat{\bm{e}}\in\mathbb{R}^3$ such that
\begin{equation}
\bm{B}
=
\left(
-\bm{I}
+
2\hat{\bm{e}}\otimes\hat{\bm{e}}
\right)
\bm{A}
\left(
-\bm{I}
+
2\hat{\bm{e}}\otimes\hat{\bm{e}}
\right).
\label{A.2}
\end{equation}

Conversely, if $3\times3$ matrices $\bm{A}$ and $\bm{B}$ satisfy \eqref{A.2} for some unit vector $\hat{\bm{e}}$, then there is $\bm{Q \in SO(3)}$ so that \eqref{A.1} is satisfied. A formula for $\hat{\bm{e}}$ can be given as follows. Under the hypotheses, there is an orthonormal basis $\{\bm{e_1},\bm{e_2},\bm{e_3}\}$ such that
\begin{equation}
\bm{A^{-1}}\bm{B^2}\bm{A^{-1}}
=
\mu_1\,\bm{e_1}\otimes\bm{e_1}
+
\bm{e_2}\otimes\bm{e_2}
+
\mu_3\,\bm{e_3}\otimes\bm{e_3},
\label{A.3}
\end{equation}
where $0<\mu_1\le1\le\mu_3$ and
\begin{equation}
\mu_1\mu_3=1,\qquad
\bm{e_1}\cdot\bm{A^2}\bm{e_1}
=
\mu_3\,\bm{e_3}\cdot\bm{A^2}\bm{e_3},
\qquad
(\bm{e_2}\cdot\bm{A^2}\bm{e_1})^2
=
\mu_3(\bm{e_2}\cdot\bm{A^2}\bm{e_3})^2.
\label{A.4}
\end{equation}

If $\mu_3>1$, all unit vectors $\hat{\bm{e}}$ satisfying \eqref{A.2} are given by
\begin{equation}
\hat{\bm{e}}
=
\pm
\left(
\delta_1\bm{A}\bm{e_1}
+
\delta_3\bm{A}\bm{e_3}
\right),
\label{A.5}
\end{equation}
where
\begin{equation}
\delta_1
=
\left(
2\left(
\bm{e_1}\cdot\bm{A^2}\bm{e_1}
+
s\sqrt{\mu_3}\,
\bm{e_3}\cdot\bm{A^2}\bm{e_1}
\right)
\right)^{-1/2},
\qquad
\delta_3
=
s\sqrt{\mu_3}\,\delta_1,
\label{A.6}
\end{equation}
and $s\in\{\pm1\}$ satisfies
\[
s\sqrt{\mu_3}
\left(
\bm{e_2}\cdot\bm{A^2}\bm{e_3}
\right)
=
-
\bm{e_2}\cdot\bm{A^2}\bm{e_1}.
\]
In the case $\mu_3=1$ necessarily $\bm{B}=\bm{A}$ and $\hat{\bm{e}}$ can be chosen as a normalized eigenvector of $\bm{A}$.
\end{theorem}

\begin{corollary}
\label{thm:compound_domains_check}
(Chen et al \cite{CHEN20132566}, Corollary 13) Assume the hypotheses of Proposition~12. There are two unit vectors
$\hat{\bm{e}}_+\nparallel\hat{\bm{e}}_-$ satisfying \eqref{A.2} if and only if
\begin{equation}
\bm{e_2}\cdot\bm{A^2}\bm{e_3}
=
\bm{e_2}\cdot\bm{A^2}\bm{e_1}
=
0.
\label{A.25}
\end{equation}

If \eqref{A.25} is satisfied and $\mu_3>1$, there are precisely two such nonparallel unit vectors (up to a premultiplied $\pm$) satisfying \eqref{A.2}, and these vectors are orthonormal,
\[
\hat{\bm{e}}_+\cdot\hat{\bm{e}}_-=0.
\]
They are given by
\begin{equation}
\hat{\bm{e}}_\sigma
=
\delta_1^\sigma\bm{A}\bm{e_1}
+
\delta_3^\sigma\bm{A}\bm{e_3},
\qquad
\sigma=\pm,
\label{A.26}
\end{equation}
where
\begin{equation}
\delta_1^\sigma
=
\left(
2\left(
\bm{e_1}\cdot\bm{A^2}\bm{e_1}
+
\sigma\sqrt{\mu_3}\,
\bm{e_3}\cdot\bm{A^2}\bm{e_1}
\right)
\right)^{-1/2},
\qquad
\delta_3^\sigma
=
\sigma\sqrt{\mu_3}\,\delta_1^\sigma,
\qquad
\sigma=\pm.
\label{A.27}
\end{equation}

In the case $\mu_3=1$, necessarily $\bm{B}=\bm{A}$ and the solutions $\hat{\bm{e}}$ of \eqref{A.2} consist of unit vectors in the eigenspace of $\bm{A}$.
\end{corollary}

\bigskip

\subsection{Crystallographic theory of martensite}
\label{app:ptmc}

\bigskip

The crystallographic theory of martensite explains the formation of the interface between austenite and a finely twinned laminate of martensite. The twinned laminate is understood as a sequence of deformations $\bm{y}^{(k)}, k = 1,2,3 ,....., $ of the two twin variants, such that
\begin{equation}
\label{eq:15}
    \bm{\int_{\Omega}}\Phi(\bm{\nabla y}^{(k)}\bm{(x)},\theta)\bm{dx} \to 0 \;\;\text{as}\;\; k\bm{\to}\infty,
\end{equation}
where $k$ is related to the inverse of the width of the interpolation layer between austenite and martensite. Ball and James \cite{Ball1987} show that as $k\bm{\to}\infty$, the sequence $\bm{y}^{(k)}\bm{(x)}$ converges to a deformation $\bm{y(x)}$ such that
\begin{equation}
\label{eq:16}
    \bm{\nabla y} = \mu \bm{(QV)} \bm{+ (}1\bm{-}\mu\bm{)U }= \mu \bm{(U + a\otimes \hat{n}) + (}1\bm{-}\mu)\bm{U}.
\end{equation}

Then assuming \eqref{eq:9} holds, the equation of the crystallographic theory of martensite is written as
\begin{equation}
\label{eq:17}
    \bm{\tilde{Q}}[\mu \bm{(U + a\otimes \hat{n})} \bm{+ (}1-\mu\bm{)U}] \bm{- I} = \bm { b\otimes \hat{m}},
\end{equation}

which is to be solved for $0\leq\mu\leq1$, the volume fraction of the laminate, $\bm{\tilde{Q}\in SO(3)}$ and vectors $\bm{b}$ and $\bm{\hat{m}}$.

\section{Transformation matrices}

\label{app:matrices}

\subsection{Cubic to tetragonal transformation}
\label{app:cubictetra}
\bigskip
The cubic to tetragonal transformation is characterized by the lattice correspondence, $\langle100\rangle_{\text{cubic}} \to \langle100\rangle_{\text{tetragonal}}$. This transformation admits three martensitic variants with stretch tensors

\begin{gather}
\label{eq:18}
\bm{U}_1 = 
    \begin{bmatrix}
        a & 0& 0 \\ 0 & a & 0 \\ 0 & 0 & b
    \end{bmatrix},\qquad
    \bm{U}_2 = 
    \begin{bmatrix}
        a & 0& 0 \\ 0 & b & 0 \\ 0 & 0 & a
    \end{bmatrix},\qquad
    \bm{U}_3 = 
    \begin{bmatrix}
        b & 0& 0 \\ 0 & a & 0 \\ 0 & 0 & a
    \end{bmatrix},
\end{gather}

where, 
\begin{equation*}
b =\frac{c_t}{a_c}, \quad \quad a =\frac{a_t}{a_c}.
\end{equation*}

All the pairs of variants $\bm{(U_1, U_2)}$, $\bm{(U_1, U_3)}$ and $\bm{(U_2, U_3)}$ form compound twins.

\bigskip

\subsection{Cubic to orthorhombic transformation}
\label{app:cubicortho}
\bigskip
In cubic to orthorhombic transformations, we have six variants in the martensitic phase. In this transformation, two of the three axes of orthorhombic symmetry are obtained from $\langle$110$\rangle$\textsubscript{cubic} axes and third axis coincides with the invariant $\langle100\rangle_\text{cubic}$ axis. The stretch tensors are of the form

\begin{equation}
\begin{aligned}
\bm{U}_1 &=
\begin{bmatrix}
d & 0 & 0\\
0 & a & b\\
0 & b & a
\end{bmatrix},
&
\bm{U}_2 &=
\begin{bmatrix}
d & 0 & 0\\
0 & a & -b\\
0 & -b & a
\end{bmatrix},
&
\bm{U}_3 &=
\begin{bmatrix}
a & 0 & -b\\
0 & d & 0\\
-b & 0 & a
\end{bmatrix},
\\[1em]
\bm{U}_4 &=
\begin{bmatrix}
a & 0 & b\\
0 & d & 0\\
b & 0 & a
\end{bmatrix},
&
\bm{U}_5 &=
\begin{bmatrix}
a & -b & 0\\
-b & a & 0\\
0 & 0 & d
\end{bmatrix},
&
\bm{U}_6 &=
\begin{bmatrix}
a & b & 0\\
b & a & 0\\
0 & 0 & d
\end{bmatrix},
\end{aligned}
\label{eq:19}
\end{equation}

where,

\begin{equation*}
    a =\frac{(a_o+c_o)}{\sqrt{2}a_c},\quad\quad b = \frac{(a_o-c_o)}{\sqrt{2}a_c}, \quad \quad d=\frac{b_o}{a_c}.
\end{equation*}

Variant pairs $(\bm{U}_1,\bm{U}_2),(\bm{U}_3,\bm{U}_4)$ and $(\bm{U}_5,\bm{U}_6)$ form compound twins. All other pairs form Type-I/II twins. 

\bigskip
\subsection{Cubic to monoclinic transformation}
\bigskip
\label{app:cubicmono}

\subsubsection{Cubic to monoclinic-I transformation}
\label{app:cubicmonoI}
\bigskip
For this transformation, we have twelve variants in the martensitic phase. Two different transformation pathways are possible here. In the first type of transformation, the axis of the monoclinic symmetry corresponds to $\langle110\rangle$\textsubscript{cubic} direction. This transformation is also called cubic to monoclinic-I transformation. The stretch tensors are of the form

\begin{equation}
\begin{aligned}
\bm{U}_1 &= \begin{bmatrix}
d & -c & -c\\
-c & a & b\\
-c & b & a
\end{bmatrix},
&
\bm{U}_2 &= \begin{bmatrix}
d & -c & c\\
-c & a & -b\\
c & -b & a
\end{bmatrix},
&
\bm{U}_3 &= \begin{bmatrix}
d & c & -c\\
c & a & -b\\
-c & -b & a
\end{bmatrix},
\\[1em]
\bm{U}_4 &= \begin{bmatrix}
a & c & -b\\
c & d & -c\\
-b & -c & a
\end{bmatrix},
&
\bm{U}_5 &= \begin{bmatrix}
a & c & b\\
c & d & c\\
b & c & a
\end{bmatrix},
&
\bm{U}_6 &= \begin{bmatrix}
d & c & c\\
c & a & b\\
c & b & a
\end{bmatrix},
\\[1em]
\bm{U}_7 &= \begin{bmatrix}
a & -b & c\\
-b & a & -c\\
c & -c & d
\end{bmatrix},
&
\bm{U}_8 &= \begin{bmatrix}
a & b & c\\
b & a & c\\
c & c & d
\end{bmatrix},
&
\bm{U}_9 &= \begin{bmatrix}
a & -c & b\\
-c & d & -c\\
b & -c & a
\end{bmatrix},
\\[1em]
\bm{U}_{10} &= \begin{bmatrix}
a & -c & -b\\
-c & d & c\\
-b & c & a
\end{bmatrix},
&
\bm{U}_{11} &= \begin{bmatrix}
a & -b & -c\\
-b & a & c\\
-c & c & d
\end{bmatrix},
&
\bm{U}_{12} &= \begin{bmatrix}
a & b & -c\\
b & a & -c\\
-c & -c & d
\end{bmatrix},
\end{aligned}
\label{eq:21}
\end{equation}

where,

\begin{equation*}
    \begin{aligned}
    d & =\frac{a_m(\sqrt{2}\;a_m\;+c_m\sin\beta)}{a_c\sqrt{2a_m^2\;+c_m^2\;+2\sqrt{2}a_mc_m\sin\beta}}, \\
    c &= \frac{a_mc_m\cos\beta}{\sqrt{2}a_c\sqrt{2a_m^2\;+c_m^2\;+2\sqrt{2}a_mc_m\sin\beta}},\\
    a & =\frac{1}{2\sqrt{2}a_c}\left(\frac{c_m(c_m +\sqrt{2}\;a_m\sin\beta)}{\sqrt{2a_m^2\;+c_m^2\;+2\sqrt{2}a_mc_m\sin\beta}}+b_m\right),\\
    b &=\frac{1}{2\sqrt{2}a_c}\left(\frac{c_m(c_m +\sqrt{2}\;a_m\sin\beta)}{\sqrt{2a_m^2\;+c_m^2\;+2\sqrt{2}a_mc_m\sin\beta}}-b_m\right).
    \end{aligned}
\end{equation*}

Each variant in \eqref{eq:21} forms compound twins with one other variant, Type-I/II twins with six variants and non-conventional,non-generic twins with the remaining four variants\cite{PITTERI1998225}.

\bigskip

\subsubsection{Cubic to monoclinic-II transformation}
\label{app:cubicmonoII}
\bigskip

For the second kind of cubic to monoclinic transformations, the axis of the monoclinic symmetry corresponds to $\langle100\rangle$\textsubscript{cubic} direction. This transformation is also called cubic to monoclinic-II transformation. The stretch tensors are of the form

\begin{equation}
\begin{aligned}
\bm{U}_1 &=
\begin{bmatrix}
d & 0 & 0\\
0 & a & b\\
0 & b & c
\end{bmatrix},
&
\bm{U}_2 &=
\begin{bmatrix}
d & 0 & 0\\
0 & a & -b\\
0 & -b & c
\end{bmatrix},
&
\bm{U}_3 &=
\begin{bmatrix}
a & 0 & -b\\
0 & d & 0\\
-b & 0 & c
\end{bmatrix},
&
\bm{U}_4 &=
\begin{bmatrix}
a & 0 & b\\
0 & d & 0\\
b & 0 & c
\end{bmatrix},
\\[1em]
\bm{U}_5 &=
\begin{bmatrix}
d & 0 & 0\\
0 & c & b\\
0 & b & a
\end{bmatrix},
&
\bm{U}_6 &=
\begin{bmatrix}
c & -b & 0\\
-b & a & 0\\
0 & 0 & d
\end{bmatrix},
&
\bm{U}_7 &=
\begin{bmatrix}
c & b & 0\\
b & a & 0\\
0 & 0 & d
\end{bmatrix},
&
\bm{U}_8 &=
\begin{bmatrix}
a & b & 0\\
b & c & 0\\
0 & 0 & d
\end{bmatrix},
\\[1em]
\bm{U}_9 &=
\begin{bmatrix}
c & 0 & b\\
0 & d & 0\\
b & 0 & a
\end{bmatrix},
&
\bm{U}_{10} &=
\begin{bmatrix}
c & 0 & -b\\
0 & d & 0\\
-b & 0 & a
\end{bmatrix},
&
\bm{U}_{11} &=
\begin{bmatrix}
a & -b & 0\\
-b & c & 0\\
0 & 0 & d
\end{bmatrix},
&
\bm{U}_{12} &=
\begin{bmatrix}
d & 0 & 0\\
0 & c & -b\\
0 & -b & a
\end{bmatrix}.
\end{aligned}
\label{eq:22}
\end{equation}

The explicit expressions of $a, b, c,$ and $d$ are too lengthy to be included here. Each variant in \eqref{eq:22} forms compound twins with two other variants, Type-I/II twins with four variants,  non-conventional but generic twins with one variant and non-conventional, non-generic twins with the remaining four variants\cite{PITTERI1998225}.

\section{Crossing twins}
\label{app:crossing_twins}

Consider four martensitic variants
\[
\bm{U_I},\quad \bm{U_J},\quad \bm{U_K},\quad \bm{U_L}
\]
forming a four-fold junction. Let
\(\bm {Q_1, Q_2,Q_3\in SO(3)}\). Kinematic compatibility across the four interfaces requires
\begin{equation}
\begin{aligned}
\bm{Q_1U_J-U_I}
&=\bm{b_1\otimes\hat{n}_1},\\
\bm{Q_1Q_2U_K-Q_1U_J}
&=\bm{b_2\otimes\hat{n}_2},\\
\bm{Q_1Q_2Q_3U_L-Q_1Q_2U_K}
&=\bm{b_3\otimes\hat{n}_3},\\
\bm{U_I-Q_1Q_2Q_3U_L}
&=\bm{b_4\otimes\hat{n}_4}.
\end{aligned}
\label{app:crossing_rank_one}
\end{equation}
Moreover, since the four interfaces meet along a common line,
\begin{equation}
\bm{\hat{n}_1,\hat{n}_2,\hat{n}_3,\hat{n}_4}
\quad\text{are coplanar}.
\label{app:crossing_coplanar}
\end{equation}

Summing the four rank-one relations in
\eqref{app:crossing_rank_one} gives the additional closure condition
\begin{equation}
\bm{b_1\otimes\hat{n}_1}+\bm{b_2\otimes\hat{n}_2}+\bm{b_3\otimes\hat{n}_3}+\bm{b_4\otimes\hat{n}_4}=\bm{0}.
\label{app:crossing_closure}
\end{equation}
Thus, pairwise rank-one compatibility of the four interfaces is not
sufficient for a compatible four-fold junction; the closure condition
\eqref{app:crossing_closure} is additionally required to avoid a
dislocation at the crossing.

\section{Triplet condition}
\label{app:triplet_condition}

We recall the triplet condition (TC) for three distinct symmetry-related martensitic variants~\cite{DellaPorta2022Triplet}. Let
\[
\mathbf U,\mathbf V,\mathbf W\bm{\in\mathbb R_{\mathrm{sym}+}^{3\times3}}
\]
be three distinct variants. Let
\[
\mathbf R_{uv},\mathbf R_{uw},\mathbf R_{vw}\in \bm{SO(3)},
\qquad
\mathbf m_{uv},\mathbf m_{uw},\mathbf m_{vw}\in\mathbb R^3\setminus\{\mathbf 0\},
\]
and
\[
\mathbf b_{uv},\mathbf b_{uw},\mathbf b_{vw}\in\mathbb R^3\setminus\{\mathbf 0\}
\]
satisfy
\begin{equation}
\begin{aligned}
\mathbf R_{uv}\mathbf V
&=\mathbf U+\mathbf b_{uv}\otimes\mathbf m_{uv},\\
\mathbf R_{uw}\mathbf W
&=\mathbf U+\mathbf b_{uw}\otimes\mathbf m_{uw},\\
\mathbf R_{vw}\mathbf W
&=\mathbf V+\mathbf b_{vw}\otimes\mathbf m_{vw}.
\end{aligned}
\label{app:rank_one_connections}
\end{equation}
We say that the triplet $(\mathbf U,\mathbf V,\mathbf W)$ satisfies the \emph{triplet condition (TC)} if
\begin{equation}
\mathbf R_{uv}\mathbf R_{vw}\mathbf R_{uw}^{T}
=\mathbf I.
\label{app:TC}
\end{equation}
The condition is a property of the triplet and is independent of its ordering.

The following lemma determines the possible types of the three pairwise twins.

\begin{lemma}
\label{lem:triplet}
Let $\mathbf U,\mathbf V,\mathbf W$ satisfy TC. Then, whenever
\[
\mathbf b_{uv}\parallel\mathbf b_{uw}
\parallel\mathbf R_{uv}\mathbf b_{vw},
\]
at least two of the three twins
\[
(\mathbf U,\mathbf b_{uv},\mathbf m_{uv}),\qquad
(\mathbf U,\mathbf b_{uw},\mathbf m_{uw}),\qquad
(\mathbf V,\mathbf b_{vw},\mathbf m_{vw})
\]
are type I twins. Conversely, whenever
\[
\mathbf m_{uv}\parallel\mathbf m_{uw}\parallel\mathbf m_{vw},
\]
at least two of the three twins are type II twins.

Consequently, the conditions
\[
\mathbf b_{uv}\parallel\mathbf b_{uw}
\parallel\mathbf R_{uv}\mathbf b_{vw}
\qquad\text{and}\qquad
\mathbf m_{uv}\parallel\mathbf m_{uw}\parallel\mathbf m_{vw}
\]
cannot hold simultaneously for three distinct variants. Thus, the triplets satisfying TC fall into four cases.
\end{lemma}

The four possible cases are as follows.

\paragraph{(TC I): Three type I twins}
All three pairwise twins are type I. In this case,
\begin{equation}
\boxed{
\mathbf b_{uv}\parallel\mathbf b_{uw}
\parallel\mathbf R_{uv}\mathbf b_{vw},
\qquad
(\mathbf m_{uv}\times\mathbf m_{uw})\cdot\mathbf m_{vw}=0.
}
\label{app:TCI}
\end{equation}

\paragraph{(TC Ih): Two type I twins and one type II or compound twin}
Two of the three pairwise twins are type I, while the remaining twin is type II or compound. In this case,
\begin{equation}
\boxed{
\mathbf b_{uv}\parallel\mathbf b_{uw}
\parallel\mathbf R_{uv}\mathbf b_{vw},
\qquad
(\mathbf m_{uv}\times\mathbf m_{uw})\cdot\mathbf m_{vw}=0.
}
\label{app:TCIh}
\end{equation}

\paragraph{(TC II): Three type II twins}
All three pairwise twins are type II. In this case,
\begin{equation}
\boxed{
\mathbf m_{uv}\parallel\mathbf m_{uw}\parallel\mathbf m_{vw},
\qquad
(\mathbf b_{uv}\times\mathbf b_{uw})\cdot\mathbf b_{vw}=0.
}
\label{app:TCII}
\end{equation}

\paragraph{(TC IIh): Two type II twins and one type I or compound twin}
Two of the three pairwise twins are type II, while the remaining twin is type I or compound. In this case,
\begin{equation}
\boxed{
\mathbf m_{uv}\parallel\mathbf m_{uw}\parallel\mathbf m_{vw},
\qquad
(\mathbf b_{uv}\times\mathbf b_{uw})\cdot\mathbf b_{vw}=0.
}
\label{app:TCIIh}
\end{equation}

Thus, TC imposes a strong restriction on the possible combinations of pairwise interfaces in a triplet: at least two of the three twins must be of the same type. The four possibilities are therefore TC I, TC Ih, TC II, and TC IIh.

 \section{Minors relations}
\label{app:minors}
For a microstructure to be kinematically admissible, the deformation gradients comprising it must satisfy the minors relations of Morrey~\cite{bhattacharya2003microstructure, Morrey1966}.
  
 \begin{theorem}
 \label{thm:morrey}
     \cite{bhattacharya2003microstructure, Morrey1966} Suppose we have a microstructure involving the matrices $\bm{F}_1,\bm{F}_2,\ldots, \bm{F}_M$ in the volume fraction $\mu_1,\mu_2,\ldots,\mu_M$ respectively. Suppose further that the microstructure satisfies the boundary condition corresponding to the matrix $\bm{F}$. Then the following are true

     \begin{equation}
         \begin{aligned}
             \bm{F} &= \sum_{m=1}^{M}\mu_m\bm{F}_m,\\
             \operatorname{cof}\bm{(F)} &= \sum_{m=1}^{M}\mu_m\:\operatorname{cof}(\bm{F}_m),\\
             \det\bm{(F)} &= \sum_{m=1}^{M}\mu_m \: \det(\bm{F}_m).           
         \end{aligned}
     \end{equation}
 \end{theorem}

\section{Vitali Covering Theorem}
\label{app:vitali}

\begin{theorem}[Vitali Covering Theorem]
Let $E\subset\mathbb{R}^n$ be a measurable set with
$\mu(E)<\infty$, where $\mu$ denotes the Lebesgue measure on
$\mathbb{R}^n$. A collection of sets $\mathcal{V}$ is called a
\emph{Vitali cover} of $E$ if, for every $x\in E$ and every
$\delta>0$, there exists a set $I\in\mathcal{V}$ such that
\[
x\in I
\qquad\text{and}\qquad
\operatorname{diam}(I)<\delta.
\]
If $\mathcal{V}$ is a Vitali cover of $E$, then there exists a
countable collection of pairwise disjoint sets
\[
I_1,I_2,\ldots\in\mathcal{V}
\]
such that
\[
\mu\!\left(E\setminus\bigcup_{i=1}^{\infty}I_i\right)=0.
\]
\end{theorem}

\end{document}